\documentstyle[preprint,aps,epsf,floats]{revtex}
\renewcommand{\baselinestretch}{1.5}
\begin{document}
\begin{titlepage} 
\renewcommand{\baselinestretch}{1}
\small\normalsize
\begin{flushright}
hep-th/0108040\\
MZ-TH/01-21     \\
\end{flushright}

\vspace{0.5cm}

\begin{center}   

{\LARGE \sc Ultraviolet Fixed Point and\\
 %[3mm]   
Generalized Flow Equation of\\[3mm] 
Quantum Gravity}

\vspace{1cm}
{\large O. Lauscher and M. Reuter}\\

\vspace{1cm}
\noindent
{\it Institute of Physics, University of Mainz\\
Staudingerweg 7, D-55099 Mainz, Germany}\\

\end{center}   
 
\vspace*{0.5cm}
\begin{abstract}
A new exact renormalization group equation for the effective average action of
Euclidean quantum gravity is constructed. It is formulated in terms of the 
component fields appearing in the transverse-traceless decomposition of the
metric. It facilitates both the construction of an appropriate infrared cutoff
and the projection of the renormalization group flow onto a large class of 
truncated parameter spaces. The Einstein-Hilbert truncation is investigated in
detail and the fixed point structure of the resulting flow is analyzed. Both
a Gaussian and a non-Gaussian fixed point are found. If the non-Gaussian fixed
point is present in the exact theory, quantum Einstein gravity is likely to be
renormalizable at the nonperturbative level. In order to assess the reliability
of the truncation a comprehensive analysis of the scheme dependence of 
universal quantities is performed. We find strong evidence supporting the 
hypothesis that 4-dimensional Einstein gravity is asymptotically safe, i.e.
nonperturbatively renormalizable. The renormalization group improvement of the
graviton propagator suggests a kind of dimensional reduction from 4 to 2 
dimensions when spacetime is probed at sub-Planckian length scales.
\end{abstract}
\end{titlepage}
%
%\newpage
\section{Introduction}
\renewcommand{\theequation}{1.\arabic{equation}}
\setcounter{equation}{0}
During the past decade, exact renormalization group equations \cite{bag}, in
particular in the context of the effective average action \cite{ber}, have
become a powerful tool for the investigation of nonperturbative phenomena
in both quantum field theory and in statistical physics. Those renormalization
group (RG), or flow equations may be regarded the counterpart for the
continuum of Wilson's renormalization group of iterated Kadanoff block spin
transformations which had been formulated for discrete spin systems originally
\cite{wilson}. In both cases the central idea is to ``integrate out'' all
fluctuations with momenta larger than some cutoff $k$, and to take account of
them by means of a modified dynamics for the remaining fluctuation modes with
momenta smaller than $k$. This modified dynamics is governed by a scale
dependent effective Hamiltonian or effective action, $\Gamma_k$, whose
$k$-dependence is described by a functional differential equation, the exact
RG equation.

In quantum field theory this general strategy can be applied to both
``effective'' and ``fundamental'' theories. By definition, an effective theory
is valid only if all relevant momenta of the process under consideration are
close to some specific scale $k$ which characterizes the theory. If $\Gamma_k$
is the action of an effective theory at scale $k$ we can compute cross
sections for the scattering of particles with momenta (or relevant momentum
transfers) of the order of $k$, with all quantum effects included, by simply
evaluating the {\it tree} diagrams of $\Gamma_k$. Exact RG equations
can be used in order to evolve $\Gamma_k$ to a smaller scale $k'<k$ by further
``coarse graining''.

Flow equations may also be used for a complete quantization of fundamental
theories. If the latter has the classical action $S$ one imposes the initial
condition $\Gamma_{\widehat{k}}=S$ at the ultraviolet (UV) cutoff scale 
$\widehat{k}$, uses
the exact RG equation to compute $\Gamma_k$ for all $k<\widehat{k}$, and then
sends $k\rightarrow 0$ and $\widehat{k}\rightarrow\infty$. Loosely speaking,
the defining property of a fundamental theory is that the ``continuum limit''
$\widehat{k}\rightarrow\infty$ actually exists after the ``renormalization''
- in the traditional sense of the word - of finitely many parameters in
the action; only a finite number of generalized couplings in $\Gamma_0$
is undetermined and has to be taken from the experiment. This is the case for
perturbatively renormalizable theories \cite{polch}, but there are also
perturbatively nonrenormalizable theories which admit a limit $\widehat{k}
\rightarrow\infty$. The ``continuum'' limit of those nonperturbatively
renormalizable theories is taken at a non-Gaussian fixed point of the RG
flow. It replaces the Gaussian fixed point which, at least implicitly,
underlies the construction of perturbatively renormalizable theories
\cite{bag}. Thus knowing its fixed point structure is crucial if one wants to
assess whether a given model qualifies as a {\it fundamental} theory.

In this paper we shall use a formulation where $\Gamma_k$ is the ``effective
average action'' \cite{ber}. It is a coarse grained free energy functional
which is constructed in close analogy with the standard effective action
$\Gamma$ to which it reduces in the limit of a vanishing infrared (IR) cutoff, 
$k\rightarrow 0$. The Euclidean functional integral for the generating
functional $W$ is modified by adding an IR cutoff term $\Delta_k S$ to the
classical action. It supplies a momentum dependent $({\rm mass})^2$-term
${\cal R}_k(p^2)$ for a mode of the quantum field with momentum $p$.
The cutoff function ${\cal R}_k(p^2)$ vanishes for $p^2\gg k^2$; hence the
high-momentum modes get integrated out in the usual way. For $p^2\ll k^2$ it
behaves as ${\cal R}_k(p^2)\propto k^2$ so that the small-momentum modes get
suppressed in the path integral by a mass term $\propto k^2$ \cite{ber}.
The scale dependent action $\Gamma_k$ is closely related to the
Legendre transform of the modified generating functional $W_k$. When regarded
as a function of $k$, $\Gamma_k$ runs along a RG trajectory in the space of
all actions which starts at $\Gamma_{\widehat{k}}=S$ and ends at $\Gamma_0=
\Gamma$. In the simplest case, the exact RG equation which describes this
trajectory has the following structure:
\begin{eqnarray}
\label{in1}
k\,\partial_k\Gamma_k=\frac{1}{2}\;{\rm Tr}\left[\left(\Gamma_k^{(2)}
+{\cal R}_k\right)^{-1}\,k\,\partial_k{\cal R}_k\right]\;.
\end{eqnarray}
Here $\Gamma_k^{(2)}$ denotes the infinite dimensional matrix of second
functional derivatives of $\Gamma_k$ with respect to all dynamical fields.

This construction is fairly straightforward for matter fields, the inclusion
of gauge fields introduces additional complications though. Using background
gauge techniques, a solution to this problem was given in refs. \cite{ym}
for Yang-Mills theory and in refs. \cite{Reu96,tif} for gravity. Leaving the
Faddeev-Popov ghosts aside, the effective average action for gravity,
$\Gamma_k[g_{\mu\nu},\bar{g}_{\mu\nu}]$, depends not only on the ``ordinary''
dynamical metric $g_{\mu\nu}$ but also on a background metric $\bar{g}_{\mu
\nu}$. The conventional effective action $\Gamma[g]$ is obtained as the
$k\rightarrow 0$ limit of the functional
$\Gamma_k[g]\equiv\Gamma_k[g,\bar{g}=g]$ with the two metrics identified
\cite{Ab81,back}. The motivation for this construction is that in this manner
$\Gamma_k[g]$ becomes invariant under general coordinate transformations.

Nonperturbative solutions to the above RG equation which do not
require a small expansion parameter can be obtained by the method of
``truncations''. This means that one projects the RG flow from the infinite
dimensional space of all actions onto some finite dimensional subspace which
is particularly relevant for the problem at hand. In this manner the
functional RG equation becomes an ordinary differential equation for a finite
set of generalized couplings which serve as coordinates on this subspace.
In ref. \cite{Reu96} the RG flow of quantum General Relativity was projected
on the 2-dimensional subspace spanned by the invariants $\int d^dx\sqrt{g}$
and $\int d^dx\sqrt{g}R$. This so-called Einstein-Hilbert truncation amounts
to considering only functionals of the form
\begin{eqnarray}
\label{in2}
\Gamma_k[g,\bar{g}]=\left(16\pi G_k\right)^{-1}\int d^dx\,\sqrt{g}\left\{
-R(g)+2\bar{\lambda}_k\right\}+\mbox{classical gauge fixing}.
\end{eqnarray}
Here $G_k$ and $\bar{\lambda}_k$ are the running Newton constant and 
cosmological constant, respectively. More general and, therefore, more 
precise truncations would include higher powers of the curvature tensor as 
well as nonlocal terms \cite{nonloc} which are not present classically.

Quantum gravity is certainly a particularly interesting topic where exact RG
equations can lead to important new insights. As quantized Einstein gravity
is perturbatively nonrenormalizable a natural option is to consider it an
effective field theory \cite{don}. Already within this setting quantum effects
can be studied in a consistent and predictive way. In fact, in refs.
\cite{bh1,bh2} the running couplings $G_k$ and $\bar{\lambda}_k$ obtained in
\cite{Reu96} were used to investigate how quantum gravity effects modify the
structure of black holes, and in \cite{cosmo} the implications for the
cosmology of the Planck era in the very early Universe were studied. Along
a different line of research it has been proposed \cite{tsamis} that there
are strong quantum gravitational effects also in the later stages of the
cosmological evolution which even might drive the cosmological constant to
zero dynamically; the effective average action would be an ideal tool for
exploring such infrared effects.

An even more intriguing possibility is that, despite its perturbative
nonrenormalizability, quantized gravity exists {\it nonperturbatively} as a
fundamental theory. It would then be mathematically consistent down to 
arbitrarily small length scales. A proposal along these lines is Weinberg's 
``asymptotic safety'' scenario \cite{wein}. It assumes that there exists a 
non-Gaussian RG fixed point at which the $\widehat{k}\rightarrow\infty$ limit 
can be taken, i.e. that the theory is ``nonperturbatively renormalizable'' in 
Wilson's sense. Asymptotic safety requires that the non-Gaussian fixed point is
UV attractive (i.e. attractive for $k\rightarrow\infty$) for finitely many
parameters in the action, i.e. that its UV critical hypersurface is finite
dimensional. This means that the RG trajectories along which the theory can
flow as we send the cutoff $\widehat{k}$ to infinity are labeled by only
finitely many parameters. Therefore the theory is as predictive as any
conventionally renormalizable theory; it is not plagued by the notorious
increase of free parameters which is typical of effective theories. The set
of generalized couplings for which the non-Gaussian fixed point is UV
attractive should include the dimensionless Newton constant, $g$, and 
cosmological constant, $\lambda$.

Using the $\varepsilon$-expansion, Weinberg showed already long ago that 
gravity in $2+\varepsilon$ dimensions $(0<\varepsilon\ll 1)$ is indeed 
asymptotically safe \cite{wein}. Further progress in this direction, 
in particular for $d=4$, was hampered by the lack of an efficient 
calculational scheme which could be used to search for nonperturbative fixed 
points.

As a solution to this problem which does not rely on the 
$\varepsilon$-expansion we propose to use the effective average action in 
order to find nontrivial fixed points $(\lambda_*,g_*,\cdots)$ of the 
gravitational RG flow. (The dots stand for the infinitely many other couplings
which parametrize a generic action functional.) Using this approach, the case
$d=2+\varepsilon$ was reanalyzed in a more general setting and, more 
importantly, it was shown that the Einstein-Hilbert truncation predicts the 
existence of a non-Gaussian fixed point $(\lambda_*,g_*)$ also in dimensions
$d>2$, in particular for $d=4$ \cite{Reu96,bh2,souma1}.

The crucial question which arises is whether this result is an artifact of the
truncation used, or if it correctly reflects a property of the full theory.
It is clear that in order to answer this question one would like to include
further invariants into the truncation and to check whether the predictions 
stabilize. 

From the technical point of view such calculations are extremely 
complicated so that in the present paper we shall use a different method in
order to get a first idea about the reliability of the nontrivial fixed point.
We are going to analyze to what extent its location in the 
$\lambda$-$g$ plane and its attractivity properties (critical exponents) are
scheme dependent. Here ``scheme dependence'' refers to the dependence on the
cutoff operator ${\cal R}_k$ used in the derivation of the RG equation. 

First of all, ${\cal R}_k$ is a matrix in the space of irreducible
component fields (see below) which is not uniquely determined by the general
principles. Hence we can vary it to some extent. In fact, in the present 
paper we shall introduce a new cutoff (``cutoff of type B'') whose matrix 
structure is different from the original one of ref. \cite{Reu96} (``cutoff of 
type A''). Either of these cutoffs is proportional to a 
``shape function'' $R^{(0)}(p^2/k^2)$ which describes the ``thinning out'' of
degrees of freedom as we pass the threshold $p^2=k^2$. Also this function can 
be varied in order to assess the scheme dependence of the fixed point 
properties. 

While in general only the critical exponents but not the location
$(\lambda_*,g_*,\cdots)$ of the fixed point are expected to be universal, i.e.
scheme independent \cite{kana}, we shall argue that the product $g_*\lambda_*$
is an observable quantity as well. For observables the ${\cal R}_k$-dependence
is a pure truncation artifact; in an exact treatment all 
${\cal R}_k$-dependencies cancel.  The status of the Einstein-Hilbert 
truncation would be rather questionable if the fixed point was present for 
some cutoffs but absent for others. Instead, we find that it is actually there
for all admissible cutoffs, and moreover that the observable $g_*\lambda_*$
is scheme independent with a quite unexpected precision. 

Our results strongly support the conjecture
that the non-Gaussian fixed point is present in the exact theory and is not a 
truncation artifact. Also another prerequisite of asymptotic safety turns out 
to be satisfied: we find that, for any cutoff, the fixed point is UV attractive
in both directions of the $\lambda$-$g$ plane.

Ultimately one would like to use more general truncations than (\ref{in2}) in
order to study the RG flow in a larger subspace. Typically this requires
computations whose algebraic complexity is quite formidable. Assume we make an 
ansatz $\Gamma_k=\sum\limits_{i=1}^n {\rm g}_i(k)I_i$ containing $n$ 
diffeomorphism invariant functionals $I_i[g_{\mu\nu},\bar{g}_{\mu\nu}]$. In 
order to project the RG flow on the $n$-dimensional space with coordinates 
${\rm g}_i$ we must insert the ansatz into the RHS of the flow equation 
(\ref{in1}). At this point the nontrivial problem, both conceptually and 
computationally, is to expand the trace with respect to a complete set of 
actions, $\{\gamma_j[g_{\mu\nu},\bar{g}_{\mu\nu}]\}$, in such a way
that the $I_i$'s retained in the ansatz are a subset of the $\gamma_j$'s. The
coefficients of the remaining $\gamma_j$'s, those not present in $\Gamma_k$, 
are set to zero by the truncation. In practice the projection on the 
$\{I_i\}$-subspace is done by inserting a set of metrics $g_{\mu\nu},
\bar{g}_{\mu\nu}$ on both sides of eq. $\!$(\ref{in1}) which give a nonzero 
value only to specific linear combinations of the $I_i$'s. Provided one 
manages to compute the functional trace for sufficiently many 
$g_{\mu\nu},\bar{g}_{\mu\nu}$-pairs one can then deduce the ordinary 
differential equations for the generalized couplings ${\rm g}_i(k)$.

For the Einstein-Hilbert truncation this procedure is fairly simple since
(ignoring the running of the gauge parameter) it is sufficient to insert for 
$g_{\mu\nu}=\bar{g}_{\mu\nu}$ the metric of a family of spheres $S^d$ 
parametrized by their 
radius $r$. Their maximal symmetry facilitates the calculations considerably.
With $r$ kept as a free parameter, these metrics are general enough to 
disentangle $\int d^dx\sqrt{g}\propto r^d$ and $\int d^dx\sqrt{g}R\propto 
r^{d-2}$. But already when we include invariants with four derivatives of the
metric this method fails: the spheres cannot distinguish 
$\int d^dx\sqrt{g}R^2\propto r^{d-4}$ from $\int d^dx\sqrt{g}R_{\mu\nu}
R^{\mu\nu}\propto r^{d-4}$, for instance. 

These remarks hint at (at least) two major problems which one faces in 
generalizations of the exact RG approach to gravity. (i) The momentum dependent
``mass'' term $\Delta_k S[h_{\mu\nu},\bar{g}_{\mu\nu}]$ depends quadratically
on the metric fluctuation $h_{\mu\nu}$, but also, via ${\cal R}_k\equiv
{\cal R}_k[\bar{g}]$ on the background $\bar{g}_{\mu\nu}$. In general it is a
quite nontrivial task to construct a cutoff operator ${\cal R}_k[\bar{g}]$
which has the desired properties mentioned above for a class of background 
metrics $\bar{g}_{\mu\nu}$ general enough for the projection on the truncation
subspace. (ii) Assume we found an appropriate ${\cal R}_k[\bar{g}]$. Then 
there arises the computational problem of evaluating the trace on the RHS of 
the RG equation for various $g_{\mu\nu}$'s and $\bar{g}_{\mu\nu}$'s. These 
metrics do not coincide when we allow for an evolution of the gauge fixing
sector. Even if we ignore this complication, the Hessian $\Gamma_k^{(2)}$ 
under the trace is an extremely complicated nonminimal covariant matrix 
differential operator constructed from the curvature tensor and covariant 
derivatives $D_\mu$. A priori, even for maximally symmetric backgrounds, not 
all derivatives $D_\mu$ are contracted to form powers of the covariant 
Laplacian $D^2\equiv D_\mu D^\mu$, and $\Gamma_k^{(2)}$ is not diagonal in the
space of fields with a definite helicity therefore. Hence standard heat kernel 
techniques or perhaps information about the spectrum of $D^2$ are of no help 
at this point.

In this paper we outline a general strategy for tackling these problems. 
It is based upon York's ``TT-decomposition'' \cite{York} which is available on
(almost) every spacetime manifold needed for our projection method. The idea 
is to decompose the fluctuation $h_{\mu\nu}$ into a transverse, traceless 
tensor $h^T_{\mu\nu}$, a longitudinal-transverse tensor (parametrized by a
transverse vector $\widehat{\xi}^\mu$), a longitudinal-longitudinal tensor 
(parametrized by a scalar $\widehat{\sigma}$), and a trace part (parametrized 
by another scalar $\phi$). In the basis of the component fields 
$\{h^T_{\mu\nu},\widehat{\xi}^\mu,\widehat{\sigma},
\phi\}$, all $D_\mu$'s appear in powers of the Laplacian only, at least for
the class of maximally symmetric backgrounds. The important point is that this
decomposition can be used in order to simplify the structure of 
$\Gamma_k^{(2)}$ on essentially all backgrounds, not just on spheres. 
(On $S^d$ the TT-decomposition boils down to the familiar decomposition of
$h_{\mu\nu}$ with respect to pieces which are irreducible under the isometry
group {\sf SO($d+1$)}. In some of the work following the original paper
\cite{Reu96} this decomposition on $S^d$ had been used already 
\cite{souma2,DP97,sven,odintsov}.) 
Compared to $S^d$ a certain complication arises, however, because the 
TT-decomposition is nonorthogonal in general.

The TT-decomposition also helps in solving the first problem, the construction
of ${\cal R}_k$, because $\Delta_k S$ has a much simpler structure when 
expressed in terms of the component fields rather than the original 
$h_{\mu\nu}$.

This paper is organized as follows.

In the first part (sections \ref{S2} and \ref{S3}) we describe the 
construction of a new RG equation where the component fields $\{h^T_{\mu\nu},
\widehat{\xi}^\mu,\widehat{\sigma},\phi\}$ are used from the outset. Along 
the way we discuss the problems related to a proper identification of 
$\Delta_k S$. This part of the paper is meant to supply a set of tools which
will become indispensable in future investigations when one 
includes further invariants into the truncation ($R^2$-terms \cite{LR3}, for 
instance), if one adds matter fields, or if one allows for a running gauge 
fixing.

As a first application, we revisit the Einstein-Hilbert truncation in the
second part of the paper (sections \ref{S4} and \ref{S5}). We introduce a new 
cutoff ${\cal R}_k$ which
is natural in the TT-language, and we use an arbitrary gauge parameter.
This allows for a nontrivial comparison of the resulting RG equations and their
fixed point properties with those of \cite{Reu96} whose cutoff operator 
${\cal R}_k$ has a rather different structure. We find both a Gaussian and a
non-Gaussian fixed point in the $(g,\lambda)$-system and we perform a detailed
analysis of their properties, in particular of their scheme dependence.
The chances for realizing the asymptotic safety scenario in 4 dimensions will
be discussed in detail.

In section \ref{S6} we investigate the implications of the non-Gaussian fixed 
point for the effective graviton propagator at large momenta. A kind of 
dimensional reduction from 4 to 2 dimensions takes place in the vicinity of 
this fixed point. The asymptotic form of the propagator suggests that when 
4-dimensional spacetime is probed by a very high-energetic graviton it appears
to be effectively 2-dimensional.

Various technical results, needed in the present paper but presented also with
an eye towards future applications \cite{LR3}, are relegated to a set of
appendices.

At this point the reader who is mostly interested in the results rather than
their derivation can proceed directly to section \ref{S5}. 

\section{The exact evolution equation}
\renewcommand{\theequation}{2.\arabic{equation}}
\setcounter{equation}{0}
\label{S2}
\subsection{Gauge fixing}
\label{2A}
Following \cite{Reu96} we define a scale dependent 
modification of the Euclidean functional integral for the generating
functional ${\bf Z}_k$ by using the background gauge fixing technique 
\cite{Ab81,back}. For this purpose we decompose the integration variable in 
the functional integral over all metrics, $\gamma_{\mu\nu}$, into a fixed 
background metric $\bar{g}_{\mu\nu}$ and a fluctuation field $h_{\mu\nu}$,
\begin{eqnarray}
\label{01}
\gamma_{\mu\nu}(x)=\bar{g}_{\mu\nu}(x)+h_{\mu\nu}(x)\;.
\end{eqnarray}
Then we replace the integration over $\gamma_{\mu\nu}$ by an integration over
$h_{\mu\nu}$. With the Faddeev-Popov ghosts $C^\mu$ and $\bar{C}_\mu$ the 
generating functional ${\bf Z}_k$ may be written as
\begin{eqnarray}
\label{a}
\lefteqn{{\bf Z}_k[{\rm sources}]=\int{\cal D}h_{\mu\nu}\,
{\cal D}C^\mu\,{\cal D}\bar{C}_\mu}
\nonumber\\
& &\times\exp\left[-S[\bar{g}+h]-S_{\rm gf}[h;\bar{g}]-S_{\rm gh}
[h,C,\bar{C};\bar{g}]-\Delta_k S[h,C,\bar{C};\bar{g}]-S_{\rm source}\right]\;.
\end{eqnarray}
The first term in the exponential, $S[\gamma]=S[\bar{g}+h]$, is the
classical action which, for the moment, is assumed to be positive definite.
It is invariant under arbitrary general coordinate transformations. 
$S_{\rm gf}$ denotes the gauge fixing term 
\begin{eqnarray}
\label{02}
S_{\rm gf}[h;\bar{g}]=\frac{1}{2\alpha}\int d^dx\,\sqrt{\bar{g}}\,
\bar{g}^{\mu\nu}\,F_\mu[\bar{g},h]\,F_\nu[\bar{g},h]\;.
\end{eqnarray}
It corresponds to the gauge condition $F_\mu[\bar{g},h]=0$. Linear gauge 
conditions, 
\begin{eqnarray}
\label{03}
F_\mu[\bar{g},h]=\sqrt{2}\kappa\,{\cal F}^{\alpha\beta}_\mu[\bar{g}]\,
h_{\alpha\beta}\;,
\end{eqnarray}
are particularly convenient. In the present paper we use the harmonic 
gauge\renewcommand{\baselinestretch}{1}\small\normalsize\footnote{For the flow
equation in the conformal gauge (2D Liouville quantum gravity) see refs.
\cite{RW97,liouv}.}
\renewcommand{\baselinestretch}{1.5}\small\normalsize 
for which
\begin{eqnarray}
\label{04}
{\cal F}^{\alpha\beta}_\mu[\bar{g}]=\delta^\beta_\mu\bar{g}^{\alpha\gamma}
\bar{D}_\gamma-\frac{1}{2}\bar{g}^{\alpha\beta}\bar{D}_\mu\;.
\end{eqnarray}
Here $\bar{D}_\mu$ denotes the covariant derivative constructed from the
background metric $\bar{g}_{\mu\nu}$, while we shall write $D_\mu$ for the
covariant derivative involving the quantum metric $\gamma_{\mu\nu}$. In eq. 
(\ref{03}) we introduced the constant
\begin{eqnarray}
\label{086}
\kappa\equiv(32\pi\bar{G})^{-\frac{1}{2}}
\end{eqnarray}
where $\bar{G}$ denotes the bare Newton constant. The Faddeev-Popov operator 
associated with the gauge fixing (\ref{03}) with (\ref{04}) takes the form
\begin{eqnarray}
\label{06}
{\cal M}[\gamma,\bar{g}]^\mu_{\;\;\;\nu}=\bar{g}^{\mu\rho}\bar{g}^{\sigma
\lambda}\bar{D}_\lambda\left(\gamma_{\rho\nu}D_\sigma+\gamma_{\sigma\nu}
D_\rho\right)-\bar{g}^{\rho\sigma}\bar{g}^{\mu\lambda}\bar{D}_\lambda
\gamma_{\sigma\nu}D_\rho\;.
\end{eqnarray}
It enters the functional integral (\ref{a}) via the ghost action
\begin{eqnarray}
\label{05}
S_{\rm gh}[h,C,\bar{C};\bar{g}]=-\sqrt{2}\int d^dx\,\sqrt{\bar{g}}\,
\bar{C}_\mu\,{\cal M}[\bar{g}+h,\bar{g}]^\mu_{\;\;\;\nu}\,C^\nu\;.
\end{eqnarray}
Furthermore, $\Delta_k S$ and $S_{\rm source}$ are the cutoff and the
source action, respectively. $\Delta_k S$ provides an appropriate infrared
cutoff for the integration variables and $S_{\rm source}$ introduces sources
for the fields $h_{\mu\nu}$, $C^\mu$ and $\bar{C}_\mu$. Their explicit
structure will be discussed later on.
\subsection{Decomposition of the quantum fields}
\label{2B}
For the calculations in the following sections it turns out to be
convenient to decompose the gravitational field $h_{\mu\nu}$ according to
(see e.g. \cite{York})
\begin{eqnarray}
\label{f}
h_{\mu\nu}=h^{T}_{\mu\nu}
+\bar{D}_\mu\widehat{\xi}_\nu+\bar{D}_\nu\widehat{\xi}_\mu
+\bar{D}_\mu \bar{D}_\nu\widehat{\sigma}
-\frac{1}{d}\bar{g}_{\mu\nu}\bar{D}^2\widehat{\sigma}
+\frac{1}{d}\bar{g}_{\mu\nu}\phi\,.
\end{eqnarray}
To obtain this ``TT-decomposition'' one starts by splitting off the trace part
$h^{Tr}_{\mu\nu}\equiv\bar{g}_{\mu\nu}\phi/d$ from $h_{\mu\nu}$. It involves
a scalar field $\phi$. The remaining symmetric traceless 
tensor may be decomposed further into a transverse component $h_{\mu\nu}^{T}$ 
and a longitudinal component $h_{\mu\nu}^{L}$. Introducing a transverse vector
field $\widehat{\xi}_\mu$ and another scalar $\widehat{\sigma}$, the
longitudinal tensor can be expressed by $h_{\mu\nu}^{L}=h_{\mu\nu}^{LT}
+h_{\mu\nu}^{LL}$ with $h_{\mu\nu}^{LT}\equiv\bar{D}_\mu\widehat{\xi}_\nu
+\bar{D}_\nu\widehat{\xi}_\mu$ and $h_{\mu\nu}^{LL}\equiv\bar{D}_\mu\bar{D}_\nu
\widehat{\sigma}-\bar{g}_{\mu\nu}\bar{D}^2\widehat{\sigma}/d$ thereby ending
up with eq. (\ref{f}). Thus the components of $h_{\mu\nu}$ 
introduced by this transverse-traceless (TT-)decomposition obey the relations
\begin{eqnarray}
\label{c}
\bar{g}^{\mu\nu}h_{\mu\nu}^{T}=0\;,\;\;\;\bar{D}^\mu h_{\mu\nu}^{T}=0
\;,\;\;\;\bar{D}^\mu\widehat{\xi}_\mu=0\;,\;\;\;\phi=\bar{g}_{\mu\nu}
h^{\mu\nu}\;.
\end{eqnarray}

This decomposition is valid for complete, closed Riemannian $d$-spaces (i.e. 
compact Riemannian manifolds without boundary). As argued in
\cite{York}, its domain of validity can be extended to open, asymptotically
flat $d$-spaces, certain assumptions concerning the asymptotic behavior of
the fields being made. From now on we assume that the gravitational background
belongs to one of these classes of spaces. 

Obviously $h_{\mu\nu}$ receives no contribution from those 
$\widehat{\xi}_\mu$- and $\widehat{\sigma}$-modes which satisfy the Killing
equation
\begin{eqnarray}
\label{09}
\bar{D}_\mu\widehat{\xi}_\nu+\bar{D}_\nu\widehat{\xi}_\mu=0
\end{eqnarray}
and the scalar equation
\begin{eqnarray}
\label{010}
\bar{D}_\mu \bar{D}_\nu\widehat{\sigma}-\frac{1}{d}\bar{g}_{\mu\nu}\bar{D}^2
\widehat{\sigma}=0\;,
\end{eqnarray}
respectively. Therefore such modes, referred to as unphysical 
$\widehat{\xi}_\mu$- and $\widehat{\sigma}$-modes, have to be 
excluded from the functional integral. Considering the conformal Killing 
equation
\begin{eqnarray}
\label{011}
\bar{D}_\mu{\cal C}_\nu+\bar{D}_\nu{\cal C}_\mu-\frac{2}{d}\bar{g}_{\mu\nu}
\,\bar{D}_\lambda{\cal C}^\lambda=0
\end{eqnarray}
we recognize that the unphysical $\widehat{\sigma}$-modes correspond 
to constants or are related via ${\cal C}_\mu=\bar{D}_\mu\widehat{\sigma}$ to
proper conformal Killing vectors (PCKV's), i.e. solutions of eq. (\ref{011}) 
which are not at the same time ordinary Killing vectors (KV's), 
\cite{Mot95}.\renewcommand{\baselinestretch}{1}\small\normalsize\footnote{As a
consequence of the 
linearity of the conformal Killing equation we may add any Killing vector 
(which is always transversal) to its solutions and obtain another solution. 
For definiteness we therefore define the PCKV's to be purely longitudinal. 
Then there is a one-to-one correspondence between the PCKV's and the 
nonconstant solutions of (\ref{010}).} 
\renewcommand{\baselinestretch}{1.5}\small\normalsize

By virtue of the decomposition (\ref{f}) the inner product on the space of 
symmetric tensor fields may be decomposed according 
to\renewcommand{\baselinestretch}{1}\small\normalsize\footnote{A remark 
concerning our notation: If not indicated otherwise each covariant derivative 
acts on everything that stands on the 
right of it.}
\renewcommand{\baselinestretch}{1.5}\small\normalsize
\begin{eqnarray}
\label{08}
\lefteqn{\left<h^{(1)},h^{(2)}\right>\equiv\int d^dx\,\sqrt{\bar{g}}
\,h^{(1)}_{\mu\nu}\,\bar{g}^{\mu\rho}\bar{g}^{\nu\sigma}\,
h^{(2)}_{\rho\sigma}}\nonumber\\
&=&\int d^dx\,\sqrt{\bar{g}}
\Bigg\{h^{(1)T}_{\mu\nu}h^{(2)\mu\nu}-2\widehat{\xi}_\mu^{(1)}
\left(\bar{g}^{\mu\nu}\bar{D}^2+\bar{R}^{\mu\nu}\right)
\widehat{\xi}_\nu^{(2)}-2\widehat{\xi}_\mu^{(1)}\,\bar{R}^{\mu\nu}
\,\bar{D}_\nu\widehat{\sigma}^{(2)}\nonumber\\
& &-2\widehat{\xi}_\mu^{(2)}\,\bar{R}^{\mu\nu}
\,\bar{D}_\nu\widehat{\sigma}^{(1)}+\widehat{\sigma}^{(1)}\left(
\frac{d-1}{d}(\bar{D}^2)^2+\bar{D}_\mu\bar{R}^{\mu\nu}\bar{D}_\nu\right)
\widehat{\sigma}^{(2)}+\frac{1}{d}\phi^{(1)}\phi^{(2)}\Bigg\}\;.
\end{eqnarray}
From eq. (\ref{08}) we see that, for a general background 
metric, only $h_{\mu\nu}^{T}$, $h_{\mu\nu}^{L}$ and $h^{Tr}_{\mu\nu}$ 
form an orthogonal set, whereas $h_{\mu\nu}^{LT}$ and $h_{\mu\nu}^{LL}$ are 
not orthogonal in general. This nonorthogonality manifests itself in the
appearance of terms where the components $\widehat{\xi}_\mu$ and 
$\widehat{\sigma}$ mix. But at least for Einstein spaces, where 
$\bar{R}_{\mu\nu}=C\bar{g}_{\mu\nu}$ with $C$ a constant, we find
$\bar{D}_\mu\bar{R}^{\mu\nu}=C\bar{D}_\mu\bar{g}^{\mu\nu}\equiv0$ and 
therefore $\left<h^{LT},h^{LL}\right>=4C\int d^dx\sqrt{\bar{g}}\,
\widehat{\sigma}\,\bar{D}^\mu\widehat{\xi}_\mu=0$. Thus 
$\{h_{\mu\nu}^{T},h_{\mu\nu}^{LT},h_{\mu\nu}^{LL},h^{Tr}_{\mu\nu}\}$ 
represents an orthogonal set of field components in this case. 

In order to determine the Jacobian $J_1$ which appears in the functional
integral (\ref{a}) after performing the transformation of integration 
variables  $h_{\mu\nu}\longrightarrow \{h_{\mu\nu}^T,\widehat{\xi}_\mu,
\widehat{\sigma},\phi\}$ we proceed as follows. We consider a Gaussian
integral over $h_{\mu\nu}$ and reexpress it in terms of the component
fields \cite{Mot95}:
\begin{eqnarray}
\label{i}
\int{\cal D}h_{\mu\nu}\exp\left[-\frac{1}{2}\left<h,h\right>\right]
&=&J_1\int{\cal D}h^T_{\mu\nu}\,{\cal D}\widehat{\xi}_\mu\,{\cal D}
\widehat{\sigma}\,{\cal D}\phi\,\exp\Bigg[
-\frac{1}{2}\int d^dx\,\sqrt{\bar{g}}\nonumber\\
& &\times
\Bigg\{h^T_{\mu\nu}h^{T\mu\nu}+\frac{1}{d}\phi^2+\left(\widehat{\xi}_\mu,
\widehat{\sigma}\right)
M^{(\mu,\nu)}\left(\begin{array}{c}\widehat{\xi}_\nu\\\widehat{\sigma}
\end{array}\right)\Bigg\}\Bigg]\;.
\end{eqnarray}
Here
\begin{eqnarray}
\label{j}
M^{(\mu,\nu)}\equiv\left(\begin{array}{cc}
-2\left(\bar{g}^{\mu\nu}\bar{D}^2+\bar{R}^{\mu\nu}\right) &
-2\bar{R}^{\mu\lambda}\bar{D}_\lambda\\
2\bar{D}_\lambda\bar{R}^{\lambda\nu}
& \frac{d-1}{d}(\bar{D}^2)^2+\bar{D}_\lambda \bar{R}^{\lambda\rho}\bar{D}_\rho
\end{array}\right)
\end{eqnarray}
is a Hermitian matrix differential operator. Since all functional integrals 
appearing in eq. (\ref{i}) are Gaussian they are easily evaluated. This
leads to  the Jacobian
\begin{eqnarray}
\label{k}
J_1=N\sqrt{\det\nolimits_{(1T,0)}' M^{(\mu,\nu)}}\;.
\end{eqnarray}
Here $N$ represents an infinite constant which may be absorbed
into the normalization of the measure ${\cal D}h_{\mu\nu}$.

The notation adopted in eq. (\ref{k}) has to be interpreted as
follows. A prime at the determinant or the trace of an operator ${\cal A}$ 
indicates that all unphysical $\widehat{\xi}_\mu$- and 
$\widehat{\sigma}$-eigenmodes of ${\cal A}$, characterized by eqs. (\ref{09})
and (\ref{010}), are to be excluded from the calculation. A
subscript at determinants or traces describes on which kind of field the 
operator ${\cal A}$ acts. We use the subscripts $(0)$, $(1T)$ and $(2ST^2)$ 
for spin-0 fields $\widehat{\sigma}$, transverse spin-1 fields 
$\widehat{\xi}_\mu$ and symmetric transverse traceless spin-2 fields 
$h_{\mu\nu}^T$, repectively. The subscript $(1T,0)$ appearing in eq. (\ref{k})
refers to a $(d+1)\times (d+1)$-matrix differential operator whose first $d$
columns act on traceless spin-1 fields $\widehat{\xi}^\nu$ whereas the last
column acts on spin-0 fields $\widehat{\sigma}$.

Likewise we decompose
the ghost and the antighost into their orthogonal components according to
\begin{eqnarray}
\label{p}
\bar{C}_\mu=\bar{C}_\mu^T+\bar{D}_\mu\widehat{\bar{\eta}}\;,\;\;\;
C^\mu=C^{T\mu}+\bar{D}^\mu\widehat{\eta}
\end{eqnarray}
where $\bar{C}_\mu^T$ and $C^{T\mu}$ are the transverse components of 
$\bar{C}_\mu$ and $C^\mu$: $\bar{D}^\mu\bar{C}_\mu^T=0$, 
$\bar{D}_\mu C^{T\mu}=0$.
In order to compute the Jacobian $J_2$ induced by the change of variables
$\bar{C}_\mu\longrightarrow \{\bar{C}^T_\mu,\widehat{\bar{\eta}}\}$,
$C^\mu\longrightarrow \{C^{T\mu},\widehat{\eta}\}$ we write
\begin{eqnarray}
\label{s}
\lefteqn{\int{{\cal D}C^\mu\,\cal D}\bar{C}_\mu\,\exp\left[-\left<\bar{C},C
\right>\right]}\nonumber\\
&=&J_2\int{\cal D}C^{T\mu}\,{\cal D}\bar{C}^T_\mu\,{\cal D}\widehat{\eta}\,
{\cal D}\widehat{\bar{\eta}}\,\exp
\left[-\int d^dx\,\sqrt{\bar{g}}\,\left\{\bar{C}^T_\mu C^{T\mu}
+\widehat{\bar{\eta}}(-\bar{D}^2)\widehat{\eta}\right\}\right]\;.
\end{eqnarray}
and perform the Grassmann functional integrals. The result is
\begin{eqnarray}
\label{t}
J_2=\left[\det\nolimits_{(0)}'\left(-\bar{D}^2\right)\right]^{-1}\;.
\end{eqnarray}
In this case the constant $\widehat{\eta}$-mode represents an unphysical mode
which has to be excluded.
\subsection{Momentum dependent redefinition of the component fields}
\label{2C}
It will prove convenient to introduce new variables of integration, 
$\xi_\mu$, $\sigma$, $\bar{\eta}$ and $\eta$, by means of the momentum 
dependent (nonlocal) redefinitions
\begin{eqnarray}
\label{y}
\xi^\mu &\equiv &\sqrt{-\bar{D}^2-\overline{{\rm Ric}}}\;\,
\widehat{\xi}^\mu\nonumber\\\sigma&\equiv&\sqrt{
(\bar{D}^2)^2+\frac{d}{d-1}\bar{D}_\mu\bar{R}^{\mu\nu}\bar{D}_\nu}\;\,
\widehat{\sigma}\nonumber\\
\bar{\eta}&\equiv &\sqrt{-\bar{D}^2}\;\,
\widehat{\bar{\eta}}\;,\;\;\;\eta\equiv\sqrt{-\bar{D}^2}
\;\,\widehat{\eta}\;.
\end{eqnarray}
Here the operator $\overline{{\rm Ric}}$ maps vectors onto vectors according to
\begin{eqnarray}
\label{ooo}
\left(\overline{{\rm Ric}}\;v\right)^\mu=\bar{R}^{\mu\nu}v_\nu\;.
\end{eqnarray}
Note that the transformations (\ref{y}) are well defined and invertible since 
for any (physical) eigenmode the operators under the square roots of (\ref{y})
have strictly positive 
eigenvalues.\renewcommand{\baselinestretch}{1}\small\normalsize\footnote{In 
order to make sure that the operators are indeed invertible we also assume 
that their eigenvalues do not have zero as an accumulation point.} 
\renewcommand{\baselinestretch}{1.5}\small\normalsize
This is due to the
fact that these operators arise from the squares of $h^{LT}_{\mu\nu}$ and
$h^{LL}_{\mu\nu}$ and from $(\bar{D}_\mu\widehat{\bar{\eta}})(\bar{D}_\mu
\widehat{\eta})$ by shifting all covariant derivatives to the right. Thus
they cannot assume negative eigenvalues. For example,
\begin{eqnarray}
\label{012}
\left<h^{LL},h^{LL}\right>&=&\int d^dx\,\sqrt{\bar{g}}\left(\bar{D}_\mu
\bar{D}_\nu\widehat{\sigma}-\frac{1}{d}\bar{g}_{\mu\nu}\bar{D}^2
\widehat{\sigma}\right)\left(\bar{D}^\mu
\bar{D}^\nu\widehat{\sigma}-\frac{1}{d}\bar{g}^{\mu\nu}\bar{D}^2
\widehat{\sigma}\right)\nonumber\\
&=&\frac{d-1}{d}\int d^dx\,\sqrt{\bar{g}}\,\widehat{\sigma}\left((\bar{D}^2)^2
+\frac{d}{d-1}\bar{D}_\mu\bar{R}^{\mu\nu}\bar{D}_\nu\right)\widehat{\sigma}\;.
\end{eqnarray}
Furthermore, the spectra of the operators in (\ref{y}) do not even contain 
zeros, since the potential zero-modes coincide precisely with the 
aforementioned unphysical modes which have to be excluded. For instance, 
$\xi^\mu$ is a zero-mode of $-\bar{D}^2-\overline{\rm Ric}$ if and only if
$\xi^\mu$ is a Killing vector. 

Along the lines outlined in the previous subsection, we now determine the
Jacobians for the transformation of integration variables. We obtain 
\begin{eqnarray}
\label{dd}
J_3&=&\left[\det\nolimits_{(1[T])}'\left(-\bar{D}^2
-\overline{{\rm Ric}}\right)\right]^{-\frac{1}{2}}\nonumber\\
J_4&=&\left[\det\nolimits_{(0)}'\left((\bar{D}^2)^2
+\frac{d}{d-1}\bar{D}_\mu\bar{R}^{\mu\nu}\bar{D}_\nu\right)
\right]^{-\frac{1}{2}}\nonumber\\
J_5&=&\det\nolimits_{(0)}'\left(-\bar{D}^2\right)
\end{eqnarray}
for the transformations $\widehat{\xi}_\mu\rightarrow
\xi_\mu$, $\widehat{\sigma}\rightarrow\sigma$ and $\widehat{\bar{\eta}}
\rightarrow\bar{\eta}$, $\widehat{\eta}\rightarrow\eta$, respectively.
(The integration measures have been chosen such that no additional infinite 
constants occur in the Jacobians.) The square brackets appearing in the 
subscript $(1,[T])$ at $J_3$ indicate that 
the operator under consideration acts on spin-one fields which are transverse 
only for certain background metrics, because the property of transversality
is not necessarily transmitted from $\widehat{\xi}^\mu$ to $\xi^\mu$. However,
at least for Einstein spaces $\xi^\mu$ is transverse as well.

After carrying out this change of integration variables, $J_1$, $J_3$ 
and $J_4$ are the only Jacobians appearing in the generating functional ${\bf
Z}_k$ since $J_2$ and $J_5$ cancel.

\subsection{The effective average action}
\label{2D}
By adding an infrared (IR) cutoff $\Delta_k S$ to the classical action under
the path integral (\ref{a}) we obtain a scale-dependent generating functional 
${\bf Z}_k$. The term $\Delta_k S$ is chosen to depend on the fluctuation 
fields in such a way that their eigenmodes with respect to $-\bar{D}^2$ which
correspond to large eigenvalues $p^2\gg k^2$ are not influenced, whereas 
contributions from 
eigenmodes with small eigenvalues $p^2\ll k^2$ are suppressed. In this sense 
${\bf Z}_k$ describes an effective theory at the scale $k<\widehat{k}$.
For technical simplicity we implement the suppression of the low-momentum
modes by momentum-dependent ``mass'' terms, i.e. by cutoffs which are 
quadratic in the fluctuation fields:
\begin{eqnarray}
\label{v}
\Delta_k S\left[h,C,\bar{C};\bar{g}\right]
=\frac{1}{2}\left<h,{\cal R}_k^{\rm grav}\,h\right>
+\left<\bar{C},{\cal R}_k^{\rm gh}\,C\right>\;.
\end{eqnarray}
Here the operators ${\cal R}_k^{\rm grav}$ and ${\cal R}_k^{\rm gh}$ are 
constructed from the covariant derivative with respect to the background 
metric, $\bar{D}_\mu$. Note that ${\cal R}_k^{\rm grav}$ and 
${\cal R}_k^{\rm gh}$ must not depend on the quantum metric $\gamma_{\mu\nu}$
but only on the background metric $\bar{g}_{\mu\nu}$ since otherwise the
cutoff cannot be quadratic. In
order to provide the desired behavior these operators must vanish for
$p^2/k^2\rightarrow\infty$ (in particular for $k\rightarrow 0$) and must 
behave  as ${\cal R}_k\rightarrow {\cal Z}_k k^2$ for $p^2/k^2\rightarrow 
0$. (The meaning of the constant ${\cal Z}_k$ will be explained later.) 
As a consequence, all modes with $p^2\ll k^2$ acquire a mass $\propto k$.

At this stage of the discussion it is not necessary to specify the explicit 
structure of the cutoff operators. We only mention the following
point. According to appendix \ref{dec}, 
${\cal R}_k^{\rm grav}$ and ${\cal R}_k^{\rm gh}$ can be chosen such that, at
the level of the component fields,
\begin{eqnarray}
\label{013}
\Delta_k S\left[h,C,\bar{C};\bar{g}\right]
=\frac{1}{2}\sum\limits_{\zeta_1,\zeta_2\in I_1}
\left<\zeta_1,\left({\cal R}_k\right)_{\zeta_1\zeta_2}\,
\zeta_2\right>+\frac{1}{2}
\sum\limits_{\psi_1,\psi_2\in I_2}
\left<\psi_1,\left({\cal R}_k\right)_{\psi_1\psi_2}\psi_2\right>
\end{eqnarray}
with the index sets $I_1\equiv
\{h^T,\xi,\sigma,\phi\}$, $I_2\equiv\{\bar{C}^T,C,\bar{\eta},\eta\}$.
In contrast to generic cutoffs which are defined in terms of the component 
fields from the outset, the structure (\ref{013}) allows us to 
return to the formulation in terms of the fundamental fields, eq. (\ref{v}), in
a straightforward way. (See appendix \ref{dec}.) The set of operators
$\left({\cal R}_k\right)_{\zeta_1\zeta_2}$, $\left({\cal R}_k
\right)_{\psi_1\psi_2}$ introduced by this realization of the cutoff
may be fixed later on. Hermiticity demands that they satisfy 
$\left({\cal R}_k\right)_{\zeta_2\zeta_1}=\left({\cal R}_k
\right)_{\zeta_1\zeta_2}^\dagger$ and $\left({\cal R}_k
\right)_{\psi_2\psi_1}=-\left({\cal R}_k
\right)_{\psi_1\psi_2}^\dagger$. Furthermore, $\left({\cal R}_k
\right)_{\psi_1\psi_2}\equiv 0$ if both $\psi_1\in\{C^T,\eta\}$ and $\psi_2\in
\{C^T,\eta\}$, or if both $\psi_1\in\{\bar{C}^T,\bar{\eta}\}$ and
$\psi_2\in\{\bar{C}^T,\bar{\eta}\}$.

A similar decomposition is applied to $S_{\rm source}$. The
source terms are defined as 
\begin{eqnarray}
\label{hh}
S_{\rm source}[h,C,\bar{C},J,K,\bar{K};\bar{g}]=-\left<J,h\right>
-\left<\bar{K},C\right>-\left<K,\bar{C}\right>
\end{eqnarray}               
with external sources $J^{\mu\nu}$, $K^\mu$ and $\bar{K}_\mu$ for the 
fundamental fields $h_{\mu\nu}$, $\bar{C}_\mu$ and $C^\mu$,
respectively.
Proceeding as described in appendix \ref{dec}, an alternative form 
of $S_{\rm source}$ may be derived from eq. (\ref{hh}) where each component 
field is coupled to a certain component of the fundamental sources. Then, in 
terms of these ``component sources'', $S_{\rm source}$ takes the form
\begin{eqnarray}
\label{071}
S_{\rm source}[h,C,\bar{C},J,K,\bar{K};\bar{g}]
=-\sum\limits_{\zeta\in I_1}\left<J_\zeta,\zeta\right>
-\sum\limits_{\psi\in\{C^T,\eta\}}\left<\bar{K}_{\psi},\psi\right>
-\sum\limits_{\psi\in\{\bar{C}^T,\bar{\eta}\}}\left<K_{\psi},\psi\right>\;.
\end{eqnarray}

As a consequence, the functional 
\begin{eqnarray}
\label{ee}
\lefteqn{{\bf Z}_k[J,K,\bar{K};\bar{g}]=J_1J_3J_4
\int{\cal D}h^{T}_{\mu\nu}\,{\cal D}\xi_\mu\,
{\cal D}\sigma\,{\cal D}\phi\,{\cal D}C^{T\mu}\,{\cal D}\bar{C}_\mu^T\,
{\cal D}\eta\,{\cal D}\bar{\eta}\,\exp\bigg[-S[h;\bar{g}]}
\nonumber\\
& &-S_{\rm gf}[h;\bar{g}]-S_{\rm gh}
[h,C,\bar{C};\bar{g}]-\Delta_k S\left[h,C,\bar{C};\bar{g}\right]
-S_{\rm source}[h,C,\bar{C},J,K,\bar{K};\bar{g}]
\bigg]
\end{eqnarray}
as well as the scale-dependent generating functional for the connected
Green's functions,
\begin{eqnarray}
\label{jj}
W_k\left[J,K,\bar{K};\bar{g}\right]=\ln {\bf Z}_k
\left[J,K,\bar{K};\bar{g}\right]\;,
\end{eqnarray}
may be viewed as functionals of either the fundamental or the component
sources. Furthermore, we may derive $k$-dependent 
classical fields for both fundamental and component fields
in terms of functional derivatives of $W_k$. In either case the 
$k$-dependent classical fields represent expectation values $\left<q\right>$ 
of quantum fields $q$, in the sense that all degrees of
freedom corresponding to momenta with $p^2>k^2$ have been averaged out. 
The classical fundamental fields are given by
\begin{eqnarray}
\label{kk}
\bar{h}_{\mu\nu}\equiv\left<h_{\mu\nu}\right>
=\frac{1}{\sqrt{\bar{g}}}\frac{\delta W_k}{\delta J^{\mu\nu}}
\;,\;\;\bar{v}_\mu\equiv\left<\bar{C}_\mu\right>
=\frac{1}{\sqrt{\bar{g}}}\frac{\delta W_k}{\delta K^\mu}
\;,\;\;v^\mu\equiv\left<C^\mu\right>=\frac{1}{\sqrt{\bar{g}}}
\frac{\delta W_k}{\delta\bar{K}_\mu}\;,
\end{eqnarray}
and the classical component fields are obtained as
\begin{eqnarray}
\label{072}
\varphi_i\equiv\left<\chi_i\right>=\frac{1}{\sqrt{\bar{g}}}
\frac{\delta W_k}{\delta {\cal J}^i}\;.
\end{eqnarray} 
Here we are making use of the shorthand notation $\chi\equiv(h^T,\xi,\sigma,
\phi,\bar{C}^T,C^T,\bar{\eta},\eta)$ for the quantum component fields, 
${\cal J}\equiv (J_{h^T},J_\xi,J_\sigma,J_\phi,K_{\bar{C}^T},\bar{K}_{C^T},
K_{\bar{\eta}},\bar{K}_\eta)$ for their sources and
$\varphi\equiv(\bar{h}^T,\bar{\xi},\bar{\sigma},\bar{\phi},\bar{v}^T,v^T,
\bar{\varrho},\varrho)$ for the classical component fields. We may  
reconstruct the classical fundamental fields from $\varphi$ according to
\begin{eqnarray}
\label{07}
\bar{h}_{\mu\nu}&=&\bar{h}_{\mu\nu}^T+\bar{D}_\mu\left[-\bar{D}^2-
\overline{\rm Ric}\right]^{-\frac{1}{2}}\bar{\xi}_\nu
+\bar{D}_\nu\left[-\bar{D}^2-
\overline{\rm Ric}\right]^{-\frac{1}{2}}\bar{\xi}_\mu
+\frac{1}{d}\bar{g}_{\mu\nu}\bar{\phi}\nonumber\\
& &+\bar{D}_\mu\bar{D}_\nu\left[(\bar{D}^2)^2+\frac{d-1}{d}\bar{D}_\rho
\bar{R}^{\rho\lambda}\bar{D}_\lambda\right]^{-\frac{1}{2}}\bar{\sigma}
-\frac{1}{d}\bar{g}_{\mu\nu}\bar{D}^2
\left[(\bar{D}^2)^2+\frac{d-1}{d}\bar{D}_\rho
\bar{R}^{\rho\lambda}\bar{D}_\lambda\right]^{-\frac{1}{2}}\bar{\sigma}\;,
\nonumber\\
\bar{v}_\mu&=&\bar{v}_\mu^T+\bar{D}_\mu\left(-\bar{D}^2\right)^{-\frac{1}{2}}
\bar{\varrho}\;,\;\;\;
v^\mu=v^{T\mu}+\bar{D}^\mu\left(-\bar{D}^2\right)^{-\frac{1}{2}}\varrho\;.
\end{eqnarray}

Performing a Legendre-transformation on $W_k$ with respect to $J_{\mu\nu}$,
$K^\mu$ and $\bar{K}_\mu$ leads to the following scale-dependent modification
of the effective action:
\begin{eqnarray}
\label{mm}
\widetilde{\Gamma}_k\left[\bar{h},v,\bar{v};\bar{g}\right]
=\left<J,\bar{h}\right>+\left<\bar{K},v\right>+\left<K,\bar{v}\right>
-W_k\left[J,K,\bar{K};\bar{g}\right]
\end{eqnarray}
Since eqs. (\ref{hh}), (\ref{071}) imply
\begin{eqnarray}
\label{079}
\left<{\cal J},\varphi\right>=\left<J,\bar{h}\right>
+\left<\bar{K},v\right>+\left<K,\bar{v}\right>
\end{eqnarray}
it is clear that eq. (\ref{mm}) also would result from Legendre-transforming 
$W_k$ with respect to the component sources. Denoting the corresponding 
Legendre-transform by $\widetilde{\Gamma}_k^{\rm comp}$ we have
$\widetilde{\Gamma}_k^{\rm comp}[\bar{h}^T,\bar{\xi},\bar{\sigma},\bar{\phi},
v^T,\bar{v}^T,\varrho,\bar{\varrho};\bar{g}]\equiv\widetilde{\Gamma}_k[\bar{h},
v,\bar{v};\bar{g}]$ where the arguments of $\widetilde{\Gamma}_k^{\rm comp}$
and $\widetilde{\Gamma}_k$ are related by (\ref{07}).

The effective average action proper, $\Gamma_k$, is defined as the difference 
between $\widetilde{\Gamma}_k$ and the cutoff action with the classical fields 
inserted \cite{W93,ym}:
\begin{eqnarray}
\label{nn}
\Gamma_k[g,\bar{g},v,\bar{v}]
\equiv\widetilde{\Gamma}_k\left[g-\bar{g},v,\bar{v};\bar{g}\right]
-\Delta_k S[g-\bar{g},v,\bar{v};\bar{g}]\;.
\end{eqnarray}
Here we expressed $\bar{h}_{\mu\nu}$ in terms of the classical counterpart 
$g_{\mu\nu}$ of
the quantum metric $\gamma_{\mu\nu}\equiv\bar{g}_{\mu\nu}+h_{\mu\nu}$ which, 
by definition, is given by
\begin{eqnarray}
\label{014}
g_{\mu\nu}\equiv\bar{g}_{\mu\nu}+\bar{h}_{\mu\nu}\;.
\end{eqnarray}

The main advantage of the background gauge is that it makes $\Gamma_k$ a gauge
invariant functional of its agruments \cite{Reu96}. It is invariant under
general coordinate transformations of the form
\begin{eqnarray}
\label{074}
\Gamma_k[\Phi]=\Gamma_k[\Phi+{\cal L}_u\Phi]\;\;,\;\;\;\Phi\equiv
\left(g_{\mu\nu},\bar{g}_{\mu\nu},v^\mu,\bar{v}_\mu\right)
\end{eqnarray}
where ${\cal L}_u$ is the Lie derivative with respect to the generating
vector field $u^\mu(x)$. Since general coordinate invariance ensures that no 
symmetry violating terms occur in the course of
the evolution of $\Gamma_k$ the class of consistent truncations is restricted
to those which involve only invariant field combinations. This is important
for practical applications of the evolution equation. 

We are mainly interested in the exclusively $g_{\mu\nu}$-dependent functional
\begin{eqnarray}
\label{024}
\bar{\Gamma}_k[g]\equiv\Gamma_k[g,g,0,0]\;.
\end{eqnarray}
In the limit $k\rightarrow 0$ it coincides with the conventional effective 
action $\Gamma[g_{\mu\nu}]$, the generator of the 1PI graviton Green's 
functions \cite{Ab81}: $\Gamma[g]=
\lim\limits_{k\rightarrow 0}\bar{\Gamma}_k[g]$. However, in order to derive
an exact evolution equation it is necessary to retain the dependence on the
ghost fields and $\bar{g}_{\mu\nu}$.

From the definition of the effective average action it follows that $\Gamma_k$
satisfies the integro-differential equation
\begin{eqnarray}
\label{022}
\lefteqn{\left.\exp\left\{-\Gamma_k[g,\bar{g},v,\bar{v}]\right\}
=\int{\cal D}h_{\mu\nu}\,{\cal D}C^\mu\,{\cal D}\bar{C}_\mu\,
\exp\right[-\widetilde{S}[h,C,\bar{C};\bar{g}]}\nonumber\\
& &\left.+\int d^dx\left\{\left(h_{\mu\nu}
-g_{\mu\nu}+\bar{g}_{\mu\nu}\right)\frac{\delta\Gamma_k}{\delta
\bar{h}_{\mu\nu}}
+\left(C^\mu-v^\mu\right)\frac{\delta\Gamma_k}{\delta v^\mu}
+\left(\bar{C}_\mu-\bar{v}_\mu\right)\frac{\delta\Gamma_k}
{\delta\bar{v}_\mu}\right\}\right]\nonumber\\
& &\times\exp\left\{-\Delta_k S\left[h-g+\bar{g},C-v,\bar{C}-\bar{v};\bar{g}
\right]\right\}
\end{eqnarray}
with
\begin{eqnarray}
\label{080}
\widetilde{S}[h,C,\bar{C};\bar{g}]\equiv S[\bar{g}+h]+S_{\rm gf}[h;\bar{g}]
+S_{\rm gh}[h,C,\bar{C};\bar{g}]\;.
\end{eqnarray}
Eq. (\ref{022}) may be derived by inserting the definition of $\Gamma_k$ into
(\ref{a}) and replacing the sources according to
\begin{eqnarray}
\label{081}
J^{\mu\nu}=\frac{1}{\sqrt{\bar{g}}}\,\frac{\delta\widetilde{\Gamma}_k}
{\delta\bar{h}_{\mu\nu}}\;,\;\;
K^\mu=-\frac{1}{\sqrt{\bar{g}}}\,\frac{\delta\widetilde{\Gamma}_k}
{\delta\bar{v}_\mu}\;,\;\;
\bar{K}_\mu=-\frac{1}{\sqrt{\bar{g}}}\,\frac{\delta\widetilde{\Gamma}_k}
{\delta v^\mu}\;.
\end{eqnarray}
\subsection{Derivation of the exact evolution equation}
\label{2E}
The exact renormalization group equation describes the change of the action
functional $\Gamma_k$ induced by a change in the scale $k$. It may be obtained
as follows. Differentiating the functional integral (\ref{ee}) with respect to
$t\equiv\ln k$ leads to
\begin{eqnarray}
\label{050}
-\partial_t W_k=\frac{1}{2}{\rm Tr}'\left[\sum\limits_{\zeta_1,\zeta_2
\in I_1}\!\!\left<\zeta_1\otimes\zeta_2\right>
\partial_t\left({\cal R}_k\right)_{\zeta_1\zeta_2}\right]
+\frac{1}{2}{\rm Tr}'\left[\sum\limits_{\psi_1,\psi_2\in I_2}\!\!
\left<\psi_1\otimes\psi_2\right>
\partial_t\left({\cal R}_k\right)_{\psi_1\psi_2}\right]
\end{eqnarray}
Here we used eq. (\ref{jj}) and adopted the matrix notation on the RHS of
(\ref{050}) which, in turn, can be expressed in terms of the Hessian
\begin{eqnarray}
\label{052}
(\widetilde{\Gamma}_k^{(2)})^{ij}(x,y)\equiv(-1)^{[j]}
\frac{1}{\sqrt{\bar{g}(x)
\bar{g}(y)}}\frac{\delta^2\widetilde{\Gamma}_k}{\delta\varphi_i(x)
\delta\varphi_j(y)}
\end{eqnarray}    
with $[j]=0$ for commuting fields $\varphi_j$ and $[j]=1$ for Grassmann fields
$\varphi_j$. Since the connected two point function
\begin{eqnarray}
\label{051}
(G_k)_{ij}(x,y)\equiv\left<\chi_i(x)\,\chi_j(y)\right>-\varphi_i(x)\,
\varphi_j(y)=\frac{1}{\sqrt{\bar{g}(x)\bar{g}(y)}}\frac{\delta^2W_k}
{\delta{\cal J}^i(x)\delta{\cal J}^j(x)}
\end{eqnarray}
and $\widetilde{\Gamma}_k^{(2)}$ are inverse matrices in the sense that
\begin{eqnarray}
\label{053}
\int d^dy\,\sqrt{\bar{g}(y)}\,(G_k)_{ij}(x,y)\,
(\widetilde{\Gamma}_k^{(2)})^{jl}(y,z)=\delta_i^l\,\frac{\delta(x-z)}
{\sqrt{\bar{g}(z)}}
\end{eqnarray}
we may replace the expectation values $\left<\chi_i(x)\chi_j(y)\right>$
appearing in eq. (\ref{050}) with $(\widetilde{\Gamma}_k^{(2)})^{-1}_{ij}(x,y)
+\varphi_i(x)\varphi_j(y)$. Then performing a Legendre-transformation
according to eq. (\ref{mm}) and subtracting the cutoff action $\Delta_k
S[\bar{h},v,\bar{v};\bar{g}]$ yields the desired exact renormalization group
equation:
\begin{eqnarray}
\label{oo}
\partial_t\Gamma_k\left[g,\bar{g},v,\bar{v}\right]
&=&\frac{1}{2}{\rm Tr}'\left[\sum\limits_{\zeta_1,\zeta_2\in \bar{I}_1}
\left(\Gamma_k^{(2)}[g,\bar{g},v,\bar{v}]
+{\cal R}_k\right)^{-1}_{\zeta_1\zeta_2}
\partial_t\left({\cal R}_k\right)_{\zeta_2\zeta_1}\right]
\nonumber\\
& &+\frac{1}{2}{\rm Tr}'\left[\sum\limits_{\psi_1,\psi_2\in \bar{I}_2}
\left(\Gamma_k^{(2)}[g,\bar{g},v,\bar{v}]
+{\cal R}_k\right)^{-1}_{\psi_1\psi_2}
\partial_t\left({\cal R}_k\right)_{\psi_2\psi_1}
\right]\;.
\end{eqnarray}
Here we wrote $({\cal R}_k)_{\zeta_1\zeta_2}
\equiv({\cal R}_k)_{\left<\zeta_1\right>\left<\zeta_2\right>}$,
$({\cal R}_k)_{\psi_1\psi_2}\equiv({\cal R}_k)_{\left<\psi_1\right>\left<
\psi_2\right>}$ 
%for $\zeta_1,\zeta_2\in I_1$,$\psi_1,\psi_2\in I_2$
and introduced the index sets
\begin{eqnarray}
\label{sets}
\bar{I}_1\equiv\{\bar{h}^T,\bar{\xi},\bar{\sigma},\bar{\phi}\}\;,\;\;\;
\bar{I}_2\equiv\{\bar{v}^T,v^T,\bar{\varrho},\varrho\}\;.
\end{eqnarray}

In a position space representation, the operators appearing on the RHS of the
flow equation are given by matrix elements whose traces are evaluated
according to
\begin{eqnarray}
\label{pp}
\int d^dx\,d^dy\,\sqrt{\bar{g}(x)}\,\sqrt{\bar{g}(y)}{\left(\left(
\Gamma_k^{(2)}+{\cal R}_k\right)^{-1}_{v^T\bar{v}^T}
\right)_{\mu x}}^{\nu y}
{\left(\partial_t\left({\cal R}_k\right)_{\bar{v}^Tv^T}
\right)_{\nu y}}^{\mu x}\;,
\end{eqnarray}
for instance. The notation adopted for the matrix elements
is similar to eq. (\ref{052}); for example,
\begin{eqnarray}
\label{qq}
{\left(\left(\Gamma_k^{(2)}\right)_{\bar{v}^Tv^T}\right)^{\mu x}}_{\nu y}=
\frac{1}{\sqrt{\bar{g}(y)}}\;\frac{\delta}{\delta v^{T\nu}(y)}\;
\frac{1}{\sqrt{\bar{g}(x)}}\frac{\delta\Gamma_k}{\delta\bar{v}^T_\mu(x)}\;.
\end{eqnarray}      
By virtue of the properties of ${\cal R}_k$ discussed above the traces 
appearing in the flow equation (\ref{oo}) are
perfectly convergent for all values of $k\le\widehat{k}$. 

Provided we impose the correct initial condition at the UV scale 
$k=\widehat{k}$ we can, in principle, determine the functional integral 
(\ref{a}) by integrating the flow
equation from $\widehat{k}$ down to $k$ and letting $k\rightarrow 0$, 
$\widehat{k}\rightarrow\infty$ after appropriate renormalizations. The initial 
condition $\Gamma_{\widehat{k}}$ can be 
obtained from the integro-differential equation (\ref{022}).
For sufficiently large values of $k$, the cutoff term in eq. (\ref{022})
strongly suppresses fluctuations with $(h,C,\bar{C})\neq (\bar{h},v,\bar{v})$
so that the main contribution to the functional integral results from small
fluctuations about $(h,C,\bar{C})=(\bar{h},v,\bar{v})$. This field
configuration corresponds to the global minimum of the total action in
the exponential of eq. (\ref{022}). Performing a saddle point expansion of the
functional integral about this minimum  leads to
\begin{eqnarray}
\label{082}
\Gamma_k[g,\bar{g},v,\bar{v}]=\widetilde{S}[g-\bar{g},v,\bar{v};\bar{g}]
-\frac{1}{2}\ln{\rm Det}'\left(\widetilde{S}^{(2)}+{\cal R}_k\right)
\end{eqnarray}
where the second term contains one-loop effects. For $k=\widehat{k}\rightarrow
\infty$ they amount to an often unimportant shift in the bare parameters of
$\widetilde{S}$ which can be ignored usually. For finite $\widehat{k}$,
additional contributions from the determinant occur which are suppressed by
inverse powers of $\widehat{k}$ \cite{RW97}. Therefore we obtain the initial
value for $\widehat{k}\rightarrow\infty$
\begin{eqnarray}
\label{023}
\Gamma_{\widehat{k}}[g,\bar{g},v,\bar{v}]=S[g]+S_{\rm gf}[g-\bar{g};
\bar{g}]+S_{\rm gh}[g-\bar{g},v,\bar{v};\bar{g}]\;.
\end{eqnarray}
At the level of the functional $\bar{\Gamma}_k[g_{\mu\nu}]$ this initial
condition boils down to
\begin{eqnarray}
\label{083}
\bar{\Gamma}_{\widehat{k}}[g]=S[g]\;.
\end{eqnarray}

So far we assumed the fundamental action to be positive definite. However,
the Einstein-Hilbert action, for instance, does not have this property which
is due to the appearance of a ``wrong-sign'' kinetic term associated with the
conformal factor. In such cases it is nevertheless possible to formulate a
well-defined evolution equation if the signs of the cutoff operators
${\cal R}_k$ are properly adjusted \cite{Reu96}. We will return to this point 
in the next section.
\subsection{A special case: Einstein backgrounds}
\label{2F}
Before continuing we summarize the simplifications that occur for Einstein
backgrounds, for which $\bar{R}_{\mu\nu}=C\bar{g}_{\mu\nu}$ with
$C$ a constant. In this case the decomposition (\ref{f}) of $h_{\mu\nu}$ is
completely orthogonal. In fact, thanks to the Einstein condition, the 
$\widehat{\xi}_\mu$-$\widehat{\sigma}$ mixing terms in the inner product 
(\ref{08}) vanish so that $h^T_{\mu\nu}$, $h^{LT}_{\mu\nu}$, $h^{LL}_{\mu\nu}$
and $h^{Tr}_{\mu\nu}$ form an orthogonal set. 

Furthermore, for Einstein spaces the Jacobians appearing in the path integral 
(\ref{ee}) cancel, at least up to an (infinite) constant which can be absorbed
into the normalization of the integration measure. This can be seen as follows:
\begin{eqnarray}
\label{aaa}
\lefteqn{J_1=N\sqrt{\det\nolimits_{(1T,0)}' M^{(\mu,\nu)}}}\nonumber\\
&=&\bigg(\int{\cal D}\widehat{\xi}_\mu\,{\cal D}\widehat{\sigma}
\,\exp\bigg[-\int d^dx\,\sqrt{\bar{g}}\,
\bigg\{-2\widehat{\xi}_\mu\left(\bar{D}^2+C
\right)\widehat{\xi}^{\mu}
+\widehat{\sigma}
\left(\frac{d-1}{d}(\bar{D}^2)^2+C\,\bar{D}^2\right)
\widehat{\sigma}\bigg\}\bigg]\bigg)^{-1}\nonumber\\
&=&N_1\sqrt{\det\nolimits_{(1T)}'\left(-\bar{D}^2-C\right)}
\sqrt{\det\nolimits_{(0)}'\left(-\bar{D}^2\right)}\sqrt{\det\nolimits_{(0)}'
\left(-\bar{D}^2-\frac{d\,C}{d-1}\right)}=N_2\,J_3^{-1}\,J_4^{-1}
\end{eqnarray}
Here $N_1$ and $N_2$ are unimportant constants so that $J_1J_3J_4$ is indeed 
field independent. 

Finally, for Einstein spaces the field redefinitions in the gravitational 
sector take the form
\begin{eqnarray}
\label{084}
\xi^\mu=\sqrt{-\bar{D}^2-C}\;\widehat{\xi}^\mu\;,\;\;
\sigma=\sqrt{-\bar{D}^2}\sqrt{-\bar{D}^2-\frac{d\,C}{d-1}}\;\widehat{\sigma}\;.
\end{eqnarray}
As a consequence, we find that $\bar{D}_\mu\xi^\mu=\sqrt{-\bar{D}^2-2C}\;
\bar{D}_\mu\widehat{\xi}^\mu$. Thus, transversality of $\widehat{\xi}^\mu$ 
implies that $\xi^\mu$ is transverse as well.
\section{Truncations and cutoffs}
\renewcommand{\theequation}{3.\arabic{equation}}
\setcounter{equation}{0}
\label{S3}
\subsection{A general class of truncations} 
\label{3A}
In practical applications of the exact evolution equation one encounters the
problem of dealing with an infinite system of coupled differential equations
since the evolution equation describes trajectories in an infinite dimensional
space of action functionals. In general it is impossible to find an exact
solution so that we are forced to rely on approximations. A powerful 
nonperturbative approximation scheme is the truncation of the parameter space,
i.e. only a finite number of couplings is considered. In this manner the
renormalization group flow of $\Gamma_k$ is projected onto a finite-dimensional
subspace of action functionals. In practice one makes an ansatz for $\Gamma_k$
that comprises only a few couplings and inserts it on both sides of 
eq. (\ref{oo}), thereby obtaining a truncated evolution equation. By 
projecting the RHS of this equation onto the space of operators appearing
on the LHS one arrives at a set of coupled differential equations
for the couplings taken into account. 

As discussed in refs. \cite{RW97,EHW}, Ward identities provide an important 
tool for judging the admissability and quantitative reliability of a given 
truncation; approximate solutions of the flow equation are not 
necessarily consistent with the Ward identities, in contrast to the exact 
solution. Therefore, only those truncations which are
indeed consistent with the Ward identities, at least up to a certain degree of
accuracy, will yield reliable results. The Ward identities
to be considered here are modified by additional
terms coming from the cutoff which are not present in the ordinary identities.
Since $\Delta_k S$ vanishes as $k\rightarrow 0$ the ordinary Ward identities
are recovered in this limit.

In ref. \cite{Reu96} the modified Ward identities were derived for the 
gravitational effective average action $\Gamma_k[g,\bar{g},v,\bar{v};\beta,
\tau]$ where $\beta$ and $\tau$ are auxiliary sources for the BRS variations of
the graviton and the ghosts which are needed in order to formulate the Ward
identities. Setting $\beta=\tau=0$ in the argument of this more general 
functional we obtain the action  $\Gamma_k[g,\bar{g},v,\bar{v}]$ discussed in
the present paper. (It would be straightforward to include the 
$\beta$,$\tau$-sources also in the new formulation of the flow equation, but
we will not need them in the following.) 

In \cite{Reu96} the Ward identities were used to test the consistency of 
truncations of the form 
\begin{eqnarray}
\label{030}
\Gamma_k[g,\bar{g},v,\bar{v}]=\bar{\Gamma}_k[g]+\widehat{\Gamma}_k[g,\bar{g}]
+S_{\rm gf}[g-\bar{g};\bar{g}]+S_{\rm gh}[g-\bar{g},v,\bar{v};\bar{g}]
\end{eqnarray}
with $\bar{\Gamma}_k[g]$ defined as in eq. (\ref{024}).
The term $\widehat{\Gamma}_k[g,\bar{g}]$ encodes  
the quantum corrections to the gauge fixing term. This interpretation of
$\widehat{\Gamma}_k[g,\bar{g}]$ is obvious because for $\bar{g}\neq g$ 
the purely gravitational part of eq. (\ref{030}) implies 
$\Gamma_k[g,\bar{g},0,0]-\Gamma_k[g,g,0,0]
=\widehat{\Gamma}_k[g,\bar{g}]+S_{\rm gf}[g-\bar{g};\bar{g}]$. 
By definition, $\widehat{\Gamma}_k[g,g]=0$. In the ansatz (\ref{030})
the ghost dependence has been extracted in terms of the classical 
$S_{\rm gh}$, thereby neglecting the evolution of the ghost action. This
guarantees that the initial condition (\ref{023}) is satisfied automatically 
in the ghost sector. In the gravitational sector it requires 
$\bar{\Gamma}_{\widehat{k}}=S$,
$\widehat{\Gamma}_{\widehat{k}}=0$. For truncations of the type (\ref{030}) 
the Ward identities demand that $\bar{\Gamma}_k[g]$ is a gauge invariant 
functional of $g_{\mu\nu}$ and they yield a constraint equation for 
$\widehat{\Gamma}_k[g,\bar{g}]$. To lowest order, this equation is solved by
$\widehat{\Gamma}_k=0$ $\forall k\le\widehat{k}$. In the Einstein-Hilbert
truncation we go
beyond this approximation and set $\widehat{\Gamma}_k\propto S_{\rm gf}$ with
a constant of proportionality which vanishes at $k=\widehat{k}$; it 
takes the running of the graviton's wave function normalization into account 
(see below).

Inserting the ansatz (\ref{030}) into the exact evolution equation (\ref{oo}) 
leads to 
a truncated renormalization group equation which describes the evolution of
$\Gamma_k$ in the subspace of action functionals spanned by (\ref{030}).
The equation governing the evolution of the purely gravitational action
\begin{eqnarray}
\label{033}
\Gamma_k[g,\bar{g}]\equiv\Gamma_k[g,\bar{g},0,0]=\bar{\Gamma}_k[g]+S_{\rm gf}
[g-\bar{g};\bar{g}]+\widehat{\Gamma}_k[g,\bar{g}]
\end{eqnarray}
reads
\begin{eqnarray}
\label{032}
\partial_t\Gamma_k\left[g,\bar{g}\right]
&=&\frac{1}{2}{\rm Tr}'\left[\sum\limits_{\zeta_1,\zeta_2\in\bar{I}_1}
\left(\Gamma_k^{(2)}[g,\bar{g}]
+{\cal R}_k\right)^{-1}_{\zeta_1\zeta_2}
\partial_t\left({\cal R}_k\right)_{\zeta_2\zeta_1}\right]\nonumber\\
& &+\frac{1}{2}{\rm Tr}'\left[\sum\limits_{\psi_1,\psi_2\in\bar{I}_2}
\left(S_{\rm gh}^{(2)}[g,\bar{g}]
+{\cal R}_k\right)^{-1}_{\psi_1\psi_2}
\partial_t\left({\cal R}_k\right)_{\psi_2\psi_1}
\right]\;.
\end{eqnarray}
Here $\Gamma^{(2)}_k$ and $S_{\rm gh}^{(2)}$ are the Hessians of $\Gamma_k[g,
\bar{g}]$ and $S_{\rm gh}[\bar{h},v,\bar{v},\bar{g}]$ with respect to the 
gravitational and the ghost component fields, respectively. They are taken 
at fixed $\bar{g}_{\mu\nu}$.
\subsection{Specification of the cutoff}
\label{3B}
In order to obtain a tractable evolution equation for a given truncation it
is convenient to use a cutoff which is
adapted to this truncation but still has the general suppression 
properties described in subsection \ref{2D}. It is desirable
to start from a definition of $\Delta_kS$ that brings about the correct 
suppression of low-momentum modes for a class of truncations and of 
gravitational backgrounds which is as large as possible. 

A convenient, adapted
cutoff can be found by the following rule \cite{Reu96,DP97}. Given a 
truncation, we assume that for $\bar{g}=g$ the kinetic operators of all modes
with a definite helicity are of the form 
$(\Gamma_k^{(2)})_{ij}=f_{ij}(-\bar{D}^2,k,\ldots)$ where $\{f_{ij}\}$ is a 
set of c-number functions and the indices $i,j$ refer to the different types of
fields. (The difficulty of bringing $\Gamma_k^{(2)}$ to this form is one of
the main reasons for using the TT-decomposition. At least for maximally 
symmetric spaces it allows us to eliminate all covariant derivatives 
which do not appear as a Laplacian $\bar{D}^2\equiv\bar{g}^{\mu\nu}\bar{D}_\mu
\bar{D}_\nu$.) Then we choose the cutoff in such a way that the structure 
\begin{eqnarray}
\label{090}
(\Gamma_k^{(2)}+{\cal R}_k)_{ij}=f_{ij}
\left(-\bar{D}^2+k^2R^{(0)}(-\bar{D}^2/k^2),k,\ldots\right)
\end{eqnarray}
is achieved. Here the function $R^{(0)}(y)$, $y=-\bar{D}^2/k^2$, describes the
details of the mode suppression; it is required to satisfy the boundary 
conditions $R^{(0)}(0)=1$ and $\lim\limits_{y\rightarrow\infty}R^{(0)}(y)
=0$, but is arbitrary otherwise. By virtue of eq. (\ref{090}), the inverse
propagator of a field mode with covariant momentum square $p^2=
-\bar{D}^2$ is given by $p^2+k^2R^{(0)}(p^2/k^2)$ which equals $p^2$ for 
$p^2\gg k^2$ and $p^2+k^2$ for $p^2\ll k^2$. This means that the small-$p^2$ 
modes, and only those, have acquired a mass $\propto k$ which leads to the 
desired suppression.

In the next section we shall see in detail that for the truncations used in
the present paper we can comply with the above rule by using the following
cutoff operator:
\begin{eqnarray}
\label{uu}
\left({\cal R}_k\right)^{\mu\nu\alpha\beta}_{\bar{h}^T\bar{h}^T}
&=&\frac{1}{2}{\cal Z}_k^{\bar{h}^T\bar{h}^T}\kappa^2\,
\left(\bar{g}^{\mu\alpha}\bar{g}^{\nu\beta}+\bar{g}^{\mu\beta}
\bar{g}^{\nu\alpha}\right)k^2 R^{(0)}\left(-\bar{D}^2/k^2\right)\;,\nonumber\\
\left({\cal R}_k\right)^{\mu\nu}_{\bar{\xi}\bar{\xi}}&=&
{\cal Z}_k^{\bar{\xi}\bar{\xi}}\kappa^2\,\bar{g}^{\mu\nu}k^2 R^{(0)}
\left(-\bar{D}^2/k^2\right)\;,\nonumber\\
\left({\cal R}_k\right)_{\bar{\sigma}\bar{\sigma}}&=&
{\cal Z}_k^{\bar{\sigma}\bar{\sigma}}\kappa^2\,k^2 R^{(0)}
\left(-\bar{D}^2/k^2\right)\;,\nonumber\\
\left({\cal R}_k\right)_{\bar{\phi}\bar{\sigma}}
&=&\left({\cal R}_k\right)_{\bar{\sigma}\bar{\phi}}^\dagger
={\cal Z}_k^{\bar{\phi}\bar{\sigma}}\kappa^2\left[\sqrt{\left(\bar{P}_k
+\frac{d}{d-1}\bar{D}_\mu\bar{R}^{\mu\nu}\bar{D}_\nu(-\bar{D}^2)^{-1}\right)
\bar{P}_k}\right.\nonumber\\
& &-\left.\sqrt{\left(\bar{D}^2\right)^2+\frac{d}{d-1}
\bar{D}_\mu\bar{R}^{\mu\nu}\bar{D}_\nu}\right]\;,\nonumber\\
\left({\cal R}_k\right)_{\bar{\phi}\bar{\phi}}&=&{\cal Z}_k^{\bar{\phi}
\bar{\phi}}\kappa^2\,k^2 R^{(0)}\left(-\bar{D}^2/k^2\right)\;,\nonumber\\
\left({\cal R}_k\right)^{\mu\nu}_{\bar{v}^Tv^T}&=&-\left({\cal R}_k
\right)^{\mu\nu}_{v^T\bar{v}^T}={\cal Z}_k^{\bar{v}^T v^T}\,
\bar{g}^{\mu\nu}\,k^2 R^{(0)}\left(-\bar{D}^2/k^2\right)\;,\nonumber\\
\left({\cal R}_k\right)_{\bar{\varrho}\varrho}
&=&-\left({\cal R}_k\right)_{\varrho\bar{\varrho}}
={\cal Z}_k^{\bar{\varrho}\varrho}\,k^2 R^{(0)}\left(-\bar{D}^2/k^2\right)\;.
\end{eqnarray}
Here $\bar{P}_k$ is defined as
\begin{eqnarray}
\label{043}
\bar{P}_k\equiv -\bar{D}^2+k^2R^{(0)}(-\bar{D}^2/k^2)\;.
\end{eqnarray}
The remaining cutoff operators not listed in eq. (\ref{uu}) are set  
to zero. The ${\cal Z}_k$'s are constants which, again by using (\ref{090}),
will be fixed in terms of the generalized couplings appearing in the ansatz
for $\Gamma_k$. The cutoff (\ref{uu}) is inspired by the ${\cal R}_k$ used in
\cite{DP97} for $S^d$.

If  eq. (\ref{090}) allows us to choose ${\cal Z}_k^{\zeta\zeta}
>0$ for all $\zeta\in\{\bar{h}^T,\bar{\xi},\bar{\sigma},\bar{\phi}\}$ and 
${\cal Z}_k^{\bar{\phi}\bar{\sigma}}=0$ one obtains a positive definite
$\Delta_k S$ in the gravitational sector. In this case $\exp(-\Delta_k S)$
is a damped exponential which indeed suppresses the contributions from the 
low-momentum modes. In the following sections we shall focus on the 
Einstein-Hilbert truncation for $\Gamma_k$ which suffers from the conformal
factor problem: its kinetic term for $\bar{\phi}$ is negative definite. As a
consequence, eq. (\ref{090}) forces us to work with a 
${\cal Z}_k^{\bar{\phi}\bar{\phi}}<0$. Hence, in the $\phi$-sector,
$\Delta_k S$ is negative definite and, at least at a naive level, 
$\exp(-\Delta_k S)$ seems to enhance rather than suppress the low-momentum
modes. As we discussed in detail in ref. \cite{Reu96} we nevertheless believe
that the rule (\ref{090}), i.e. allowing for ${\cal Z}_k^{\bar{\phi}
\bar{\phi}}<0$, is correct also in this case. We emphasize that the
RHS of the flow equation, contrary to the Euclidean path integral, is 
perfectly well-defined even if $S$ and $\Gamma_k$ are not positive definite.

At this point it should be mentioned that the situation with respect to the
positivity of the action improves considerably
by including higher-derivative terms in $S$ and the truncated $\Gamma_k$
since these actions are bounded below, provided we choose the correct sign in
front of these higher-derivative terms. Furthermore, their quadratic forms are
positive definite at least for sufficiently large momenta, and so is the
cutoff. For a study of the evolution equation for $R^2$-gravity we refer to
\cite{LR3}.

As compared to the original paper \cite{Reu96}, the cutoff (\ref{uu}) has a 
rather different structure which is due to the fact that it is formulated in 
terms of the component fields arising form the TT-decomposition. Contrary to
the original one of ref. \cite{Reu96}, the new cutoff (\ref{uu}) is defined 
for {\it all} values of $\alpha$. This is one of the main advantages of the new
approach.

Note that in refs. 
\cite{sven,souma2} where the TT-decomposition was used on $S^d$ the actual 
construction of the effective average action and its RG equation was omitted
and has been replaced by an ad hoc modification of the standard one-loop 
determinants. No $\Delta_k S$ has been specified at the component field level.
Hence the scale dependent action constructed in this manner has no reason to 
respect the general properties of an effective average action \cite{ber}. 
Despite the use of the component fields in \cite{sven,souma2} their cutoff 
seems to be more similar to the original one in \cite{Reu96} than to the new 
one of the present paper. In fact, it represents an $\alpha$-dependent 
generalization of the cutoff in \cite{Reu96}, in the sense that the latter is 
recovered from the one of refs. \cite{sven,souma2} by setting $\alpha=1$.

From now on we will refer to the cutoff used in the original paper \cite{Reu96}
and in \cite{souma1,sven,souma2} as the cutoff of {\bf type A}. However, one 
has to keep in mind that the existence of a corresponding $\Delta_k S$ is 
guaranteed only for $\alpha=1$, i.e. the case considered in 
\cite{Reu96,souma1}. Furthermore the cutoff (\ref{uu}) of the present paper 
will be referred to as the cutoff {\bf type B}; it is defined for all 
values of $\alpha$.

Each cutoff type contains the shape function $R^{(0)}$. A particularly
suitable choice is the exponential shape function
\begin{eqnarray}
\label{expcut}
R^{(0)}(y)=y\left[\exp(y)-1\right]^{-1}\;.
\end{eqnarray}
In order to check the scheme independence of universal quantities we employ a
one-parameter ``deformation'' of (\ref{expcut}), the {\bf class of
exponential shape functions},
\begin{eqnarray}
\label{H6}
R^{(0)}(y;s)=sy\left[\exp(sy)-1\right]^{-1}\;,
\end{eqnarray}
with $s$ parametrizing the profile of $R^{(0)}$ \cite{souma2}. Another 
admissible choice is
the following {\bf class of shape functions with compact support}: 
\begin{eqnarray}
\label{supp}
R^{(0)}(y;b)=\left\{\begin{array}{ll}1 & \;\;\;y\le b\\
\exp\left[(y-1.5)^{-1}\exp\left[\left(b-y\right)^{-1}\right]\right] & \;\;\;b
<y<1.5\\0 & \;\;\;y\ge 1.5
\end{array}\right.\;.
\end{eqnarray}
Here $b\in [0,1.5)$ parametrizes the profile of $R^{(0)}$.

For our analysis of the flow equation in section \ref{S5} we shall 
use both cutoff types with both classes of shape functions.

\section{The Einstein-Hilbert truncation}
\renewcommand{\theequation}{4.\arabic{equation}}
\setcounter{equation}{0}
\label{S4}
\subsection{The ansatz}
\label{4A}
In this section we use a simple truncation to derive the renormalization
group flow of the Newton and the cosmological ``constant'' by means of the 
truncated flow equation (\ref{032}). In our example we assume that, at the 
UV scale $\widehat{k}\rightarrow\infty$, gravity is described by 
the classical Einstein-Hilbert action in $d$ dimensions,
\begin{eqnarray}
\label{069}
\bar{\Gamma}_{\widehat{k}}[g]=S[g]=\frac{1}{16\pi\bar{G}}\int d^dx\,\sqrt{g}
\left\{-R(g)+2\bar{\lambda}\right\}\;.
\end{eqnarray}
For the investigation of the evolution of $\Gamma_k[g,\bar{g}]$ towards smaller
scales $k<\widehat{k}$ we consider a truncated action functional of the 
following form:
\begin{eqnarray}
\label{37}
\Gamma_k[g,\bar{g}]&=&2\kappa^2 Z_{Nk}\int d^dx\,\sqrt{g}\left\{
-R(g)+2\bar{\lambda}_k\right\}\nonumber\\
& &+\kappa^2 \frac{Z_{Nk}}{\alpha}\int d^dx\,\sqrt{\bar{g}}\,\bar{g}^{\mu\nu}
\left({\cal F}_\mu^{\alpha\beta}
g_{\alpha\beta}\right)\left({\cal F}_\nu^{\rho\sigma}g_{\rho\sigma}\right)\;.
\end{eqnarray}
Eq. (\ref{37}) is obtained from $S+S_{\rm gf}$ by replacing 
\begin{eqnarray}
\label{042}
\bar{G}\rightarrow G_k\equiv Z_{Nk}^{-1}\,\bar{G}\;,\;\;\bar{\lambda}
\rightarrow\bar{\lambda}_k\;,\;\;\alpha\rightarrow Z_{Nk}^{-1}\alpha
\end{eqnarray}
so that its form agrees with that of the gravitational sector of the ansatz
(\ref{030}) with 
\begin{eqnarray}
\label{091}
\widehat{\Gamma}_k[g,\bar{g}]=\kappa^2 \frac{Z_{Nk}-1}{\alpha}\int d^dx\,
\sqrt{\bar{g}}\,\bar{g}^{\mu\nu}\left({\cal F}_\mu^{\alpha\beta}
g_{\alpha\beta}\right)\left({\cal F}_\nu^{\rho\sigma}g_{\rho\sigma}\right)\;.
\end{eqnarray}

Generally speaking also the gauge fixing parameter $\alpha$ should be treated 
as a running quantity, $\alpha\rightarrow\alpha_k$. Fortunately there is a 
simple shortcut which avoids an explicit computation of the corresponding
$\mbox{\boldmath $\beta$}$-function. In fact, there are general arguments
showing that $\alpha$ should have a (IR attractive) fixed point at $\alpha_*
=0$. This means that the initial condition $\alpha_{\widehat{k}}=0$ leads to
$\alpha_k=0$ for all $k\le\widehat{k}$. Thus, even using the truncation with
a constant $\alpha$, we can take the correct ``flow'' of the gauge fixing term
into account simply by setting $\alpha=0$. 

In Yang-Mills theory the existence
of the fixed point $\alpha_*=0$ has been demonstrated for a truncation 
containing a covariant gauge fixing \cite{EHW}, while for the axial gauge a
nonperturbative proof is available \cite{litim}. The following general 
argument\renewcommand{\baselinestretch}{1}\small\normalsize\footnote{We are
grateful to J. M. Pawlowski for a discussion of this point.} 
\renewcommand{\baselinestretch}{1.5}\small\normalsize
suggests that this fixed point should exist in any gauge theory, including
gravity \cite{litim}. In the ordinary functional integral, the limit $\alpha
\rightarrow 0$ corresponds to a sharp implementation of the gauge fixing 
condition, i.e. $\exp(-S_{\rm gf})$ becomes proportional to $\delta[F_\mu]$.
The domain of the $\int{\cal D}h_{\mu\nu}$-integration consists of those 
$h_{\mu\nu}$'s which satisfy the gauge condition exactly, $F_\mu=0$. Adding the
IR cutoff at $k$ amounts to suppressing some of the $h_{\mu\nu}$-modes while
retaining the others. But since all of them satisfy $F_\mu=0$, it is clear that
a variation of $k$ cannot change the domain of the $h_{\mu\nu}$-integration.
The delta-functional $\delta[F_\mu]$ continues to be present for any value of
$k$ if it was there originally. Hence $\alpha$ vanishes for arbitrary $k$.

\subsection{Projecting the flow equation}
\label{4B}
The $k$-dependent couplings in eq. (\ref{042}) satisfy the initial conditions 
$\bar{\lambda}_{\widehat{k}}=\bar{\lambda}$ and $Z_{N\widehat{k}}=1$ which 
implies $G_{\widehat{k}}=\bar{G}$. Here the UV scale $\widehat{k}$ is taken to 
be large but 
finite. The evolution of $Z_{Nk}$ and $\bar{\lambda}_k$ towards
smaller scales may now be determined as follows. As a first step the ansatz
(\ref{37}) is inserted into both sides of the truncated flow equation 
(\ref{032}). Then we may set $g_{\mu\nu}=\bar{g}_{\mu\nu}$. As a consequence,
the gauge fixing term drops out from the LHS which then reads
\begin{eqnarray}
\label{040}
\partial_t\Gamma_k[\bar{g},\bar{g}]=2\kappa^2\int d^dx\,\sqrt{\bar{g}}
\left[-\bar{R}(\bar{g})\,\partial_t
Z_{Nk}+2\partial_t\left(Z_{Nk}\bar{\lambda}_k\right)\right]\;.
\end{eqnarray}
Performing a derivative expansion on the RHS we may extract those 
contributions which are proportional to the operators spanning the LHS, i.e.
$\int d^dx\sqrt{g}$ and $\int d^dx\sqrt{g}R$. Then, comparing the coefficients
of these operators yields a system of coupled differential equations for
$Z_{Nk}$ and $\bar{\lambda}_k$. It describes the projection of the 
renormalization group flow onto the two-dimensional subspace of the space of 
all action functionals which is spanned by $\int d^dx\sqrt{g}$ and 
$\int d^dx\sqrt{g}R$.

It is important to note that during this calculation we may insert any metric
$\bar{g}_{\mu\nu}$ that is general enough to allow for a unique identification
of the operators $\int d^dx\sqrt{g}$ and $\int d^dx\sqrt{g}R$. In practice it 
proves particularly convenient to exploit this freedom by choosing the  
gravitional background to be maximally symmetric. Such spaces form a special 
class of Einstein spaces and are characterized by
\begin{eqnarray}
\label{xx}
\bar{R}_{\mu\nu\rho\sigma}=\frac{\bar{R}}{d(d-1)}\left(\bar{g}_{\mu\rho}
\bar{g}_{\nu\sigma}-\bar{g}_{\mu\sigma}\bar{g}_{\nu\rho}\right)\;,
\;\;\;\bar{R}_{\mu\nu}=\frac{\bar{R}}{d}\bar{g}_{\mu\nu}
\end{eqnarray}
with the curvature scalar $\bar{R}$ considered a constant number rather than a
functional of the metric. In the following we
restrict our considerations to maximally symmetric spaces with positive 
curvature scalar $\bar{R}>0$, i.e. $d$-spheres $S^d$. For $d$ fixed, $S^d$ is 
parametrized by the radius $r$ of the spheres, which is related 
to the curvature scalar and the volume in the usual way,
\begin{eqnarray}
\label{035}
\bar{R}=\frac{d(d-1)}{r^2}\;,\;\;\;
\int d^dx\,\sqrt{\bar{g}}=\frac{\Gamma\left(\frac{d}{2}\right)}{\Gamma(d)}
(4\pi r^2)^{\frac{d}{2}}\;.
\end{eqnarray}

Before continuing with the evaluation of the RHS of the flow equation we have
to comment on the properties of fields defined on spherical backgrounds.
According to appendix \ref{harm}, we may expand the quantum and classical
component fields in terms of spherical harmonics $T^{lm}_{\mu\nu}$, 
$T^{lm}_\mu$, $T^{lm}$, which form complete sets of orthogonal eigenfunctions 
with respect to the corresponding covariant Laplacians. The expansions of 
$h^T_{\mu\nu}$, $\phi$, $C^\mu$, $\bar{C}_\mu$ and their classical counterparts
can be inferred directly from eq. (\ref{067}). The remaining component 
fields are expanded according to
\begin{eqnarray}
\label{092}
\xi_\mu(x)&=&
\sum\limits_{l=2}^{\infty}\sum\limits_{m=1}^{D_l(d,1)}\xi_{lm}\,
T^{lm}_\mu(x)\;,\;\;\;
\sigma(x)=\sum\limits_{l=2}^{\infty}\sum\limits_{m=1}^{D_l(d,0)}\sigma_{lm}\,
T^{lm}(x)\;,\nonumber\\
\eta(x)&=&\sum\limits_{l=1}^{\infty}\sum\limits_{m=1}^{D_l(d,0)}\eta_{lm}\,
T^{lm}(x)\;,\;\;\;
\bar{\eta}(x)=\sum\limits_{l=1}^{\infty}\sum\limits_{m=1}^{D_l(d,0)}
\bar{\eta}_{lm}\,T^{lm}(x)\;.
\end{eqnarray}
Similar expansions hold for the associated classical fields.

Note that in eq. (\ref{092}) the summations do {\it not} start at $l=1$ for 
vectors and at $l=0$ for scalars as in eq. (\ref{067}), but at $l=2$ for 
$\xi_\mu$ and $\sigma$, and at $l=1$ for the scalar ghost fields. The modes 
omitted here are the KV's $(T^{l=1,m}_\mu)$, the solutions of the scalar 
equation (\ref{010}) which are in one-to-one correspondence with the PCKV's 
$(T^{l=1,m})$, and the constants $(T^{l=0,m=1})$. As discussed in subsection 
\ref{2B}, the fundamental fields obtain no contribution from these modes. 
Therefore they have to be excluded from eq. (\ref{092}).

This exclusion is also of importance for the momentum dependent field
redefinitions (\ref{y}) because they would not be well-defined otherwise, as
can be seen e.g. from
\begin{eqnarray}
\label{095}
\widehat{\sigma}(x)=\sum\limits_{l=2}^{\infty}\sum\limits_{m=1}^{D_l(d,0)}
\frac{\sigma_{lm}}{\sqrt{\Lambda_l(d,0)\left(\Lambda_l(d,0)-\frac{\bar{R}}{d-1}
\right)}}\,T^{lm}(x)\;.
\end{eqnarray}
Eq. (\ref{095}) follows from inverting eq. (\ref{y}) and then inserting eq. 
(\ref{092}). The eigenvalues corresponding to the modes excluded, i.e. 
$\Lambda_0(d,0)=0$ and $\Lambda_1(d,0)=\frac{\bar{R}}{d-1}$ (see table 1 in 
appendix \ref{harm}), would lead to a vanishing denominator in eq. 
(\ref{095}). Similar arguments hold for the other fields in (\ref{y}).

We may now split the quantum field $\phi$ into
a part $\phi_1$ spanned by the same set of eigenfunctions as $\sigma$,
and a part $\phi_0$ containing the contributions from the remaining modes:
\begin{eqnarray}
\label{112}
\phi(x)=\phi_0(x)+\phi_1(x)\,,\;
\phi_0(x)=\sum\limits_{l=0}^{1}\sum\limits_{m=1}^{D_l(d,0)}\phi_{lm}
\,T^{lm}(x)\,,\;
\phi_1(x)=\sum\limits_{l=2}^{\infty}\sum\limits_{m=1}^{D_l(d,0)}
\phi_{lm}\,T^{lm}(x)
\end{eqnarray}
The orthogonality of the spherical harmonics implies $\left<\phi_1,\phi_0
\right>=\left<\sigma,\phi_0\right>=0$ so that $\left<\phi,\phi\right>
=\left<\phi_0,\phi_0\right>+\left<\phi_1,\phi_1\right>$ and in particular 
$\left<\sigma,\phi\right>=\left<\sigma,\phi_1\right>$. As a consequence, 
decomposing $\phi$ according to eq. (\ref{112}) ensures that any nonzero term 
bilinear in the scalar fields is of such a form that the scalars involved are 
spanned by the same set of eigenfunctions. The same is true for the 
corresponding classical fields $\bar{\phi}_0$ and $\bar{\phi}_1$. 
\subsection{Evaluation of the functional trace}
\label{4C}
Let us now return to the evaluation of the flow equation. On the RHS we need
the operator $\Gamma^{(2)}_k[g,\bar{g}]$. For our purposes it is sufficient
to determine this operator at $g_{\mu\nu}=\bar{g}_{\mu\nu}$. It may be
derived by expanding $\Gamma_k[g,\bar{g}]$ according to
\begin{eqnarray}
\label{041}
\Gamma_k[\bar{g}+\bar{h},\bar{g}]=\Gamma_k[\bar{g},\bar{g}]+{\cal O}(\bar{h})
+\Gamma_k^{\rm quad}[\bar{h};\bar{g}]+{\cal O}(\bar{h}^3)
\end{eqnarray}
and retaining only the part quadratic in $\bar{h}_{\mu\nu}$, i.e. 
$\Gamma_k^{\rm quad}[\bar{h};\bar{g}]$. For our truncation it takes the form
\begin{eqnarray}
\label{38}
\lefteqn{\Gamma_k^{\rm quad}\left[\bar{h};\bar{g}\right]
=\kappa^2 Z_{Nk}\int d^dx\,\sqrt{\bar{g}}\,\bar{h}_{\mu\nu}\Bigg\{
-\left[\frac{1}{2}\delta^\mu_\rho\delta_\sigma^\nu+\frac{1-2\alpha}{4\alpha}
\bar{g}^{\mu\nu}\bar{g}_{\rho\sigma}\right]\bar{D}^2}\nonumber\\
& &+\frac{1}{4}\left[2\delta^\mu_\rho\delta^\nu_\sigma-\bar{g}^{\mu\nu}
\bar{g}_{\rho\sigma}\right]\left(\bar{R}-2\bar{\lambda}_k\right)
+\bar{g}^{\mu\nu}\bar{R}_{\rho\sigma}-\delta_\sigma^\mu\bar{R}^\nu_{\;
\rho}-\bar{R}^{\nu\;\;\mu}_{\;\rho\;\;\;\sigma}\nonumber\\
& &+\frac{1-\alpha}{\alpha}
\left[\bar{g}^{\mu\nu}\bar{D}_\rho\bar{D}_\sigma-\delta^\mu_\sigma
\bar{D}^\nu\bar{D}_\rho\right]
\Bigg\}\bar{h}^{\rho\sigma}
\end{eqnarray}
where $\bar{g}_{\mu\nu}$ is fixed but still arbitrary. In order to
(partially) diagonalize this quadratic form we insert the family of spherical
background metrics into eq. (\ref{38}) and decompose $\bar{h}_{\mu\nu}$
according to eq. (\ref{07}). Then we use the classical analog of eq. 
(\ref{112}) to decompose $\bar{\phi}$ as well. This leads to
\begin{eqnarray}
\label{44}
\lefteqn{\Gamma_k^{\rm quad}\left[\bar{h};\bar{g}\right]
=\kappa^2 Z_{Nk}\int d^dx\,\sqrt{\bar{g}}\,\frac{1}{2}\Bigg\{
\bar{h}^T_{\mu\nu}\left[-\bar{D}^2+A_T(d)\bar{R}-2\bar{\lambda}_k
\right]\bar{h}^{T\mu\nu}}\nonumber\\
& &+\frac{2}{\alpha}\,\bar{\xi}_\mu\left[-\bar{D}^2+A_V(d,\alpha)\bar{R}
-2\alpha\bar{\lambda}_k\right]\bar{\xi}^\mu
+C_{S2}(d,\alpha)\Bigg(
\bar{\sigma}\left[-\bar{D}^2+A_{S2}(d,\alpha)\bar{R}
+B_{S2}(d,\alpha)\bar{\lambda}_k\right]\bar{\sigma}\nonumber\\
& &+2C_{S3}(d,\alpha)\,
\bar{\phi}_1\sqrt{-\bar{D}^2}\sqrt{-\bar{D}^2-\frac{\bar{R}}{d-1}}\bar{\sigma}
\nonumber\\
& &+C_{S1}(d,\alpha)\,\sum\limits_{\bar{\phi}\in\{\bar{\phi}_0,\bar{\phi}_1\}}
\bar{\phi}\left[-\bar{D}^2+A_{S1}(d,\alpha)\bar{R}+B_{S1}(d,\alpha)
\bar{\lambda}_k\right]\bar{\phi}\Bigg)\Bigg\}\;.
\end{eqnarray}
Here the $A$'s, $B$'s, and $C$'s are functions of the dimensionality $d$
and the gauge parameter $\alpha$. The explicit expressions for these 
coefficients can be found in appendix \ref{coeff}.

Note that this partial diagonalization simplifies further calculations
considerably, and this is the main reason for using the decomposition
(\ref{07}) and specifying a concrete background. In contrast to the case 
$\alpha=1$ considered in \cite{Reu96}, a complete diagonalization cannot be
achieved by merely
splitting off the trace part from $\bar{h}_{\mu\nu}$ since eq. (\ref{38})
contains additional terms proportional to $1-\alpha$ which introduce mixings
between the traceless part and $\phi$. To be more precise, it is the term
$\int d^dx\sqrt{\bar{g}}\bar{h}_{\mu\nu}[\bar{g}^{\mu\nu}\bar{D}_\rho
\bar{D}_\sigma-\delta_\sigma^\mu\bar{D}^\nu\bar{D}_\rho]\bar{h}^{\rho\sigma}$
that gives rise to such cross terms. 

In terms of the component fields these
cross terms boil down to a purely scalar $\bar{\sigma}$-$\bar{\phi}$ mixing
term that vanishes for the spherical harmonics $T^{l=0,m=1}$ and
$T^{l=1,m}$. Since these modes contribute to $\bar{\phi}$ (but not to
$\bar{\sigma}$) we cannot directly invert the associated matrix differential 
operator
$\left((\Gamma_k^{(2)})_{ij}\right)_{i,j\in\{\bar{h}^T,\bar{\xi},
\bar{\sigma},\bar{\phi}\}}$. As a way out, we split $\bar{\phi}$ according to 
eq. (\ref{112}) into $\bar{\phi}_0$ and $\bar{\phi}_1$. This has the effect 
that only mixings between the scalars $\bar{\sigma}$ and $\bar{\phi}_1$ 
survive, which have the same set of eigenfunctions $T^{lm}$ starting at $l=2$.
Hence the resulting matrix differential operator 
$\left((\Gamma_k^{(2)})_{ij}\right)_{i,j\in\{\bar{h}^T,\bar{\xi},\bar{\phi}_0,
\bar{\sigma},\bar{\phi}_1\}}$ is invertible, but since this additional split 
of $\phi$ affects the matrix structure of this operator it leads to a slightly
modified flow equation. In fact, on the RHS of eq. (\ref{032})
the summation in the gravitational sector now runs over the set of fields
$\{\bar{h}^T,\bar{\xi},\bar{\phi}_0,\bar{\sigma},\bar{\phi}_1\}$, with
$({\cal R}_k)_{\phi_0\phi_0}\equiv({\cal R}_k)_{\phi_1\phi_1}\equiv
({\cal R}_k)_{\phi\phi}$.

In the context of the Einstein-Hilbert truncation it is only the
$\alpha$-dependence that introduces mixings of $\bar{\phi}$ and the traceless 
part of $\bar{h}_{\mu\nu}$ and therefore necessitates the decompositions 
(\ref{f}), (\ref{112}). In general the inclusion of higher derivative terms 
like $\int d^dx\,\sqrt{g} R^2$ and of matter fields leads to similar mixings.

In order to determine the contributions from the ghosts appearing on the RHS
of eq. (\ref{032}) we set $g_{\mu\nu}=\bar{g}_{\mu\nu}$ in $S_{\rm gh}$ and
assume that $\bar{g}_{\mu\nu}$ corresponds to a spherical background. Then
we decompose the ghost fields according to eq. (\ref{07}) which leads to
\begin{eqnarray}
\label{48}
S_{\rm gh}\left[0,v,\bar{v};g\right]=\sqrt{2}\int 
d^dx\,\sqrt{g}\left\{\bar{v}_\mu^T\left[-D^2-\frac{R}{d}\right]
v^{T\mu}+\bar{\varrho}\left[-D^2-2\frac{R}{d}\right]\varrho\right\}\;.
\end{eqnarray}
From now on the bars are omitted form the metric, the curvature and the
operators $D^2$ and $P_k$. Note that the decomposition of the ghosts is not
really necessary, but it allows for a comparison with the results obtained in
\cite{DP97}.

At this point we can continue with the adaption of the cutoff to the operators
$\Gamma_k^{(2)}$ and $S^{(2)}_{\rm gh}$ of eqs. (\ref{44}), (\ref{48}). 
According to the rule 
(\ref{090}) the ${\cal Z}_k$'s have to be chosen as
\begin{eqnarray}
\label{49}
& &{\cal Z}_k^{\bar{h}^T\bar{h}^T}=Z_{Nk}\;,\;\;{\cal Z}_k^{\bar{\xi}
\bar{\xi}}=\frac{2}{\alpha}Z_{Nk}\;,\;\;
{\cal Z}_k^{\bar{\phi}_1\bar{\sigma}}
=C_{S2}(d,\alpha)C_{S3}(d,\alpha)Z_{Nk}\;,\;\;
{\cal Z}_k^{\bar{\sigma}\bar{\sigma}}=C_{S2}(d,\alpha)Z_{Nk}\;,\nonumber\\
& &{\cal Z}_k^{\bar{\phi}_0\bar{\phi}_0}
={\cal Z}_k^{\bar{\phi}_1\bar{\phi}_1}=C_{S2}(d,\alpha)C_{S1}(d,\alpha)
Z_{Nk}\;,\;\;{\cal Z}_k^{\bar{v}^Tv^T}={\cal Z}_k^{\bar{\varrho}\varrho}
=\sqrt{2}\;.
\end{eqnarray}
Thus, for $g=\bar{g}$ the nonvanishing entries of the
matrix differential operators $\Gamma_k^{(2)}+{\cal R}_k$ and
$S^{(2)}_{\rm gh}+{\cal R}_k$ take the form
\begin{eqnarray}
\label{50}
\left(\Gamma^{(2)}_k[g,g]+{\cal R}_k\right)_{\bar{h}^T\bar{h}^T}
&=&Z_{Nk}\kappa^2\left[P_k+A_T(d)R-2\bar{\lambda}_k\right]\;,\nonumber\\
\left(\Gamma^{(2)}_k[g,g]+{\cal R}_k\right)_{\bar{\xi}\bar{\xi}}
&=&Z_{Nk}\kappa^2\,\frac{2}{\alpha}\left[P_k+A_V(d,\alpha)R-2\alpha
\bar{\lambda}_k\right]\;,\nonumber\\
\left(\Gamma^{(2)}_k[g,g]+{\cal R}_k\right)_{\bar{\sigma}\bar{\sigma}}
&=&Z_{Nk}\kappa^2\,C_{S2}(d,\alpha)\left[P_k+A_{S2}(d,\alpha)R
+B_{S2}(d,\alpha)\bar{\lambda}_k\right]\;,\nonumber\\
\left(\Gamma^{(2)}_k[g,g]+{\cal R}_k\right)_{\bar{\phi}_1\bar{\sigma}}
&=&\left(\Gamma^{(2)}_k[g,g]+{\cal R}_k\right)_{\bar{\sigma}\bar{\phi}_1}
\nonumber\\
&=&Z_{Nk}\kappa^2\,C_{S2}(d,\alpha)C_{S3}(d,\alpha)\,\sqrt{P_k}
\sqrt{P_k-\frac{R}{d-1}}\;,\nonumber\\
\left(\Gamma^{(2)}_k[g,g]+{\cal R}_k\right)_{\bar{\phi}_0\bar{\phi}_0}
&=&\left(\Gamma^{(2)}_k[g,g]+{\cal R}_k\right)_{\bar{\phi}_1\bar{\phi}_1}
\nonumber\\
&=&Z_{Nk}\kappa^2\,C_{S2}(d,\alpha)C_{S1}(d,\alpha)\left[P_k+A_{S1}(d,\alpha)
R+B_{S1}(d,\alpha)\bar{\lambda}_k\right]\;,\nonumber\\
\left(S_{\rm gh}^{(2)}[g,g]+{\cal R}_k\right)_{\bar{v}^Tv^T}
&=&-\left(S_{\rm gh}^{(2)}[g,g]+{\cal R}_k\right)_{v^T\bar{v}^T}
=\sqrt{2}\left[P_k-\frac{R}{d}\right]\;,\nonumber\\
\left(S_{\rm gh}^{(2)}[g,g]+{\cal R}_k\right)_{\bar{\varrho}\varrho}
&=&-\left(S_{\rm gh}^{(2)}[g,g]+{\cal R}_k\right)_{\varrho\bar{\varrho}}
=\sqrt{2}\left[P_k-2\frac{R}{d}\right]\;.
\end{eqnarray}
Here we set $\left(S_{\rm gh}^{(2)}[0,v,\bar{v};g]\right)_{\psi_1\psi_2}\equiv
\left(S_{\rm gh}^{(2)}[g,g]\right)_{\psi_1\psi_2}$ for $\psi_1,\psi_2\in
\bar{I}_2$.

Now we are in a position to write down the RHS of the flow equation with
$g=\bar{g}$. We shall denote it ${\cal S}_k(R)$ in the following. In 
${\cal S}_k(R)$ we need the inverse operators
$(\Gamma_k^{(2)}+{\cal R}_k)^{-1}$ and $(S^{(2)}_{\rm gh}+{\cal R}_k)^{-1}$. 
The inversion is carried out in appendix \ref{inv}. Inserting the inverse 
operators into ${\cal S}_k(R)$ leads to
\begin{eqnarray}
\label{52}
\lefteqn{{\cal S}_k(R)=}\nonumber\\
& &{\rm Tr}_{(2ST^2)}\left[\left(P_k+A_T(d)R-2\bar{\lambda}_k
\right)^{-1}{\cal N}\right]
+{\rm Tr}_{(1T)}'\left[\left(P_k+A_V(d,\alpha)R-2\alpha
\bar{\lambda}_k\right)^{-1}{\cal N}\right]\nonumber\\
& &+{\rm Tr}_{(0)}''\Bigg[
\left(P_k+A_{S3}(d)R-2\bar{\lambda}_k\right)^{-1}
\left(P_k+A_{S4}(d,\alpha)R-2\alpha\bar{\lambda}_k\right)^{-1}\nonumber\\
& &\times\Bigg\{\left(F_{S1}(d,\alpha)P_k+A_{S5}(d,\alpha)R-2
(\alpha+1)\bar{\lambda}_k\right){\cal N}\nonumber\\
& &+F_{S2}(d,\alpha)\sqrt{P_k}\sqrt{P_k-\frac{R}{d-1}}\nonumber\\
& &\times\frac{1}{2Z_{Nk}}\partial_t\left[Z_{Nk}\left(
\sqrt{P_k}\sqrt{P_k-\frac{R}{d-1}}
-\sqrt{-D^2}\sqrt{-D^2-\frac{R}{d-1}}\right)\right]
\Bigg\}\Bigg]
\nonumber\\
& &-2{\rm Tr}_{(1T)}\left[\left(P_k-\frac{R}{d}\right)^{-1}{\cal N}_0\right]
-2{\rm Tr'}_{(0)}\left[\left(P_k-2\frac{R}{d}\right)^{-1}{\cal N}_0\right]
\nonumber\\
& &+\frac{1}{2Z_{Nk}}\sum\limits_{l=0}^{1}\left[D_l(d,0)\frac{\partial_t
\left[Z_{Nk}
k^2R^{(0)}(\Lambda_l(d,0)/k^2)\right]}{\Lambda_l(d,0)+k^2R^{(0)}
(\Lambda_l(d,0)/k^2)+A_{S1}(d,\alpha) R+B_{S1}(d,\alpha)\bar{\lambda}_k}
\right]\;.
\end{eqnarray}
Here we set
\begin{eqnarray}
\label{53}
{\cal N}&=&\left(2 Z_{Nk}\right)^{-1}
\partial_t \left[Z_{Nk}k^2R^{(0)}(-D^2/k^2)\right]\nonumber\\
&=&\left[1-\frac{1}{2}\eta_N(k)\right]k^2R^{(0)}(-D^2/k^2)
+D^2R^{(0)'}(-D^2/k^2)\;,\nonumber\\
{\cal N}_0&=&2^{-1}
\partial_t \left[k^2R^{(0)}(-D^2/k^2)\right]
=k^2R^{(0)}(-D^2/k^2)+D^2R^{(0)'}(-D^2/k^2)
\end{eqnarray}
where
\begin{eqnarray}
\label{71}
\eta_N(k)\equiv -\partial_t \ln Z_{Nk}
\end{eqnarray}
is the anomalous dimension of the operator $\int d^dx\sqrt{g}R$ 
and the prime at $R^{(0)}$ denotes the derivative with respect to the argument.
Furthermore, the new $A$'s and $F$'s introduced above are again functions of 
$d$ and $\alpha$, tabulated in appendix \ref{coeff}.

In eq. (\ref{52}) we refined our notation concerning the primes at the 
traces. From now on one prime indicates the subtraction of the contribution
from the lowest eigenvalue, while two primes indicate that the
modes corresponding to the lowest two eigenvalues have to be excluded.

The next step is to extract the contributions proportional to 
$\int d^dx\sqrt{g}$ and $\int d^dx\sqrt{g}R$ by expanding ${\cal S}_k(R)$ 
with respect to $R$ or $r$, respectively. Since
$\int d^dx\sqrt{g}\propto r^d$, $\int d^dx\sqrt{g}R\propto r^{d-2}$, only 
terms of order $r^d$ and $r^{d-2}$ are needed. This leads to
\begin{eqnarray}
\label{55}
{\cal S}_k(R)&=&
{\rm Tr}_{(2ST^2)}\left[\left(P_k-2\bar{\lambda}_k\right)^{-1}{\cal N}
\right]
+{\rm Tr}_{(1T)}'\left[\left(P_k-2\alpha\bar{\lambda}_k
\right)^{-1}{\cal N}\right]\nonumber\\
& &+{\rm Tr}_{(0)}''\left[\left(P_k-2\bar{\lambda}_k\right)^{-1}{\cal N}
\right]
+{\rm Tr}_{(0)}''\left[\left(P_k-2\alpha\bar{\lambda}_k
\right)^{-1}{\cal N}\right]\nonumber\\
& &-2{\rm Tr}_{(1T)}\left[P_k^{-1}{\cal N}_0\right]
-2{\rm Tr}_{(0)}'\left[P_k^{-1}{\cal N}_0\right]
-R\Bigg\{A_T(d){\rm Tr}_{(2ST^2)}\left[\left(P_k-2\bar{\lambda}_k
\right)^{-2}{\cal N}\right]\nonumber\\
& &+A_V(d,\alpha){\rm Tr}_{(1T)}'\left[\left(P_k-2\alpha
\bar{\lambda}_k\right)^{-2}{\cal N}\right]
+A_{S3}(d){\rm Tr}_{(0)}''\left[\left(P_k-2\bar{\lambda}_k
\right)^{-2}{\cal N}\right]
\nonumber\\
& &+A_{S4}(d,\alpha){\rm Tr}_{(0)}''\left[\left(P_k-2\alpha
\bar{\lambda}_k\right)^{-2}{\cal N}\right]
+\frac{2}{d}{\rm Tr}_{(1T)}\left[P_k^{-2}{\cal N}_0\right]\nonumber\\
& &
+\frac{4}{d}{\rm Tr}_{(0)}'\left[P_k^{-2}{\cal N}_0\right]-\frac{\delta_{d,2}}
{4\pi}\int d^2x\,\sqrt{g}\frac{\partial_t\left(Z_{Nk}k^2\right)}
{Z_{Nk}\left(k^2-2\bar{\lambda}_k\right)}
\Bigg\}+{\cal O}(r^{< d-2})\;.
\end{eqnarray}
Here ${\cal O}(r^{<d-2})$ means that terms $\propto r^n$ with powers 
$n<d-2$ are neglected. 

The term in eq. (\ref{55}) proportional to $\delta_{d,2}$ arises from
the last term in eq. (\ref{52}). Contrary to the other terms of eq. 
(\ref{52}), its expansion does not contain $d$-dependent powers of $r$, but is
of the form $\sum \limits_{m=0}^\infty b_{2m}r^{-2m}$ with $\{b_{2m}\}$ a set 
of $r$-independent coefficients. As for comparing powers of $r$, this has the 
following consequence. Since, for all $m\ge 0$ and $d>0$, $-2m=d-2$ is 
satisfied only if $m=0$ and $d=2$, and since $-2m=d$ cannot be satisfied at 
all, this term contributes to the evolution equation only in the 
two-dimensional case. Using eq. (\ref{035}) the 
piece contributing, i.e. $b_{2m=0}\;r^0$, may be expressed in terms of the 
operator $\int d^dx\sqrt{g}R$ which yields the last term in eq. (\ref{55}).

The traces appearing in eq. (\ref{55}) are evaluated in appendix \ref{trace} 
using heat kernel techniques. Then combining the result with the LHS of the 
flow equation, eq. (\ref{040}), and comparing the coefficients of the 
invariants $\int d^dx\sqrt{g}$ and $\int d^dx\sqrt{g} R$ leads to the desired 
system of coupled differential equations for $Z_{NK}$ and $\bar{\lambda}_k$.
We obtain
\begin{eqnarray}
\label{74}
\partial_t\left(Z_{Nk}\bar{\lambda}_k\right)
& &=(4\kappa^2)^{-1}(4\pi)^{-d/2}k^d\bigg\{
\frac{1}{2}d(d-1)\,\Phi^1_{d/2}(-2\bar{\lambda}_k/k^2)
+d\,\Phi^1_{d/2}(-2\alpha\bar{\lambda}_k/k^2)\\
& &-\frac{1}{2}\eta_N(k)\bigg[
\frac{1}{2}d(d-1)\,\widetilde{\Phi}^1_{d/2}(-2\bar{\lambda}_k/k^2)
+d\,\widetilde{\Phi}^1_{d/2}(-2\alpha\bar{\lambda}_k/k^2)
\bigg]-2d\,\Phi^1_{d/2}(0)\bigg\}\;,\nonumber
\end{eqnarray}
\begin{eqnarray}
\label{75}
\partial_t Z_{Nk}&=&-(2\kappa^2)^{-1}(4\pi)^{-d/2}k^{d-2}\Bigg\{
c_1(d)\,\Phi^1_{d/2-1}(-2\bar{\lambda}_k/k^2)
+c_2(d)\,\Phi^1_{d/2-1}(-2\alpha\bar{\lambda}_k/k^2)\nonumber\\
& &+c_3(d)\,\Phi^2_{d/2}(-2\bar{\lambda}_k/k^2)
+c_4(d,\alpha)\,\Phi^2_{d/2}(-2\alpha\bar{\lambda}_k/k^2)\nonumber\\
& &-\frac{1}{2}\eta_N(k)\bigg[c_1(d)\,\widetilde{\Phi}^1_{d/2-1}
(-2\bar{\lambda}_k/k^2)
+c_2(d)\,\widetilde{\Phi}^1_{d/2-1}(-2\alpha\bar{\lambda}_k/k^2)
\nonumber\\
& &+c_3(d)\,\widetilde{\Phi}^2_{d/2}(-2\bar{\lambda}_k/k^2)
+c_4(d,\alpha)\,\widetilde{\Phi}^2_{d/2}(-2\alpha\bar{\lambda}_k/k^2)\bigg]
\nonumber\\
& &-2c_2(d)\,\Phi^1_{d/2-1}(0)+c_5(d)\,\Phi^2_{d/2}(0)\nonumber\\
& &+3\delta_{d,2}\left(1-\frac{1}{2}\eta_N(k)\right)\left[\frac{1}{1
-2\bar{\lambda}_k/k^2}-\frac{1}{1
-2\alpha\bar{\lambda}_k/k^2}\right]\Bigg\}\;.
\end{eqnarray}
Here $\Phi^p_n$, $\widetilde{\Phi}^p_n$ are cutoff-dependent ``threshold'' 
functions defined as
\begin{eqnarray}
\label{68}
\Phi^p_n(w)&\equiv&\left\{\begin{array}{ll}\frac{1}{\Gamma(n)}
\int\limits_{0}^{\infty} dy\,y^{n-1}\frac{R^{(0)}(y)-yR^{(0)'}(y)}
{\left(y+R^{(0)}(y)+w\right)^p} &,\;\;n>0\\
(1+w)^{-p}&,\;\; n=0\end{array}\right.\;,\nonumber\\
\widetilde{\Phi}^p_n(w)&\equiv&\left\{\begin{array}{ll}\frac{1}{\Gamma(n)}
\int\limits_{0}^{\infty} dy\,y^{n-1}\frac{R^{(0)}(y)}{\left(y+R^{(0)}(y)
+w\right)^p}&,\;\; n>0\\(1+w)^{-p}&,\;\;n=0\end{array}\right.\;.
\end{eqnarray}
The coefficients $c_i$ are given by
\begin{eqnarray}
\label{044}
& &c_1(d)\equiv\frac{d^3-2d^2-11d-12}{12(d-1)}\;,\;\;
c_2(d)\equiv\frac{d^2-6}{6d}
\;,\;\;c_3(d)\equiv-\frac{d^3-4d^2+7d-8}{2(d-1)}\;,\nonumber\\
& &c_4(d,\alpha)\equiv
-\frac{\alpha d(d-2)-d-1}{d}\;,\;\,c_5(d)\equiv-\frac{2(d+1)}{d}\;.
\end{eqnarray}

In eq. (\ref{75}) the terms proportional to $\delta_{d,2}$
arise not only from the last term of eq. (\ref{55}), but also by evaluating
the ``primed'' traces, i.e. by subtracting the contributions coming from 
unphysical modes, see appendix \ref{trace} for details.
All these contributions are obtained by expanding various functions $f(R)$ 
with respect to $R$ and retaining only the term of zeroth order, $f(0)$. As we
argued above, these are the only pieces of $f$ which may contribute
to the evolution in the truncated parameter space. Furthermore, the heat
kernel expansions of the traces corresponding to differentially constrained 
fields introduce additional contributions $\propto\delta_{d,2}$ into eq. 
(\ref{75}).

In appendix \ref{d=4} we concentrate on the 4-dimensional case and compare our 
result for ${\cal S}_k(R)$ and for the corresponding RG flow of $Z_{Nk}$ and 
$\bar{\lambda}_k$ with the one of ref. \cite{DP97} where a cutoff of type B is 
used, too.
\subsection{The system of flow equations for $g_k$ and $\lambda_k$}
\label{4D}
Now we introduce the dimensionless, renormalized Newton constant
\begin{eqnarray}
\label{003}
g_k\equiv k^{d-2}\,G_k\equiv k^{d-2}\,Z_{Nk}^{-1}\,\bar{G}
\end{eqnarray}
and the dimensionless, renormalized cosmological constant
\begin{eqnarray}
\label{007}
\lambda_k\equiv k^{-2}\,\bar{\lambda}_k
\end{eqnarray}
where $G_k$ denotes the corresponding dimensionful, renormalized Newton
constant at scale $k$. Inserting eq. (\ref{007}) into $\partial_t(Z_{Nk}
\bar{\lambda}_k)$ leads to the relation
\begin{eqnarray}
\label{045}
\partial_t\lambda_k=-\left(2-\eta_N(k)\right)\lambda_k+32\pi\,g_k\,
\kappa^2\,k^{-d}\,\partial_t\left(Z_{Nk}\bar{\lambda}_k\right)\;.
\end{eqnarray}
Then, by using eq. (\ref{74}), we obtain the following differential equation
for the dimensionless cosmological 
constant:
\begin{eqnarray}
\label{001}
\fbox{$\displaystyle\partial_t\lambda_k
=\mbox{\boldmath$\beta$}_\lambda(\lambda_k,g_k;\alpha,d)
\equiv A_1(\lambda_k,g_k;\alpha,d)+\eta_N(k)\,A_2(\lambda_k,g_k;\alpha,d)$}
\end{eqnarray}
The $\mbox{\boldmath$\beta$}$-function $\mbox{\boldmath$\beta$}_\lambda$
contains the quantities $A_1$ and $A_2$ which are defined as
\begin{eqnarray}
\label{002}
A_1(\lambda_k,g_k;\alpha,d)&\equiv& -2\lambda_k+(4\pi)^{1-d/2}\,g_k
\bigg\{d(d-1)\,\Phi^1_{d/2}(-2\lambda_k)\nonumber\\
& &+2d\,\Phi^1_{d/2}(-2\alpha\lambda_k)-4d\,\Phi^1_{d/2}(0)\bigg\}\;,
\nonumber\\
A_2(\lambda_k,g_k;\alpha,d)&\equiv & \lambda_k-(4\pi)^{1-d/2}\,g_k
\bigg\{\frac{1}{2}d(d-1)\,\widetilde{\Phi}^1_{d/2}(-2\lambda_k)
+d\,\widetilde{\Phi}^1_{d/2}(-2\alpha\lambda_k)\bigg\}\;.
\end{eqnarray}
The corresponding $\mbox{\boldmath$\beta$}$-function for $g_k$ may be 
determined as follows. Taking the scale derivative of eq. (\ref{003}) leads to
\begin{eqnarray}
\label{004}
\fbox{$\partial_t g_k=\mbox{\boldmath$\beta$}_g(\lambda_k,g_k;\alpha,d)\equiv
\left[d-2+\eta_N(k)\right]g_k$}\;.
\end{eqnarray}
For the anomalous dimension $\eta_N(k)$ we obtain from eq. (\ref{75}) 
\begin{eqnarray}
\label{005}
\eta_N(k)=g_k\,B_1(\lambda_k;\alpha,d)+\eta_N(k)\,g_k\,B_2(\lambda_k;\alpha,d)
\end{eqnarray}
with $B_1$, $B_2$ functions of $\lambda_k$, $d$ and $\alpha$ given by
\begin{eqnarray}
\label{006}
B_1(\lambda_k;\alpha,d)&\equiv &4(4\pi)^{1-d/2}
\bigg\{c_1(d)\,\Phi^1_{d/2-1}(-2\lambda_k)
+c_2(d)\,\Phi^1_{d/2-1}(-2\alpha\lambda_k)
+c_3(d)\,\Phi^2_{d/2}(-2\lambda_k)\nonumber\\
& &+c_4(d,\alpha)\,\Phi^2_{d/2}(-2\alpha\lambda_k)
-2c_2(d)\,\Phi^1_{d/2-1}(0)+c_5(d)\,\Phi^2_{d/2}(0)\nonumber\\
& &+3\delta_{d,2}\left[\frac{1}{1
-2\lambda_k}-\frac{1}{1
-2\alpha\lambda_k}\right]\bigg\}\;,\nonumber\\
B_2(\lambda_k;\alpha,d)&\equiv& -2(4\pi)^{1-d/2}\bigg\{
c_1(d)\,\widetilde{\Phi}^1_{d/2-1}(-2\lambda_k)
+c_2(d)\,\widetilde{\Phi}^1_{d/2-1}(-2\alpha\lambda_k)\nonumber\\
& &+c_3(d)\,\widetilde{\Phi}^2_{d/2}(-2\lambda_k)
+c_4(d,\alpha)\,\widetilde{\Phi}^2_{d/2}(-2\alpha\lambda_k)\nonumber\\
& &+3\delta_{d,2}\left[\frac{1}{1-2\lambda_k}-\frac{1}{1
-2\alpha\lambda_k}\right]\bigg\}\;.
\end{eqnarray}
Eq. (\ref{005}) may now be solved for the anomalous dimension in terms of
$\lambda_k$, $g_k$, $\alpha$ and $d$:
\begin{eqnarray}
\label{008}
\eta_N(k)=\frac{g_k\,B_1(\lambda_k;\alpha,d)}{1-g_k\,B_2(\lambda_k;\alpha,
d)}\;.
\end{eqnarray}

The system of coupled flow equations (\ref{001}), (\ref{004}) is the main 
result of this section.
\subsection{Comparing the cutoffs A and B}
\label{4E}
Let us compare the flow equations (\ref{001}), 
(\ref{004}) of the present paper with those obtained in ref. \cite{Reu96}.
Ref. \cite{Reu96} covers the case $\alpha=1$ only, and the cutoff used there 
has a different structure than the present one. In ref. \cite{Reu96}, the
$\Delta_k S$ for the cutoff A is formulated at the level of the
{\it complete} field $h_{\mu\nu}$, i.e., symbolically, $\Delta_k S\propto
\int h_{\mu\nu}{\cal R}_k h^{\mu\nu}$, while the cutoff B of the present
paper is based upon a similar action for the component fields:
$\Delta_k S\propto\int h^T_{\mu\nu}{\cal R}_k h^{T\mu\nu}
+\int \xi_\mu{\cal R}_k\xi^\mu+\cdots$.

For $\alpha=1$ the new $\mbox{\boldmath$\beta$}$-function for $\lambda_k$, 
$\mbox{\boldmath$\beta$}_\lambda$, agrees perfectly with the result in 
\cite{Reu96}, whereas the coefficients $B_1$, $B_2$ in the $\mbox{\boldmath$
\beta$}$-function for $g_k$, $\mbox{\boldmath$\beta$}_g$, do not coincide with
the corresponding results
derived there. However, with both cutoffs these coefficients are of the form
\begin{eqnarray}
\label{78}
B_1(\lambda_k;\alpha=1,d)&=&\frac{1}{3}(4\pi)^{1-d/2}k^{d-2}
\Bigg\{e_1(d)\,\Phi^1_{d/2-1}(-2\lambda_k)
+e_2(d)\,\Phi^2_{d/2}(-2\lambda_k)\nonumber\\
& &+e_3(d)\Phi^1_{d/2-1}(0)+e_4(d)\Phi^2_{d/2}(0)\Bigg\}\;,\nonumber\\
B_2(\lambda_k;\alpha=1,d)&=&-\frac{1}{6}(4\pi)^{1-d/2}k^{d-2}
\Bigg\{e_1(d)\,\widetilde{\Phi}^1_{d/2-1}(-2\lambda_k)
+e_2(d)\,\widetilde{\Phi}^2_{d/2}(-2\lambda_k)\Bigg\}\;.
\end{eqnarray}
In the present paper (cutoff B) the coefficients $e_i$ are obtained as
\begin{eqnarray}
\label{065}
e_1(d)&=&\frac{d^4-13d^2-24d+12}{d(d-1)}\;,\;\,
e_2(d)=-6\frac{d^4-2d^3-d^2-4d+2}{d(d-1)}\;,\nonumber\\
e_3(d)&=&-4\frac{d^2-6}{d}\;,\;\;e_4(d)=-24\frac{d+1}{d}
\end{eqnarray}
while in \cite{Reu96} (cutoff A) they are given by
\begin{eqnarray}
\label{066}
e_1(d)=d(d+1)\;,\;\,e_2(d)=-6d(d-1)\;,\;\;
e_3(d)=-4d\;,\;\;e_4(d)=-24\;.
\end{eqnarray}
Upon subtracting the coefficients in eq. (\ref{065}) from those in eq. 
(\ref{066}) we obtain $\Delta e_1=-\Delta e_2=12(d^2+2d-1)/(d(d-1))$ and 
$\Delta e_3=-\Delta e_4=-24/d$. Quite remarkably, the sum of the deviations
$\Delta e_i$ vanishes not only in total but also separately for the
gravitational contributions, involving $e_1$ and $e_2$, and
the contributions from the ghost, which contain $e_3$ and $e_4$.
Obviously, this amounts to a shift from the gravitational 
$(p=1,n=d/2-1)$-sector to the $(p=2,n=d/2)$-sector as well as to a shift 
between the corresponding ghost-sectors. The simplicity of this result is
somewhat mysterious.
\section{The fixed points}
\label{S5}
\renewcommand{\theequation}{5.\arabic{equation}}
\setcounter{equation}{0}
\subsection{Fixed points, critical exponents, and nonperturbative 
renormalizability}
\label{5A}
Because of its complexity it is impossible to solve the system of flow
equations for $g_k$ and $\lambda_k$, eqs. (\ref{001}) and (\ref{004}), exactly.
Even a numerical solution would be a formidable task. However, it is possible 
to gain important information about the general structure of the RG flow 
by looking at its fixed point structure.

Given a set of $\mbox{\boldmath $\beta$}$-functions corresponding to an 
arbitrary set of dimensionless essential couplings ${\rm g}_i(k)$, it is 
often possible to predict the scale dependence of the couplings for very small
and/or very large scales $k$ by investigating their fixed points. The
fixed points are those points in the space spanned by the ${\rm g}_i$ where
all $\mbox{\boldmath $\beta$}$-functions vanish. (The essential couplings are
 those combinations of the couplings 
appearing in the action functional that are invariant under point 
transformations of the fields.) Fixed points are characterized by their
stability properties. A given  eigendirection of the linearized flow is said to
be UV or IR attractive (or stable) if, for $k\rightarrow\infty$ or 
$k\rightarrow 0$, respectively, the trajectories are attracted towards the 
fixed point along this direction. The UV critical hypersurface in the space of
all couplings is defined to consist of all trajectories that run into the fixed
point for $k\rightarrow\infty$.

In quantum field theory, fixed points play an important role in the modern 
approach to renormalization theory\cite{wilson}. At a UV fixed point the 
infinite
cutoff limit can be taken in a controlled way. As for gravity, Weinberg
\cite{wein} argued that a theory described by a trajectory lying on a 
{\it finite-dimensional} UV critical hypersurface of some fixed point is 
presumably free from unphysical singularities. It is predictive since it 
depends only on a {\it finite} number of free (essential) parameters. In 
Weinberg's words, such a theory is {\it asymptotically safe}. Asymptotic 
safety has to be regarded as a generalized, nonperturbative version of 
renormalizability. It covers the class of perturbatively renormalizable 
theories, whose infinite cutoff limit is taken at the Gaussian fixed point 
${\rm g}_{*i}=0$, as well as those perturbatively nonrenormalizable theories 
which are described by a RG trajectory on a finite-dimensional UV critical 
hypersurface of a non-Gaussian fixed point ${\rm g}_{*i}\neq 0$ and are
nonperturbatively renormalizable therefore \cite{wein}.

Let us now consider the system of differential equations 
\begin{eqnarray}
\label{fp0}
k\,\partial_k{\rm g}_i(k)=\mbox{\boldmath$\beta$}_i({\rm g})
\end{eqnarray} 
for a set of dimensionless essential couplings ${\rm g}(k)\equiv\{{\rm g}_1(k),
\ldots,{\rm g}_n(k)\}$. We assume that ${\rm g}_*$ is a fixed point of eq. 
(\ref{fp0}), i.e. $\mbox{\boldmath$\beta$}_i({\rm g}_*)=0$ for all $i=1,\ldots,
 n$. We linearize the RG flow about ${\rm g}_*$ which leads to
\begin{eqnarray}
\label{gfp4}
k\,\partial_k\,{\rm g}_i(k)=\sum\limits_{i=1}^n B_{ij}\,\left({\rm g}_j(k)
-{\rm g}_{*j}\right)
\end{eqnarray}
where $B_{ij}\equiv\partial_j\mbox{\boldmath$\beta$}_i({\rm g}_*)$ are the 
entries of the stability matrix ${\bf B}=(B_{ij})$. Diagonalizing ${\bf B}$ 
according to $S^{-1}{\bf B}S=-{\rm diag}(\theta_1,\dots,\theta_n)$, 
$S=(V^1,\ldots,V^n)$, where $V^I$ is the right-eigenvector of ${\bf B}$ with 
eigenvalue $-\theta_I$ we have
\begin{eqnarray}
\label{gfp7}
\sum\limits_{j=1}^n B_{ij}\,V^I_j=-\theta_I\,V^I_i\;,\;\; I=1,\ldots ,n\;.
\end{eqnarray}
The general solution to eq. (\ref{gfp4}) may be written as 
\begin{eqnarray}
\label{gfp5}
{\rm g}_i(k)={\rm g}_{*i}+\sum\limits_{I=1}^n C_I\,V^I_i\,
\left(\frac{k_0}{k}\right)^{\theta_I}\;.
\end{eqnarray}
Here
\begin{eqnarray}
\label{fp1}
C_I\equiv\sum\limits_{j=1}^n(S^{-1})_{Ij}\,{\rm g}_j(k_0)
\end{eqnarray}
are arbitrary real parameters and $k_0$ is a reference scale. 

Obviously the 
fixed point $g_*$ is UV attractive (i.e. attractive for $k\rightarrow\infty$) 
only if all $C_I$ corresponding to negative $\theta_I<0$ are set to
zero. Therefore the dimensionality of the UV critical hypersurface equals the
number of positive $\theta_I>0$. Conversely, setting to zero all $C_I$ 
corresponding to positive $\theta_I$, ${\rm g}_*$ becomes an IR 
attractive fixed point (approached in the limit $k\rightarrow 0$) with an IR 
critical hypersurface whose dimensionality
equals the number of negative $\theta_I$. 

In a slight abuse of language we 
shall refer to the $\theta_I$'s as the {\it critical exponents}. 

Strictly speaking, the solution (\ref{gfp5}) and its above interpretation
is valid only in such cases where all eigenvalues $-\theta_I$ are real, 
which is
not  guaranteed since the matrix $\bf B$ is not symmetric in general. If 
complex eigenvalues occur one has to consider complex $C_I$'s and to take the 
real part of eq. (\ref{gfp5}), see below. Then the real parts of the critical 
exponents determine which directions in coupling constant space 
are attractive or repulsive.

At this point it is necessary to discuss the impact a change of the cutoff
scheme has on the scaling behavior. Since the path integral for $\Gamma_k$ 
depends on the cutoff scheme, i.e. on the $\Delta_k S$
chosen, it is clear that the couplings and their fixed point values are scheme
dependent. Hence a variation of the cutoff scheme, i.e. of ${\cal R}_k$, 
induces a change in the 
corresponding ${\bf B}$-matrix. So one might naively expect that also its 
eigenvalues, the critical exponents, are scheme dependent. In fact, this is 
not the case. According to the general theory of critical phenomena and a 
recent reanalysis in the framework of the exact RG equations \cite{kana} any 
variation of the cutoff scheme can be generated by a specific coordinate 
transformation in the space of couplings with the cutoff held fixed. Such 
transformations leave the eigenvalues of the ${\bf B}$-matrix invariant,
so that the critical behavior near the corresponding fixed point 
is universal. The positions of fixed points are scheme dependent but their
(non)existence and the qualitative structure of the RG flow are universal 
features. Therefore a truncation can be considered reliable only if it predicts
the same fixed point structure for all admissible choices of ${\cal R}_k$.

In the context of the Einstein-Hilbert truncation the space of couplings is
parametrized by ${\rm g}_1=\lambda$ and ${\rm g}_2=g$. The $\mbox{\boldmath
$\beta$}$-functions occuring in the two flow equations
\begin{eqnarray}
\label{fp2}
\partial_t\lambda_k=\mbox{\boldmath$\beta$}_\lambda(\lambda_k,g_k)\;,
\;\;\;\partial_t g_k=\mbox{\boldmath$\beta$}_g(\lambda_k,g_k)
\end{eqnarray}
are given in eqs. (\ref{001}) and (\ref{004}), respectively. As we shall see in
subsection B, they have a trivial zero at $\lambda_*=g_*=0$, referred 
to as the {\bf Gaussian fixed point.} The analysis of subsection C 
reveals that there exists also a {\bf non-Gaussian fixed point} at 
$\lambda_*\neq 0$, $g_*\neq 0$. In subsection C we study its cutoff dependence
and the cutoff dependence of the associated critical exponents employing the 
above $\mbox{\boldmath$\beta$}$-functions of type
B as well as those of refs. \cite{sven,souma2} based on the cutoff type A, 
with the families of shape functions (\ref{H6}) or (\ref{supp}) inserted. 
\subsection{The Gaussian fixed point}
\label{5B}
In this subsection we discuss the features of the Gaussian fixed point
$(\lambda_*,g_*)=(0,0)$. In order to investigate the RG flow in its vicinity
 we expand the $\mbox{\boldmath$\beta$}$-functions in powers of $\lambda$ and
$g$ according to eq. (\ref{gfp2}) of appendix \ref{approx} and read off the 
${\bf B}$-matrix. It takes the form
\begin{eqnarray}
\label{gfp6}
{\bf B}=\left(\begin{array}{rr}
-2 \;& \nu_d\,d\\ 0 \;& d-2\end{array}\right)\;.
\end{eqnarray}
Here $\nu_d$ is a $d$-dependent parameter defined as
\begin{eqnarray}
\label{gfp3}
\nu_d\equiv(d-3)(4\pi)^{1-\frac{d}{2}}\,
\Phi^1_{d/2}(0)\;.
\end{eqnarray}
Diagonalizing the matrix (\ref{gfp6}) yields the (obviously universal) critical
exponents $\theta_1=2$ and $\theta_2=2-d$ which are associated with the 
eigenvectors $V^1=(1,0)^{\bf T}$ and $V^2=(\nu_d,1)^{\bf T}$. Hence, 
for the linearized system obtained from (\ref{gfp2}) the solution (\ref{gfp5})
assumes the 
following form:
\begin{eqnarray}
\label{gfp8}
\lambda_k&=&\left(\lambda_{k_0}-\nu_d\,g_{k_0}\right)\left(\frac{k_0}{k}
\right)^2+\nu_d\,g_{k_0}\,\left(\frac{k}{k_0}\right)^{d-2}\;,\nonumber\\
g_k&=&g_{k_0}\,\left(\frac{k}{k_0}\right)^{d-2}\;.
\end{eqnarray}

Since the expanded $\mbox{\boldmath$\beta$}$-function 
$\mbox{\boldmath$\beta$}_g$ of eq. (\ref{gfp2}) is $\lambda_k$-independent up
to terms of third order in the couplings we can easily  
calculate also the next-to-leading approximation for $g_k$ near the 
fixed point. In terms of the dimensionful 
quantity $G_k$ this improved solution reads
\begin{eqnarray}
\label{gfp9}
G_k=G_{k_0}\left[1-\omega_d\,G_{k_0}\left(k_0^{d-2}-k^{d-2}\right)\right]^{-1}
\end{eqnarray}
with
\begin{eqnarray}
\label{gfp13}
\omega_d\equiv-\frac{1}{d-2}\,B_1(0;\alpha,d)&=&-2(4\pi)^{1-\frac{d}{2}}
\left\{
\frac{(d+2)(d^3-6d^2+3d-6)}{6d(d-1)(d-2)}\,\Phi^1_{d/2-1}(0)\right.\nonumber\\
& &-\left.\left(\frac{d^4-4d^3+9d^2-8d-2}{d(d-1)(d-2)}+2\alpha\right)\,
\Phi^2_{d/2}(0)\right\}
\end{eqnarray}
a $d$- and $\alpha$-dependent parameter. For $k\ll |\omega_d
G_{k_0}|^{-1/(d-2)}$ and with the reference scale $k_0=0$ (which is admissible 
only for trajectories lying on the IR critical hypersurface of the fixed
point) eq. (\ref{gfp9}) 
yields
\begin{eqnarray}
\label{gfp10}
G_k=G_0\left[1-\omega_d G_0 k^{d-2}+{\cal O}\left(G_0^2k^{2(d-2)}\right)\right]
\;.
\end{eqnarray}
For the dimensionful cosmological constant we obtain from eq. (\ref{gfp8})
\begin{eqnarray}
\label{gfp11}
\bar{\lambda}_k=\bar{\lambda}_{k_0}+\nu_d G_{k_0}\left(k^d-k_0^d\right)\;.
\end{eqnarray}
Apart from the different expression of $\omega_d$ due to the new cutoff, 
the solutions (\ref{gfp10}) and (\ref{gfp11}) coincide in $d=4$ dimensions 
with those derived in ref. \cite{Reu96} by using a similar approximation 
scheme, see eqs. (5.18) and (5.25) of this reference.

Let us now analyze the scaling behavior near $(\lambda_*,g_*)=(0,0)$. Since
$\theta_1=2>0$ the $V^1$-eigendirection, which coincides with the 
$\lambda$-direction, is IR repulsive (and thus UV attractive). For $d<2$, 
$\theta_2$ is positive which implies that the Gaussian fixed point is UV
attractive for any direction in the 2-dimensional parameter space. 

For $d>2$ we have $\theta_2<0$ so that the $V^2$-eigendirection
is IR attractive (and UV repulsive). Hence, in this case both the UV and the
IR critical hypersurface of the Gaussian fixed point are one-dimensional, i.e.
they consist of a single trajectory. For the IR critical trajectory that hits 
the fixed point in the limit $k\rightarrow0$ we have
\begin{eqnarray}
\label{gfp12}
\bar{\lambda}_k=\nu_d\,G_k\,k^d\;\;\;\Longleftrightarrow\;\;\;\lambda_k=\nu_d\,
 g_k
\end{eqnarray}
for sufficiently small values of $k$ \cite{frank}, with $G_k$ given by eq. 
(\ref{gfp10}). Since $\Phi^1_{d/2}(0)$ depends on the shape function 
$R^{(0)}$, $\nu_d$
is not a universal quantity. Therefore the slope of the distinguished 
trajectory (\ref{gfp12}) is not fixed in a universal manner. This is in
accordance with the general expectation that the eigenvalues of ${\bf B}$
should be universal, but not its eigenvectors.

For $d\neq 2$ the parameters $\Phi^1_{d/2-1}(0)$ and $\Phi^2_{d/2}(0)$ 
appearing in $\omega_d$ are scheme dependent as well. Furthermore, $\omega_d$
is a function of the gauge parameter $\alpha$. Hence $\omega_d$ is a 
nonuniversal quantity, too. In the most interesting case of $d=4$ dimensions
it takes the form
\begin{eqnarray}
\label{gfp14}
\omega_4=\frac{1}{24\pi}\left[13\Phi^1_1(0)+(55+24\alpha)
\Phi^2_2(0)\right]\;.
\end{eqnarray} 
Since $\Phi^1_1(0)$ and $\Phi^2_2(0)$ are positive for any admissible shape
function we can infer from eq. (\ref{gfp14}) that $\omega_4$ is positive for 
all $\alpha>\alpha_0$ and negative for all $\alpha<\alpha_0$. Here $\alpha_0
\equiv [-13\Phi^1_1(0)/\Phi^2_2(0)-55]/24$ is a negative number of order 
unity. Thus, if we identify Einstein gravity with the
theory described by the IR critical trajectory of the Gaussian fixed point,
eq. (\ref{gfp10}) implies that Einstein gravity is antiscreening for all 
$\alpha>\alpha_0$, i.e. $G_k$ decreases as $k$ increases. On the other hand if
$\alpha<\alpha_0$ gravity would exhibit a screening behavior. As argued in 
subsection \ref{4A}, the gauge parameter should be regarded as a scale 
dependent parameter in an exact treatment where it is expected to approach the 
fixed point value $\alpha_*=0$. Setting $\alpha=0$ from the outset we may 
conclude that the physical $G_k$ displays the antiscreening behavior found 
for $\alpha>\alpha_0$. 

In ref.
\cite{sven} a similar result was obtained with a cutoff of type A, while the
above calculation employed the cutoff B. The only difference 
between our result for the behavior of $G_k$ and the one obtained in 
\cite{sven} lies in the slightly differing value of $\alpha_0$ which is a 
scheme dependent parameter. For the cutoff type A, $\omega_4\equiv
\omega_4^{(\rm A)}$ is given by (see \cite{sven})
\begin{eqnarray}
\label{gfp15}
\omega_4^{(\rm A)}=\frac{1}{6\pi}\left[(18+6\alpha)\Phi^2_2(0)-\Phi^1_1(0)
\right]
\end{eqnarray}
so that $\alpha_0=[\Phi^1_1(0)/\Phi^2_2(0)-18]/6$ in this case, while 
$\nu_4$ is the same with both cutoffs. For $\alpha=1$, eq. (\ref{gfp15}) boils
down to $\omega_4^{(\rm A)}=\left[24\Phi^2_2(0)-\Phi^1_1(0)\right]/(6\pi)$ 
which equals the result obtained in the original paper \cite{Reu96}. This is 
because for $\alpha=1$ the cutoff type A coincides with the one used in ref. 
\cite{Reu96}. For a comparison of this result with the one for the cutoff type
B we insert $\alpha=1$ into eq. (\ref{gfp14}) which yields $\omega_4^{(\rm B)}
=\left[13\Phi^1_1(0)+79\Phi^2_2(0)\right]/(24\pi)$. Using the exponential 
shape function $R^{(0)}$ with $s=1$  we have $\Phi^1_1(0)
=\pi^2/6$, $\Phi^2_2(0)=1$ so that $\omega_4^{(\rm B)}
\approx 1.33$ for the cutoff type B, which lies rather close to the value
$\omega_4^{(\rm A)}\approx 1.19$ obtained in \cite{Reu96,sven} for the cutoff 
type A. Furthermore, we have $\Phi^1_2(0)=2\zeta(3)$ where $\zeta$ denotes the
zeta function, and thus $\nu_4\approx 0.19$ with both cutoffs.

\subsection{The non-Gaussian fixed point}
\label{5C}
Now we turn to the nontrivial zeros of the set of
$\mbox{\boldmath$\beta$}$-functions $\{\mbox{\boldmath$\beta$}_\lambda,
\mbox{\boldmath$\beta$}_g\}$ given by eqs. (\ref{001}), (\ref{004}). Such 
non-Gaussian fixed points $(\lambda_*,g_*)\neq(0,0)$ satisfy the condition
\begin{eqnarray}
\label{ngfp0}
\eta_{N*}=2-d
\end{eqnarray}
which follows immediately from eq. (\ref{004}).

\subsubsection{In $2+\varepsilon$ dimensions}
As a warm up we consider the case of $d=2+\varepsilon$ dimensions with
$0<|\varepsilon|\ll 1$ which can be dealt with analytically and for which the 
existence of a non-Gaussian fixed point 
has already been shown \cite{wein,Reu96,gastmans,nin,jack}. In this case the 
condition (\ref{ngfp0}) takes the form $\eta_{N*}=-\varepsilon$ with
\begin{eqnarray}
\label{ngfp6}
\eta_{N*}=\frac{g_*(\varepsilon)\,B_1(\lambda_*(\varepsilon);\alpha,2+
\varepsilon)}{1-g_*(\varepsilon)\,B_2(\lambda_*(\varepsilon);\alpha,
2+\varepsilon)}\;.
\end{eqnarray}
Solving eq. (\ref{ngfp6}) for $g_*(\varepsilon)$ and expanding the result with
respect to $\varepsilon$ leads to 
\begin{eqnarray}
\label{ngfp7}
g_*(\varepsilon)=-\left[B_1(\lambda_*(0);\alpha,2)\right]^{-1}\,\varepsilon
+{\cal O}\left(\varepsilon^2\right)\;.
\end{eqnarray}
Furthermore, expanding also $\beta_\lambda(\lambda_*(\varepsilon),
g_*(\varepsilon);\alpha,2+\varepsilon)$ with respect to $\varepsilon$ and 
equating equal powers of $\varepsilon$ yields
\begin{eqnarray}
\label{ngfp8}
\lambda_*(\varepsilon)=\left[B_1(0;\alpha,2)\right]^{-1}
\Phi^1_1(0)\,\varepsilon+{\cal O}\left(\varepsilon^2\right)\;.
\end{eqnarray}
In particular we obtain $\lambda_*(0)=0$ which implies $g_*(\varepsilon)
=-\left[B_1(0;\alpha,2)\right]^{-1}\,\varepsilon+{\cal O}\left(\varepsilon^2
\right)$. 

The parameters $B_i(0;\alpha,2)$, $i=1,2$, may be obtained from the zeroth
order terms of the expansions $B_i(\lambda_k;\alpha,2+\varepsilon)=
B_i^{(0)}(\lambda_k;\alpha)+B_i^{(1)}(\lambda_k;\alpha)\varepsilon
+{\cal O}(\varepsilon^2)$ which take the form
\begin{eqnarray}
\label{uni1}
B_1^{(0)}(\lambda_k;\alpha)&=&-\frac{34}{3}\left(1-2\lambda_k\right)^{-1}-
\frac{2}{3}\left(1-2\alpha\lambda_k\right)^{-1}+4\Phi^2_1(-2\lambda_k)
+6\Phi^2_1(-2\alpha\lambda_k)-\frac{32}{3}\;,\nonumber\\
B_2^{(0)}(\lambda_k;\alpha)&=&\frac{17}{3}\left(1-2\lambda_k\right)^{-1}
+\frac{1}{3}\left(1-2\alpha\lambda_k\right)^{-1}-2\widetilde{\Phi}^2_1(-2
\lambda_k)-3\widetilde{\Phi}^2_1(-2\alpha\lambda_k)\;.
\end{eqnarray}
Inserting $\lambda_k=0$ into eq. (\ref{uni1}) and using that $\Phi^2_1(0)=1$ is
scheme independent \cite{Reu96} yields
\begin{eqnarray}
\label{uni2}
B_1(0;\alpha,2)&=&B_1^{(0)}(0;\alpha)=-\frac{38}{3}\;,\nonumber\\
B_2(0;\alpha,2)&=&B_2^{(0)}(0;\alpha)=6-5\widetilde{\Phi}^2_1(0)\;.
\end{eqnarray}
In contrast to the universal quantity $B_1^{(0)}(0;\alpha)$, 
$B_2^{(0)}(0;\alpha)$ depends on the shape of $R^{(0)}$ via 
$\widetilde{\Phi}^2_1(0)$. However, $B_2^{(0)}(0;\alpha)$ does not enter the
leading order term of $\lambda_*(\varepsilon)$ and $g_*(\varepsilon)$, which
may now be written as
\begin{eqnarray}
\label{ngfp9}
\lambda_*(\varepsilon)&=&-\frac{3}{38}\Phi^1_1(0)\,\varepsilon+{\cal O}\left(
\varepsilon^2\right)\;,\nonumber\\
g_*(\varepsilon)&=&\frac{3}{38}\,\varepsilon+{\cal O}\left(
\varepsilon^2\right)\;.
\end{eqnarray}
The leading order term  of $\lambda_*(\varepsilon)$ is nonuniversal since it 
contains the scheme dependent parameter $\Phi^1_1(0)$. This is not the case 
for $g_*(\varepsilon)$ whose leading order 
contribution has a universal meaning.

Let us now analyze the scaling behavior near the non-Gaussian fixed point 
(\ref{ngfp9}). One finds that the associated ${\bf B}$-matrix is of the form
\begin{eqnarray}
\label{ngfp10}
{\bf B}=\left(\begin{array}{cc}
-2+\frac{12\alpha-13}{19}\,\varepsilon+{\cal O}\left(\varepsilon^2\right) &\;\;
-2\Phi^1_1(0)+{\cal O}\left(\varepsilon\right)\\
{\cal O}\left(\varepsilon^2\right) &\;\;
-\varepsilon+{\cal O}\left(\varepsilon^2\right)
\end{array}\right)\;.
\end{eqnarray}
From (\ref{ngfp10}) we obtain the critical exponents $\theta_1=2-
\frac{12\alpha-13}{19}\,\varepsilon+{\cal O}\left(\varepsilon^2\right)$ and
$\theta_2=\varepsilon+{\cal O}\left(\varepsilon^2\right)$. $\theta_1$ and
$\theta_2$ are scheme independent up to terms of
${\cal O}\left(\varepsilon^2\right)$. $\theta_1$ depends on the gauge 
parameter. For $\alpha=1$ the critical exponents coincide with those following
from the $\mbox{\boldmath$\beta$}$-functions of ref. \cite{Reu96}, where the
cutoff type A is used. These findings nicely confirm that, to lowest order,
{\it the critical exponents are the same for the cutoffs A and B and are 
independent of $R^{(0)}$.}

For $\varepsilon>0$ both critical exponents are positive. Hence the
non-Gaussian fixed point (\ref{ngfp9}) is UV attractive for all trajectories so
that the condition for the asymptotic safety scenario is met. 
It is interesting to investigate whether this result stabilizes in the sense 
that more general truncations including higher powers of the curvature tensor
reproduce this fixed point and lead to a finite-dimensional UV critical 
hypersurface. In \cite{LR3} we will discuss this point in detail.

\subsubsection{Location of the fixed point $(d=4)$}
In $d=4$ dimensions, and for the cutoff A, the non-Gaussian fixed point of the
Einstein-Hilbert truncation was first discussed in \cite{souma1}, and in ref. 
\cite{souma2} the $\alpha$- and $R^{(0)}$-dependence of its projection 
$(0,g_*)$ onto the $g$-direction has been investigated. However, since for
$\alpha\neq 1$ the cutoff of type A is introduced by an ad hoc modification 
of the standard one-loop determinants it is not clear whether it may be 
expressed in terms of an action $\Delta_k S$, except for the case $\alpha=1$ 
\cite{Reu96}. Since a specification of $\Delta_k S$ is indispensable for the 
construction of $\Gamma_k$, the status of the results derived in 
\cite{souma2} is somewhat unclear. In the following we determine the fixed 
point properties using different cutoffs of type B, for which a $\Delta_k S$ 
is known to exist, and compare them to the analogous results for the cutoff A.

In a first attempt to determine the non-Gaussian fixed point we neglect the
cosmological constant and set $\lambda_k=\lambda_*=0$, thereby projecting
the renormalization group flow onto the one-dimensional space 
parametrized by $g$. In this case the non-Gaussian fixed point is obtained
as the nontrivial solution of $\mbox{\boldmath$\beta$}_g(0,g_*;\alpha,d)=0$.
It is determined in appendix \ref{approx} with the result given by eq. 
(\ref{H1}). In order to get a first impression of the position of $g_*$ we 
insert the exponential shape function  with $s=1$ into eq. (\ref{H1})
and set $d=4$, $\alpha=1$. We obtain $g_*\approx 0.590$.

Assuming that for the combined $\lambda$-$g$ system both $g_*$ and 
$\lambda_*$ are of the same order of magnitude as $g_*$ above we expand 
the $\mbox{\boldmath$\beta$}$-functions about $(\lambda_k,g_k)=(0,0)$ and 
neglect terms of higher orders in the couplings. Again in appendix 
\ref{approx} we determine the non-Gaussian fixed point for the corresponding
system of differential equations. Inserting the shape function (\ref{expcut}) 
and setting $d=4$, $\alpha=1$, we find $(\lambda_*,g_*)\approx(0.287,0.751)$.

\renewcommand{\baselinestretch}{1}
\small\normalsize
\begin{figure}[ht]
\begin{minipage}{7.7cm}
        \epsfxsize=7.7cm
        \epsfysize=6cm
        \centerline{\epsffile{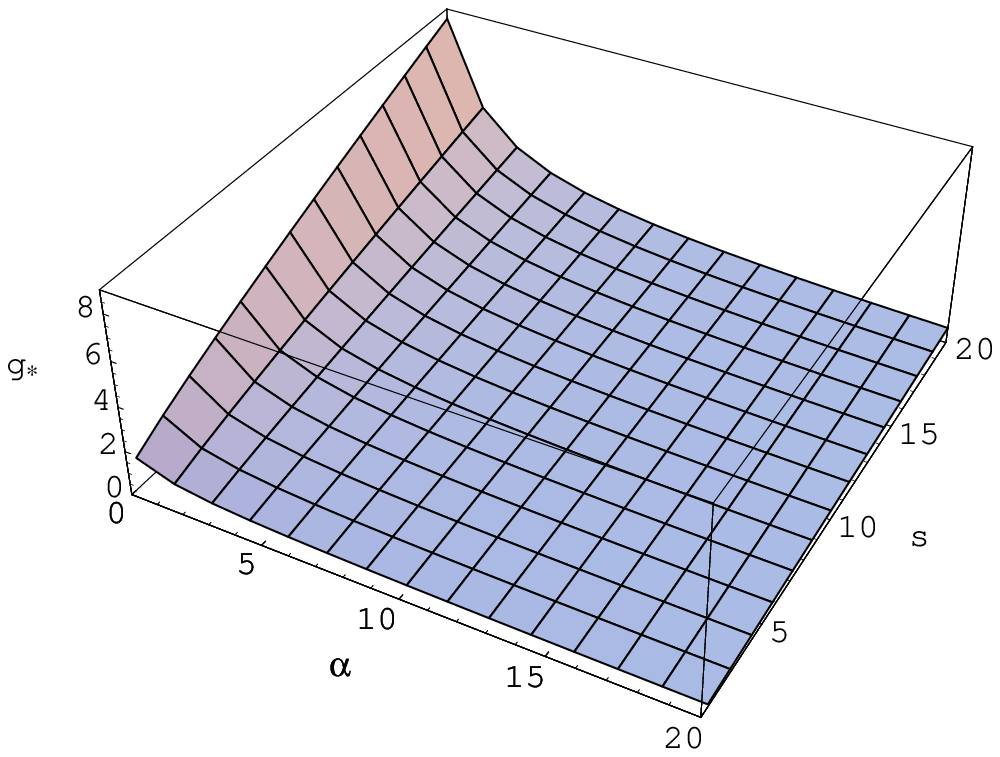}}
\centerline{(a)}
\end{minipage}
\hfill
\begin{minipage}{7.7cm}
        \epsfxsize=7.7cm
        \epsfysize=6cm
        \centerline{\epsffile{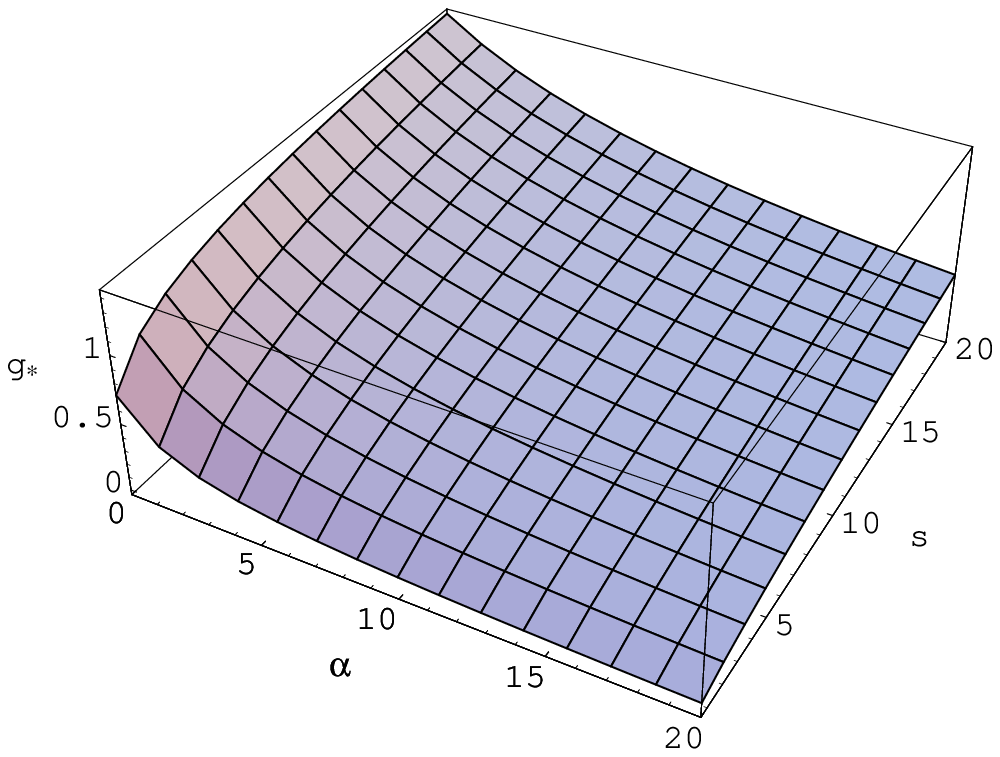}}
\centerline{(b)}
\end{minipage}
\vspace{0.3cm}
\caption{$g_*$ as a function of $s$ and $\alpha$ from the approximation 
$\lambda_k=\lambda_*=0$,
using (a) the cutoff type A, and (b) the cutoff type B, with the family of 
exponential shape functions (\ref{H6}) inserted.}  
\label{fi1}
\end{figure}
%
%\renewcommand{\baselinestretch}{1.5}
%\small\normalsize
%
%
%
%\renewcommand{\baselinestretch}{1}
%\small\normalsize
%
\begin{figure}[ht]
\begin{minipage}{7.7cm}
        \epsfxsize=7.7cm
        \epsfysize=6cm
        \centerline{\epsffile{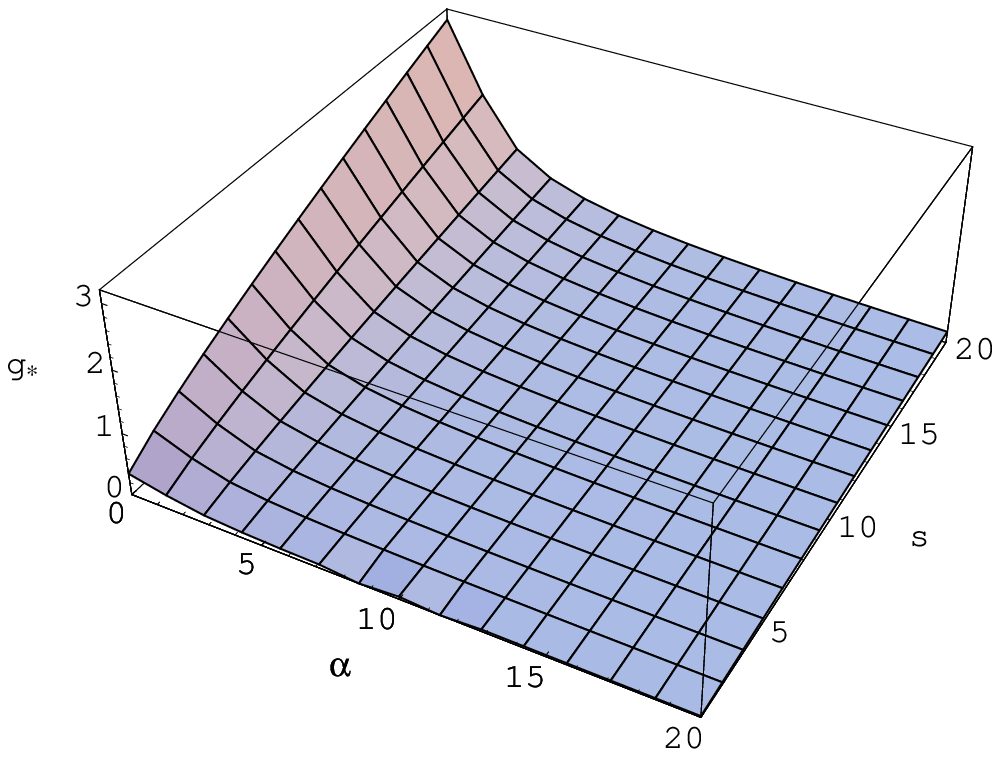}}
\centerline{(a)}
\end{minipage}
\hfill
\begin{minipage}{7.7cm}
        \epsfxsize=7.7cm
        \epsfysize=6cm
        \centerline{\epsffile{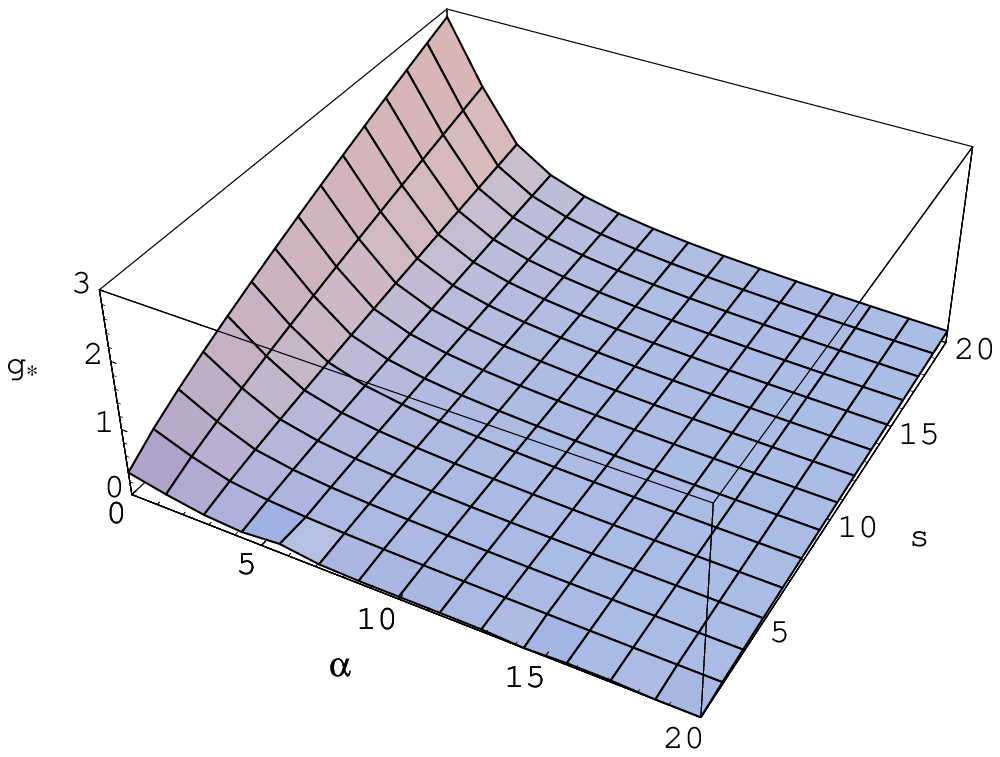}}
\centerline{(b)}
\end{minipage}
\vspace{0.3cm}
\caption{The exact $g_*$ as a function of $s$ and $\alpha$ from the combined 
$\lambda$-$g$ system, using (a) the cutoff type A, and (b) the cutoff type B, 
with the family of exponential shape functions (\ref{H6}) inserted.}  
\label{fi2}
\end{figure}
\renewcommand{\baselinestretch}{1.5}
\small\normalsize
\renewcommand{\baselinestretch}{1}
\small\normalsize
\begin{figure}[ht]
\begin{minipage}{7.7cm}
        \epsfxsize=7.7cm
        \epsfysize=5cm
        \centerline{\epsffile{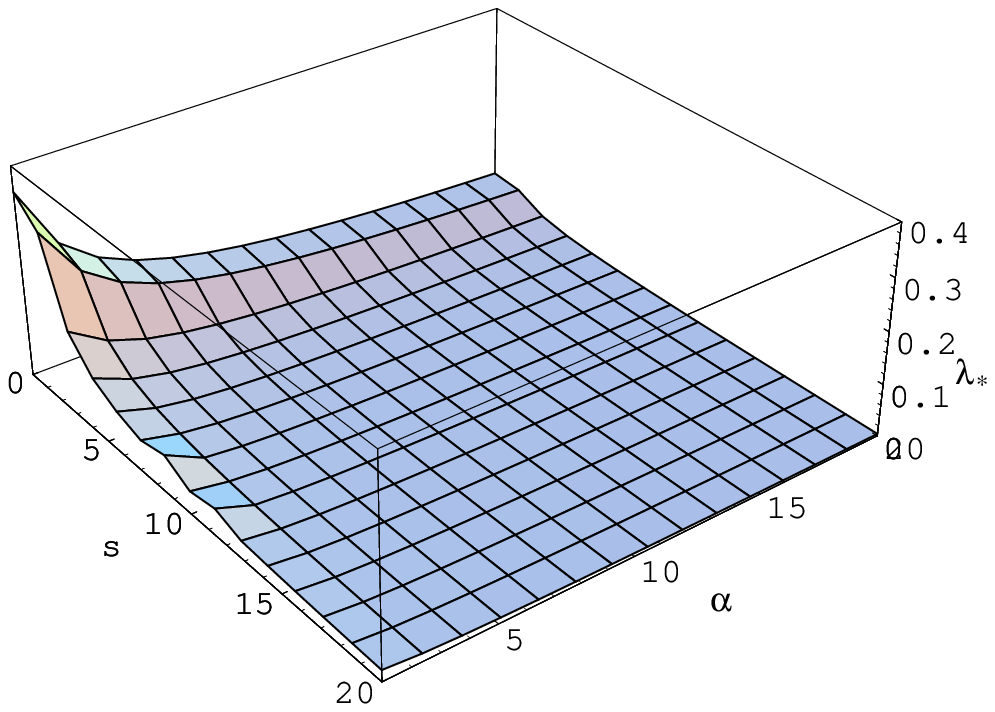}}
\centerline{(a)}
\end{minipage}
\hfill
\begin{minipage}{7.7cm}
        \epsfxsize=7.7cm
        \epsfysize=5cm
        \centerline{\epsffile{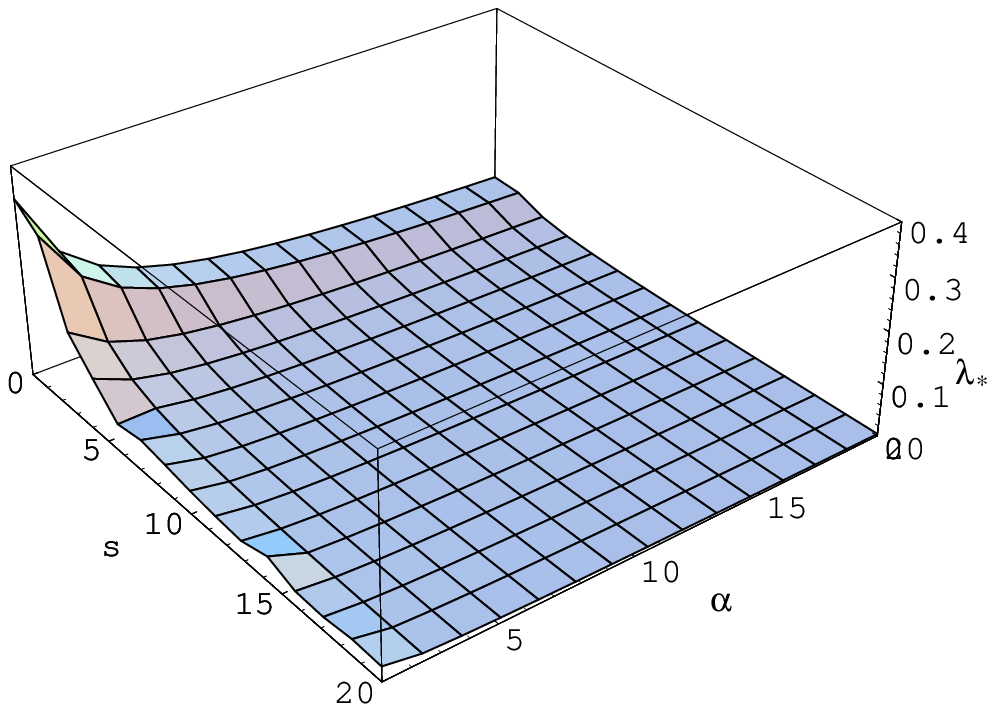}}
\centerline{(b)}
\end{minipage}
\vspace{0.3cm}
\caption{The exact $\lambda_*$ as a function of $s$ and $\alpha$ from the 
combined $\lambda$-$g$ system, using (a) the cutoff type A, and (b) the cutoff
type B, with the family of exponential shape functions (\ref{H6}) inserted.}  
\label{fi3}
\end{figure}
\renewcommand{\baselinestretch}{1.5}
\small\normalsize

In order to determine the {\it exact} position of the non-Gaussian fixed point 
$(\lambda_*,g_*)$ we have to resort to numerical methods. Given a starting 
value for the fixed point, e.g. one of the approximate solutions above, the
program we use determines a numerical solution which is exact up to an 
arbitrary degree of accuracy. Under the same conditions as above, i.e. $s=1$, 
$d=4$, we obtain
\begin{eqnarray}
\label{fixedpoint} 
(\lambda_*,g_*)=\left\{\begin{array}{l}\mbox{$(0.348,0.272)$ for $\alpha=1$} \\
\mbox{$(0.339,0.344)$ for $\alpha=0$}\end{array}\right.\;.
\end{eqnarray}

Next we study the gauge and scheme dependence of the non-Gaussian
fixed point. The scheme dependence is investigated by looking at the 
$s$-dependence introduced via the family
of exponential shape functions (\ref{H6}) where $s$ parametrizes the profile
of $R^{(0)}$. 

FIG. \ref{fi1} shows $g_*(\alpha,s)$ obtained from the 
approximation $\lambda_k=\lambda_*=0$, while FIGS. \ref{fi2} and \ref{fi3}
display the (exact) functions $g_*(\alpha,s)$ and $\lambda_*(\alpha,s)$ 
resulting from the combined $\lambda$-$g$ system. In each of these figures the
plot on the LHS (i.e. FIGS. 1,2,3(a)) is obtained from the cutoff type A and 
the one on the RHS (i.e. FIGS. 1,2,3(b)) is 
obtained from the cutoff type B used in the present paper. 

Our results establish
the existence of the non-Gaussian fixed point in a wide range of 
$\alpha$- and $s$-values. As expected, the position of the fixed point turns 
out to be $s$-, i.e. scheme dependent, but the crucial point is that it
exists for any of the cutoffs employed. This is one of the important results 
of our analysis because it gives a first hint at the reliability of the 
Einstein-Hilbert truncation.    

As for the $\alpha$-dependence, $\alpha=0$ is, in principle, the only relevant
case since according to subsection \ref{4A}, $\alpha=0$ is assumed to be 
the physical value of the gauge parameter. In practical calculations
$\alpha=1$ is often used instead, because this simplifies the evaluation of 
the flow equation considerably. In \cite{LR3}, for instance, all calculations
are performed with $\alpha=1$ for this reason. Therefore it is necessary to 
compare the gauges $\alpha=0$ and $\alpha=1$ in order to judge whether the
results obtained by using $\alpha=1$ are a sensible approximation to
the physical case $\alpha=0$. Here we see that this is indeed the case.

As for comparing different types of cutoffs, we recognize from
FIG. \ref{fi1} that, in the approximation $\lambda_k=\lambda_*=0$, the 
$s$-dependence of $g_*$ is much weaker for type B than for type A. Contrary
to this, both cutoffs yield nearly the same results for $g_*$ and
$\lambda_*$ if we consider the combined $\lambda$-$g$ system, see FIGS. 
\ref{fi2} and \ref{fi3}. Furthermore, the scheme dependence of $g_*$
in FIG. \ref{fi2} is stronger than in FIG. \ref{fi1}(b), but much weaker than
in FIG. \ref{fi1}(a). FIG. \ref{fi1}(a) reproduces the result of ref. 
\cite{souma2} obtained from the cutoff A, see FIG. 2 of this reference. 

It should be noted that we are forced to restrict our considerations to 
shape functions (\ref{H6}) with $s\ge 1$. This is because for $s<1$ the 
numerical integrations are plagued by convergence problems which is due to the
fact that in $d=4$ dimensions the threshold functions in $\mbox{\boldmath
$\beta$}_\lambda$ and $\mbox{\boldmath$\beta$}_g$ diverge in the limit 
$s\rightarrow 0$, see also \cite{souma2}.

Because the scale $k$ enters the flow equation via ${\cal R}_k$ as a purely
mathematical device it is clear that the functions $k\mapsto\lambda_k,g_k$
and their UV limits $\lambda_*,g_*$ are scheme dependent and not directly 
observable therefore. It can be argued that the product $g_*\lambda_*$ must 
be scheme independent, however. While $k$ and, at a fixed value of $k$, $G_k$
and $\bar{\lambda}_k$ cannot be measured separately, we may invert the function
$k\mapsto G_k$ and insert the result $k=k(G)$ into $\bar{\lambda}_k$. This
leads to a relationship between Newton's constant and the cosmological 
constant which, at least in principle, could be tested experimentally: 
$\bar{\lambda}=\bar{\lambda}(G)$. In general this relation depends on the RG
trajectory chosen (specified by its IR values $\bar{\lambda}_0$ and $G_0$, for
instance), but in the fixed point regime all trajectories approach 
$\bar{\lambda}_k=\lambda_*k^2$ and $G_k=g_*/k^2$ which gives rise to
\begin{eqnarray}
\label{ngfp4}
\bar{\lambda}(G)=\frac{g_*\lambda_*}{G}\;.
\end{eqnarray}
Eq. (\ref{ngfp4}) is valid if $\bar{\lambda}\gg m_{\rm Pl}^2$ and $G\ll 
m_{\rm Pl}^{-2}$. (We define the Planck mass in terms of the IR limit of $G_k$,
$m_{\rm Pl}\equiv G_0^{-1/2}$.) Assuming that $\bar{\lambda}$ and $G$ have the
status of observable quantities, eq. (\ref{ngfp4}) shows that $g_*\lambda_*$
must be observable, and hence scheme independent, too. (For a related 
discussion see \cite{nin}.) Below the
Planck regime the function $\bar{\lambda}(G)$ becomes much more complicated
than $\bar{\lambda}\propto 1/G$ (which follows already from dimensional 
analysis) because the dimensionful quantities $\bar{\lambda}_0$ and $G_0$
enter explicitly there.

\renewcommand{\baselinestretch}{1}
\small\normalsize
\begin{figure}[ht]
\begin{minipage}{7.9cm}
        \epsfxsize=7.9cm
        \epsfysize=5.2cm
        \centerline{\epsffile{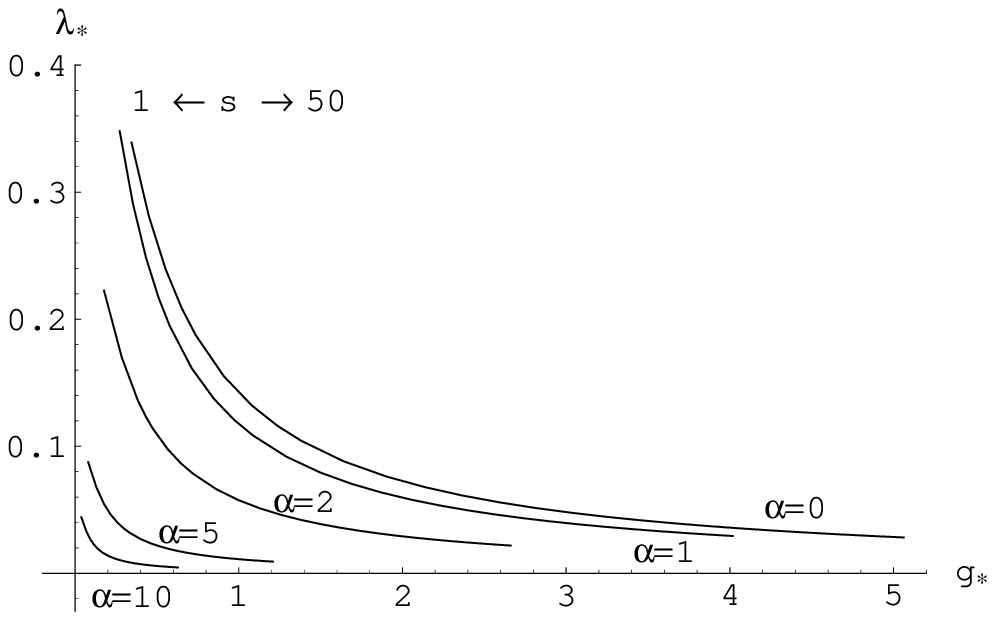}}
\centerline{(a)}
\end{minipage}
\hfill
\begin{minipage}{7.9cm}
        \epsfxsize=7.9cm
        \epsfysize=5.2cm
        \centerline{\epsffile{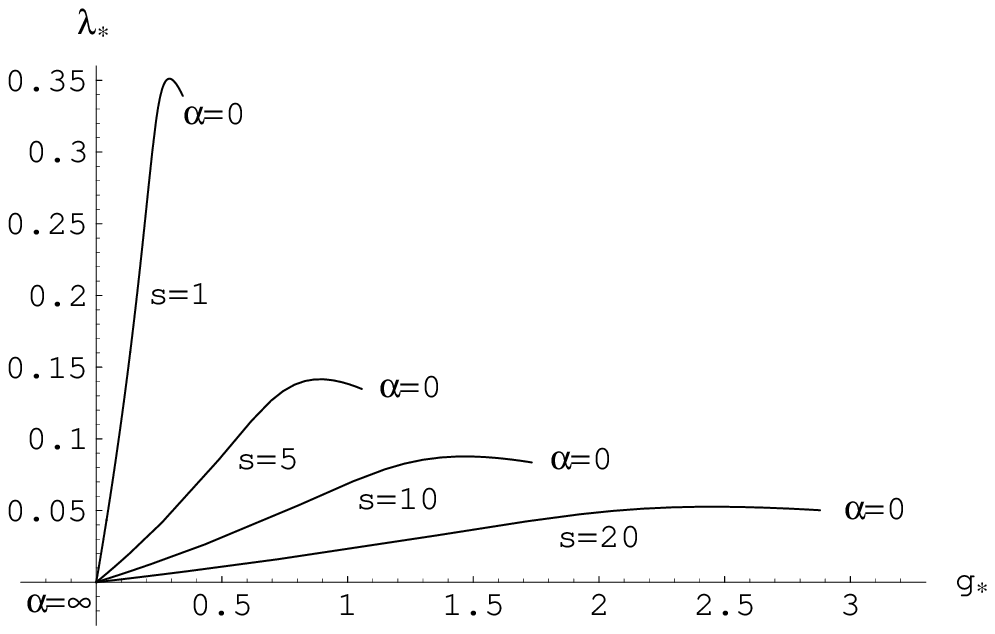}}
\centerline{(b)}
\end{minipage}
\vspace{0.3cm}
\caption{(a) $s$-parametric plot of $(\lambda_*(s),g_*(s))$ in the range 
$1\le s\le 50$ for various values of $\alpha$. Each curve starts on the left 
at $s=1$, and ends on the right at $s=50$. (b) $\alpha$-parametric plot of 
$(\lambda_*(\alpha),g_*(\alpha))$ for various values of $s$. In both (a) and
(b) the cutoff type B is used, with the family of exponential shape functions 
(\ref{H6}) inserted.}  
\label{fi4}
\end{figure}
\renewcommand{\baselinestretch}{1.5}
\small\normalsize

As for the universality of $g_*\lambda_*$, it is also interesting to note that,
for any $k$, the product $g_k\lambda_k=G_k\bar{\lambda}_k$ is essentially the
inverse of the on shell value of $\Gamma_k$. The stationary points of 
(\ref{in2}) with $\bar{g}_{\mu\nu}=g_{\mu\nu}$ satisfy Einstein's equation
$G_{\mu\nu}=-\bar{\lambda}_k\,g_{\mu\nu}$. Hence $R=4\bar{\lambda}_k$, so that
from (\ref{in2}) in four dimensions,
\begin{eqnarray}
\label{ngfp5}
\Gamma_k[\mbox{on shell}]=-\frac{v}{8\pi G_k\bar{\lambda}_k}
=-\frac{v}{8\pi g_k\lambda_k}\;.
\end{eqnarray}
Here we used that, for dimensional reasons, $\int d^4x\sqrt{g}=v/
\bar{\lambda}_k^2$ where $v$ is a finite, positive constant for any solution
with a finite four-volume.

\renewcommand{\baselinestretch}{1}
\small\normalsize
\begin{figure}[ht]
\begin{minipage}{7.9cm}
        \epsfxsize=7.9cm
        \epsfysize=5.2cm
        \centerline{\epsffile{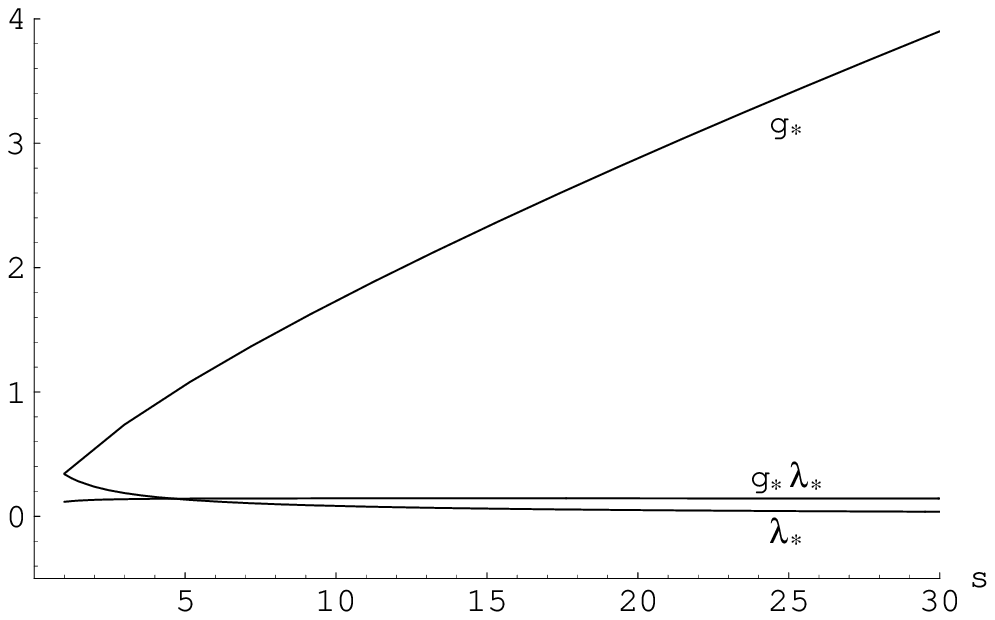}}
\centerline{(a)}
\end{minipage}
\hfill
\begin{minipage}{7.9cm}
        \epsfxsize=7.9cm
        \epsfysize=5.2cm
        \centerline{\epsffile{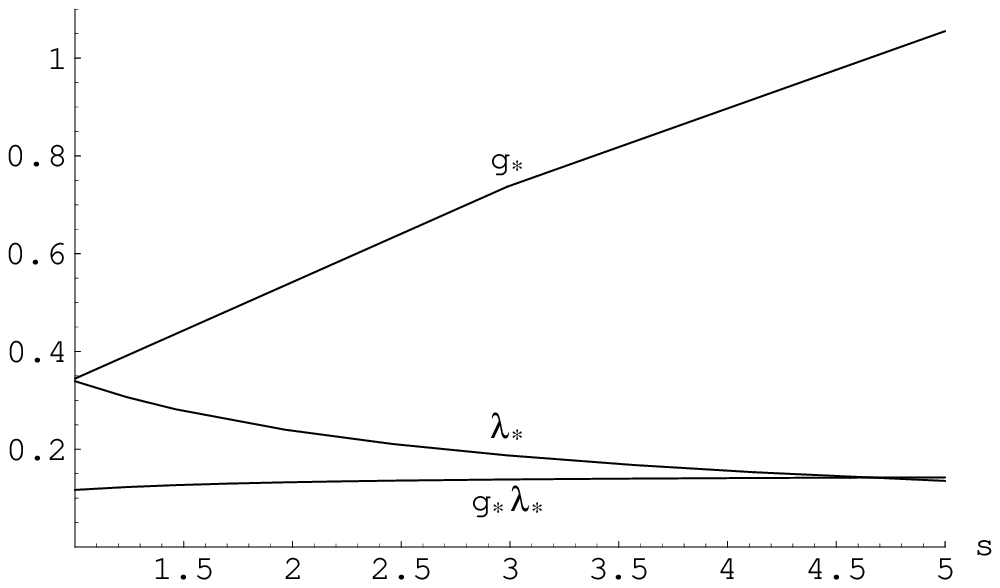}}
\centerline{(b)}
\end{minipage}
\begin{minipage}{7.9cm}
        \epsfxsize=7.9cm
        \epsfysize=5.2cm
        \centerline{\epsffile{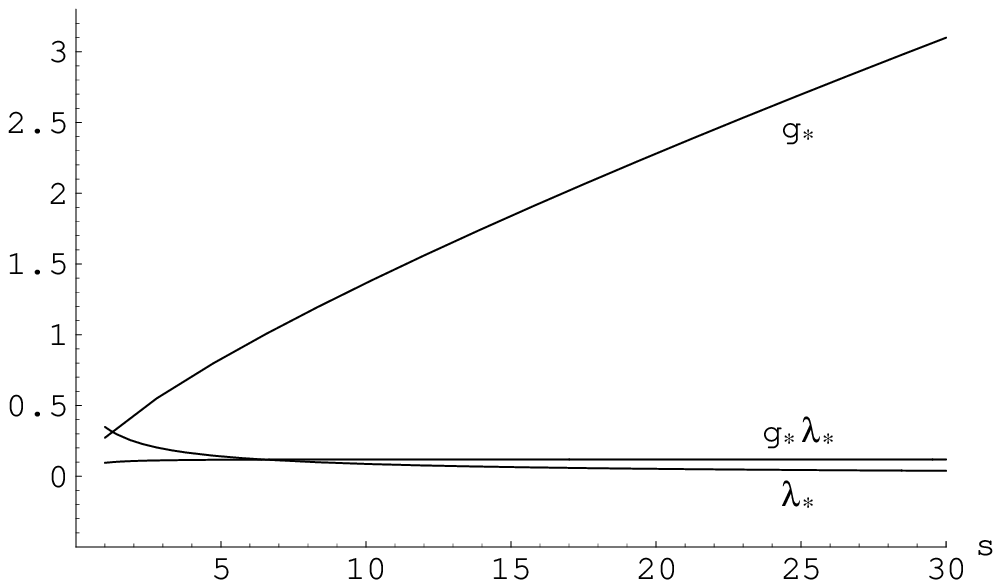}}
\centerline{(c)}
\end{minipage}
\hfill
\begin{minipage}{7.9cm}
        \epsfxsize=7.9cm
        \epsfysize=5.2cm
        \centerline{\epsffile{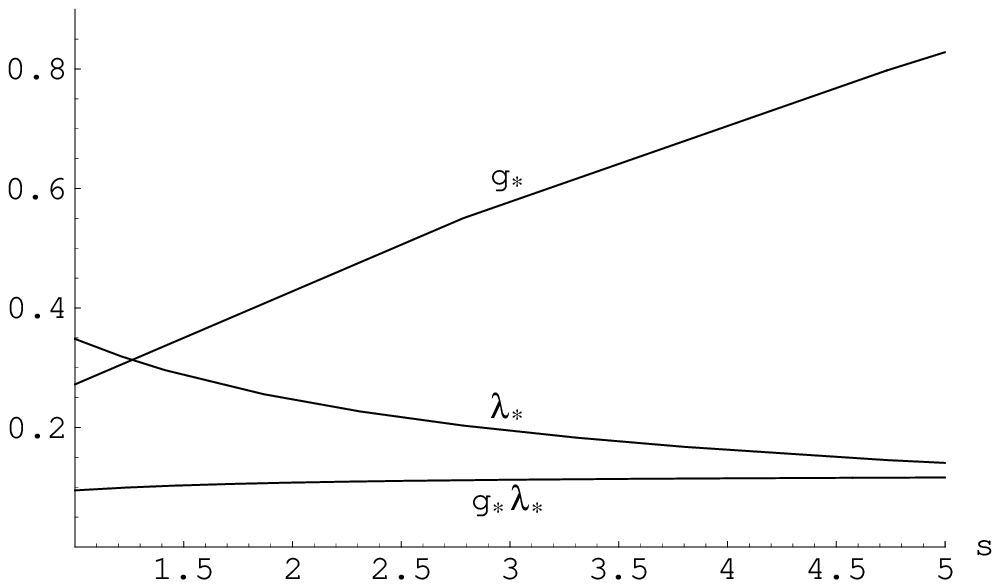}}
\centerline{(d)}
\end{minipage}
\vspace{0.3cm}
\caption{$g_*$, $\lambda_*$, and $g_*\lambda_*$ as functions of $s$ for (a)
$\alpha=0$, $1\le s\le 30$, (b) $\alpha=0$, $1\le s\le 5$, (c) $\alpha=1$, 
$1\le s\le 30$, and (d) $\alpha=1$, $1\le s\le 5$, using
the cutoff type B with the family of exponential shape functions (\ref{H6}) 
inserted.}  
\label{fi5}
\end{figure}

\renewcommand{\baselinestretch}{1.5}
\small\normalsize
\renewcommand{\baselinestretch}{1}
\small\normalsize
\begin{figure}[ht]
\begin{minipage}{7.9cm}
        \epsfxsize=7.9cm
        \epsfysize=5.2cm
        \centerline{\epsffile{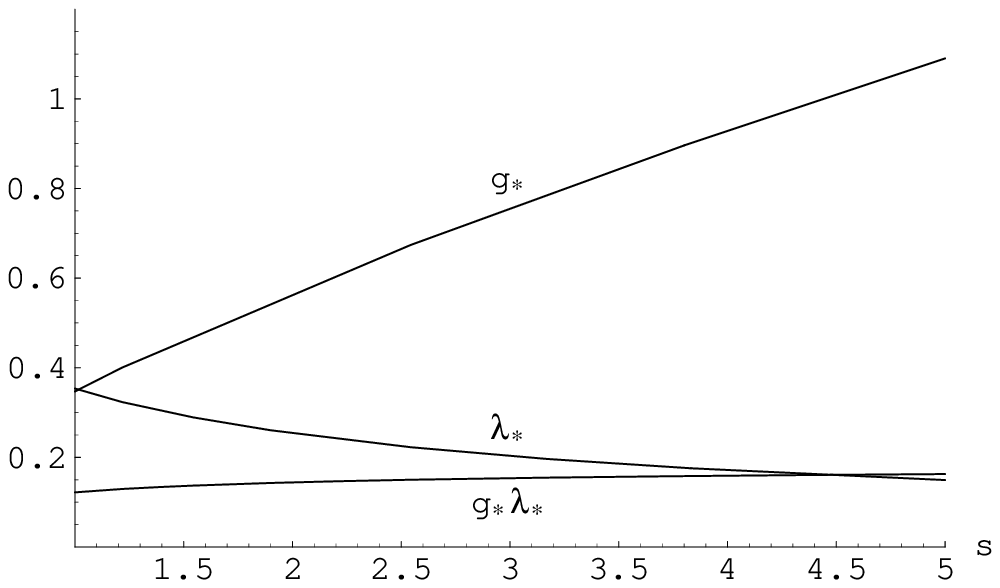}}
\centerline{(a)}
\end{minipage}
\hfill
\begin{minipage}{7.9cm}
        \epsfxsize=7.9cm
        \epsfysize=5.2cm
        \centerline{\epsffile{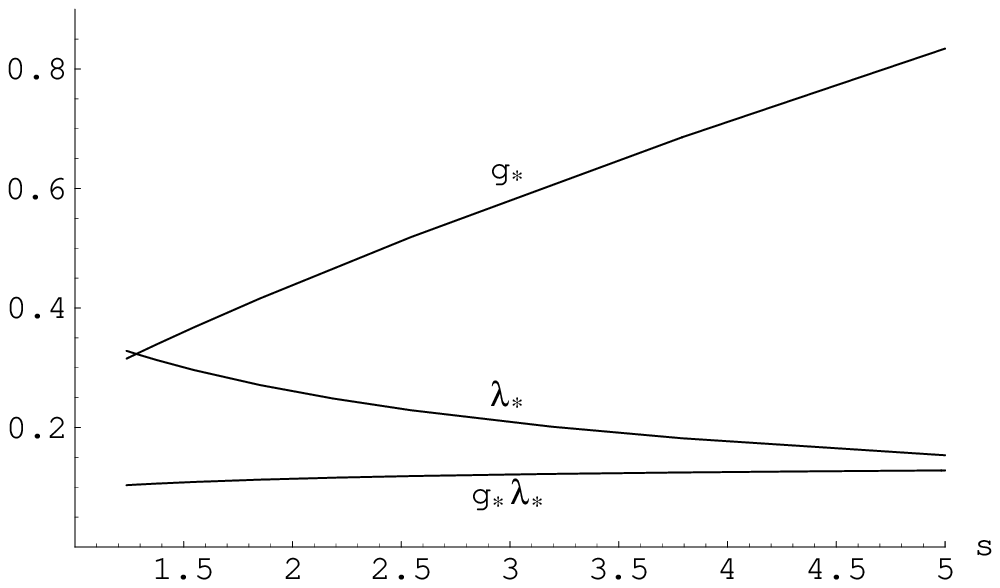}}
\centerline{(b)}
\end{minipage}
\vspace{0.3cm}
\caption{$g_*$, $\lambda_*$, and $g_*\lambda_*$ as functions of $s$ for (a)
$\alpha=0$, and  (b) $\alpha=1$, using
the cutoff type A with the family of exponential shape functions (\ref{H6}) 
inserted.}  
\label{fi6}
\end{figure}
\renewcommand{\baselinestretch}{1.5}
\small\normalsize

\begin{figure}[ht]
\begin{minipage}{7.9cm}
        \epsfxsize=7.9cm
        \epsfysize=5.2cm
        \centerline{\epsffile{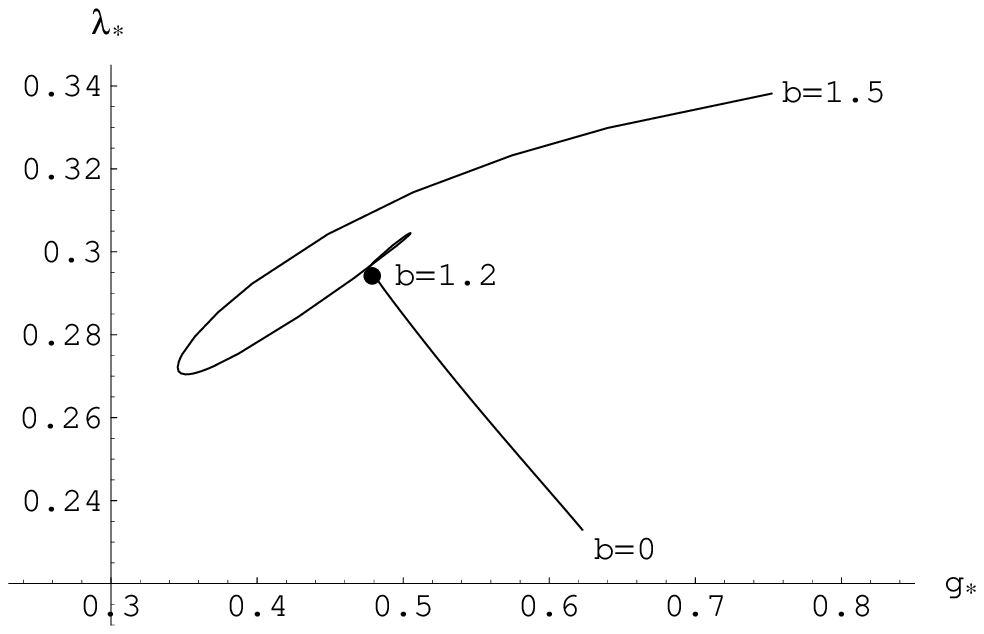}}
\centerline{(a)}
\end{minipage}
\hfill
\begin{minipage}{7.9cm}
        \epsfxsize=7.9cm
        \epsfysize=5.2cm
        \centerline{\epsffile{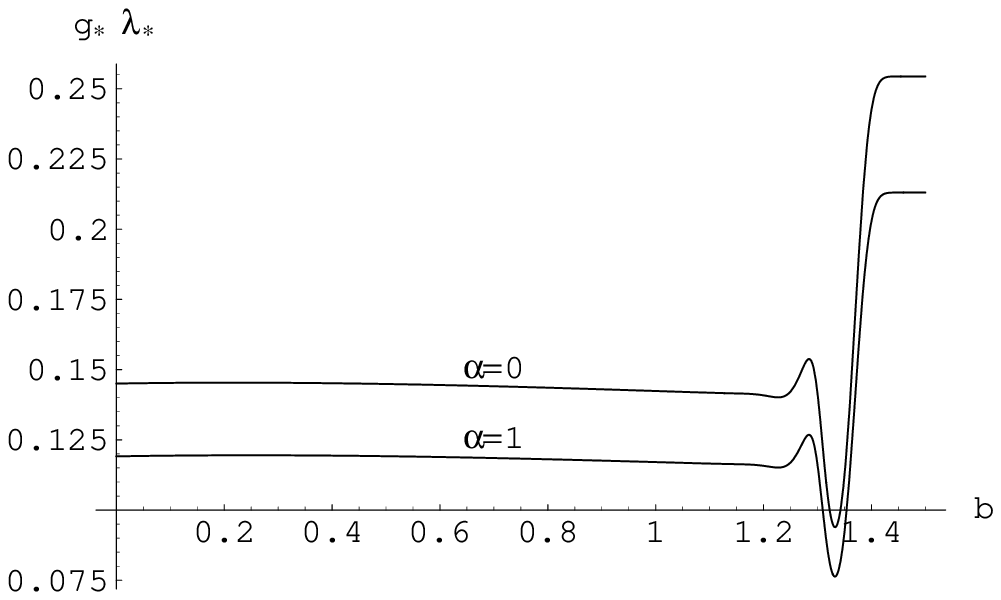}}
\centerline{(b)}
\end{minipage}
\vspace{0.3cm}
\caption{(a) $b$-parametric plot of $(\lambda_*(b),g_*(b))$ for $\alpha=0$, and
(b) $g_*\lambda_*$ as a function of $b$ for $\alpha=0$ and $\alpha=1$, using
the cutoff type B with the family of shape functions with compact support 
(\ref{supp}) inserted.}  
\label{fi7}
\end{figure}

Quite remarkably, the universality of the product $g_*\lambda_*$ is confirmed
by our results in a rather impressive manner, as is illustrated in FIGS. 
\ref{fi4}-\ref{fi7}. FIG. \ref{fi4}(a) contains several parametric plots of
$(\lambda_*(s),g_*(s))$ for various values of $\alpha$, obtained from the
$\mbox{\boldmath$\beta$}$-functions (\ref{001}) and (\ref{004}) which are based
on the cutoff type B. The hyperbolic shape of these plots is a first hint at
the $s$-independence of the product $g_*\lambda_*$. Its direct confirmation is 
supplied by FIG. \ref{fi5} which shows $g_*$, $\lambda_*$, and $g_*\lambda_*$
as functions of $s$ for $\alpha=0$ (FIG. \ref{fi5}(a),(b)) and $\alpha=1$
(FIG. \ref{fi5}(c),(d)), again using the cutoff type B. In FIG. 
\ref{fi5}(a),(c) these functions are plotted in the range of values 
$1\le s\le 30$ while 
FIG. \ref{fi5}(b),(d) contains the sector corresponding to $1\le s\le 5$
where the largest changes in $\lambda_*$ and $g_*$ occur. In any of these 
figures the product of $\lambda_*$ and $g_*$ is almost constant for the whole
range of $s$-values considered. Its universal value is 
\begin{eqnarray}
\label{product}
g_*\lambda_*\approx\left\{\begin{array}{l}\mbox{$0.12$ for $\alpha=1$}\\
\mbox{$0.14$ for $\alpha=0$}\end{array}\right.\;.
\end{eqnarray}
Obviously the difference between the physical case $\alpha=0$ and the case
preferred for technical reasons, $\alpha=1$, is rather small.     

It is reassuring to see that employing the $\mbox{\boldmath$\beta$}$-functions
of refs. \cite{sven,souma2} 
which are based on the cutoff A we obtain almost identical results. They 
are illustrated by means of  FIG. \ref{fi6} which shows $g_*$, $\lambda_*$, 
and $g_*\lambda_*$ as functions of $s$, $1\le s\le 5$, for $\alpha=0$ 
(FIG. \ref{fi6}(a)) and $\alpha=1$ (FIG. \ref{fi6}(b)). 

\renewcommand{\baselinestretch}{1}
\small\normalsize
\renewcommand{\baselinestretch}{1.5}
\small\normalsize

It is also interesting to compare the above results with those obtained from
a different shape function. FIG. \ref{fi7} displays the behavior of $g_*$,
$\lambda_*$, and their product resulting from the 
$\mbox{\boldmath$\beta$}$-functions (\ref{001}) and (\ref{004}) of the cutoff 
B, with the family of shape functions $R^{(0)}(y;b)$ with compact support, eq.
(\ref{supp}), inserted. Here $b\in[0,1.5)$ parametrizes the profile of these 
shape functions. In FIG. \ref{fi7}(a) we present a
parametric plot of $(\lambda_*(b),g_*(b))$ for $\alpha=0$ starting at $b=0$
and ending at $b=1.5$. Furthermore, FIG. \ref{fi7}(b) shows the product
$g_*\lambda_*$ as a function of $b$ for $\alpha=0$ and $\alpha=1$. For $b
\le1.2$, the parametric plot in FIG. \ref{fi7}(a) exhibits an approximately 
linear behavior which leads to a $g_*\lambda_*$-plateau in FIG. \ref{fi7}(b) 
where $g_*\lambda_*$ is nearly constant. Remarkably, the position of these 
plateaus coincides quite precisely with those of the corresponding
plateaus in FIGS. \ref{fi5} and \ref{fi6} obtained with the other cutoffs. As 
for the quality of the Einstein-Hilbert truncation this result is rather 
encouraging.

For $b>1.2$ the curves in FIG. \ref{fi7} have a rather strong and erratic
$b$-dependence. This is because $R^{(0)}(y;b)$ approaches a sharp
cutoff as $b\rightarrow 1.5$, which introduces discontinuities into the
integrands of the threshold functions $\Phi^p_n$ and $\widetilde{\Phi}_n^p$.
Already for $b\gtrsim 1.2$ the $\mbox{\boldmath$\beta$}$-functions
start to ``feel'' the sharp cutoff limit so that the results cannot be trusted
beyond this point.

As for the $\alpha$-dependence of the fixed point, in FIG. \ref{fi4}(b) we 
present parametric plots of $(\lambda_*(\alpha),g_*(\alpha))$ for various 
fixed values of $s$. Here we used the cutoff type B with the shape function
(\ref{H6}) inserted. These plots start at positions in the $\lambda$-$g$ 
plane which correspond to $\alpha=0$ and which are different for the distinct
$s$-values. As $\alpha\rightarrow\infty$ all curves run into the Gaussian fixed
point.

{\bf To summarize:}\\
a) In 4 dimensions, the Einstein-Hilbert truncation leads to a non-Gaussian
fixed point with positive values of $\lambda_*$ and $g_*$ for all admissible
cutoffs, both of type A and type B. The scheme independence of this prediction
is a nontrivial result. (To emphasize this point we mention that in higher
dimensions, where the Einstein-Hilbert truncation is less reliable, the fixed
point exists or does not exist depending on the cutoff chosen 
\cite{frank}.)\\[12pt]
b) Universal quantities are strictly cutoff independent only in an exact
treatment. Any truncation leads to a scheme dependence of these quantities.
The extent of this scheme dependence is a measure for the reliability of the
truncation. The product $g_*\lambda_*$ is an example of a universal quantity.
While we find a considerable scheme dependence of $g_*$ and $\lambda_*$
separately, their product is scheme independent at a quite amazing level of
accuracy, see FIG. \ref{fi5}. As for the reliability of the Einstein-Hilbert
truncation, we consider this result a highly nontrivial confirmation of our
assumption that the region of parameter space where the fixed point occurs is
well described by this truncation ansatz so that the fixed point also exists
in the {\it exact} theory and is not a truncation artifact.

\subsubsection{Higher and lower dimensions}
\begin{figure}[ht]
\begin{minipage}{7.9cm}
        \epsfxsize=7.9cm
        \epsfysize=5.2cm
        \centerline{\epsffile{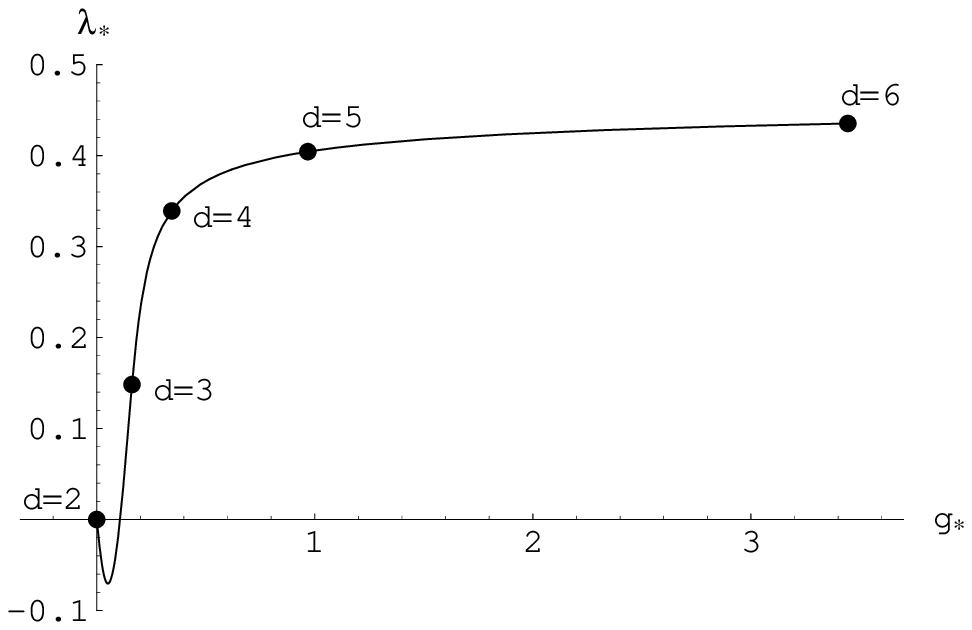}}
\centerline{(a)}
\end{minipage}
\hfill
\begin{minipage}{7.9cm}
        \epsfxsize=7.9cm
        \epsfysize=5.2cm
        \centerline{\epsffile{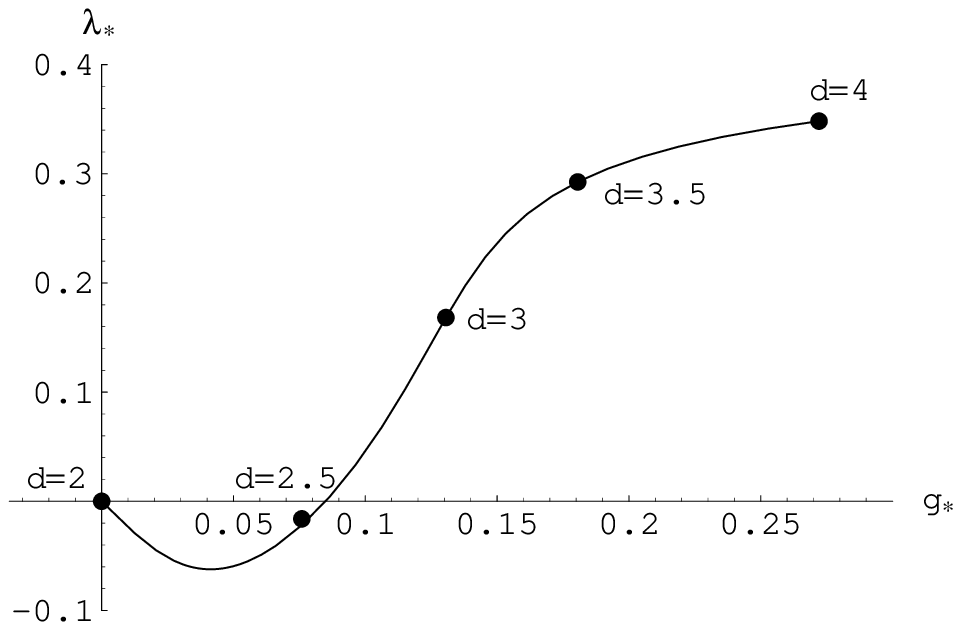}}
\centerline{(b)}
\end{minipage}
\vspace{0.3cm}
\caption{$d$-parametric plot of $(\lambda_*(d),g_*(d))$ for (a) $\alpha=0$, 
$2\le d\le 6$, and (b) $\alpha=1$, $2\le d\le 4$, using the cutoff type B 
with the exponential shape function (\ref{H6}) with $s=1$ inserted.}  
\label{fi8}
\end{figure}

Before continuing our analysis of the $d=4$ dimensional case we study the 
$d$-dependence of the non-Gaussian fixed point. This is done by means of the
parametric plots in FIG. \ref{fi8} which are obtained from the $\mbox{
\boldmath$\beta$}$-functions of type B, eqs. (\ref{001}) and (\ref{004}), with
the shape function (\ref{H6}) with $s=1$ inserted. FIG. \ref{fi8}(b) shows 
$(\lambda_*(d),g_*(d))$ in $2\le d\le 4$ for $\alpha=1$. Remarkably, this plot
is almost identical with that in FIG. 4 of ref. \cite{souma1} which was 
derived from the $\mbox{\boldmath$\beta$}$-functions based on the cutoff 
type A.

\renewcommand{\baselinestretch}{1}
\small\normalsize
\renewcommand{\baselinestretch}{1.5}
\small\normalsize

In order to gain information about the behavior of the fixed point in $d>4$
we also plotted $(\lambda_*(d),g_*(d))$ in $2\le d\le 6$, but this time for
$\alpha=0$, see FIG. \ref{fi8}(a). For $d$ beyond $d\approx 5$, the value of
$g_*$ increases significantly, whereas $\lambda_*$ seems to approach a 
constant value. 

However, this result might be a truncation artifact since it 
is plausible to assume that in higher dimensions the Einstein-Hilbert 
truncation becomes less reliable. This is because with increasing 
dimensionality $d$ the number of terms which are relevant at the non-Gaussian
fixed point and which are neglected in the Einstein-Hilbert truncation 
increases most probably. At the non-Gaussian fixed point the scaling dimensions
of local invariants such as $\int\sqrt{g}$, $\int\sqrt{g}R$, 
$\int\sqrt{g}R^2$, $\int\sqrt{g}R_{\mu\nu}R^{\mu\nu}$, etc. are not known a
priori. We only know that {\it with respect to the Gaussian fixed point} all
local monomials $R,R^2,\cdots$ are relevant or marginal if their canonical
mass dimension does not exceed $d$. A sensible truncation should retain at 
least the relevant terms, whence it is clear that the number of terms needed
increases with the dimensionality $d$. (In 4 dimensions, $\int\sqrt{g}$ and
$\int\sqrt{g}R$ are relevant and the $(\mbox{curvature})^2$-invariants are
marginal.) By analogy we expect that the description of the non-Gaussian fixed
point, too, requires increasingly high powers of the curvature when $d$ is
increased \cite{frank}. 

\subsubsection{The critical exponents (d=4)}
Let us now return to the 4-dimensional case and analyze the critical 
behavior near the non-Gaussian fixed point. In order to get a first impression
of its features we restrict our considerations to the cutoff type B
with the exponential shape function (\ref{H6}) with $s=1$ and to the gauge 
$\alpha=1$. In this case we have $(\lambda_*,g_*)=(0.348,0.272)$, see
above. The corresponding $\bf B$-matrix  assumes the form
\begin{eqnarray}
\label{ngfp1}
{\bf B}=\left(\begin{array}{lr}-0.187 & 5.129\\
-3.228 & -2.907\end{array}\right)\;.
\end{eqnarray}
It leads to a pair of {\it complex} critical exponents 
$\theta_1\equiv\theta'+{\rm i}\theta''$ and $\theta_2=\theta_1^*\equiv\theta'
-{\rm i}\theta''$. For the real quantities $\theta'$ and $\theta''$ we find
$\theta'=1.547$ and $\theta''=3.835$. (In general we define $\theta_1$ as the 
critical exponent with the positive imaginary part so that $\theta''>0$.) 
The behavior of $\lambda_k$ and $g_k$ near the fixed point is described by the
real part of eq. (\ref{gfp5}) in this case. Using that $V^2=(V^1)^*$, and 
setting $V^1\equiv V$ and $C_1\equiv C$, the general solution to the 
linearized flow equation may then be written as
\begin{eqnarray}
\label{ngfp3}
\left(\begin{array}{l}\lambda_k \\ g_k\end{array}\right)
&=&\left(\begin{array}{l}\lambda_* \\ g_*\end{array}\right)
+2\Bigg\{\left[{\rm Re}\,C\,\cos\left(\theta''\,t\right)
+{\rm Im}\,C\,\sin\left(\theta''\,t\right)\right]
{\rm Re}\,V\nonumber\\
& &+\left[{\rm Re}\,C\,\sin\left(\theta''\,t\right)-{\rm Im}\,C
\,\cos\left(\theta''\,t\right)\right]{\rm Im}\,V\Bigg\}e^{-\theta' t}\;.
\end{eqnarray}
Here $t\equiv\ln(k/k_0)$. Obviously the non-Gaussian fixed point is UV (IR) 
attractive if $\theta'\equiv{\rm Re}\,\theta_1={\rm Re}\,\theta_2>0$ $(<0)$. 
The imaginary parts $\pm\theta''$ of the critical exponents do not influence 
the stability of the fixed point. They only give rise to a rotation of the 
vector $(\lambda_k-\lambda_*,g_k-g_*)^{\bf T}$ about the fixed point.

In the case under consideration we have $\theta'>0$ which implies that the 
non-Gaussian fixed point is UV attractive in both directions of 
$(\lambda,g)$-space. All RG trajectories which reach its basin of attraction
spiral into the fixed point for $k\rightarrow\infty$. Thus, 
the Einstein-Hilbert truncation predicts all the ingredients which are 
necessary for the asymptotic safety scenario and the nonperturbative 
renormalizability of 4-dimensional quantum gravity. Clearly the dimensionality
of the UV critical hypersurface cannot be determined within the present
approach. We shall come back to this question in the framework of a more
general truncation including higher derivative terms \cite{LR3}.

\renewcommand{\baselinestretch}{1}
\small\normalsize
\begin{figure}[ht]
\begin{minipage}{7.9cm}
        \epsfxsize=7.9cm
        \epsfysize=5.2cm
        \centerline{\epsffile{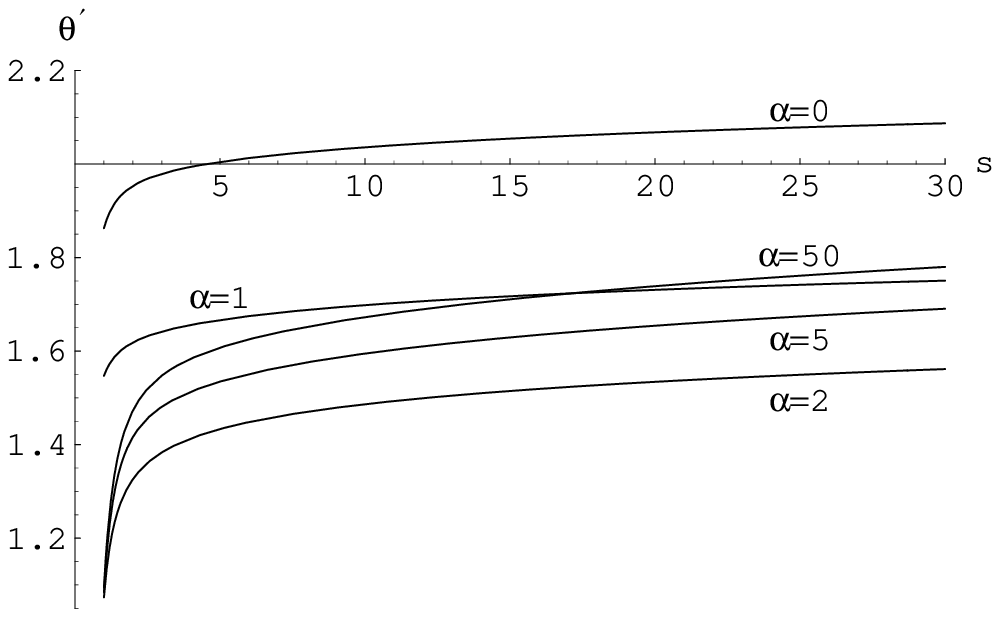}}
\centerline{(a)}
\end{minipage}
\hfill
\begin{minipage}{7.9cm}
        \epsfxsize=7.9cm
        \epsfysize=5.2cm
        \centerline{\epsffile{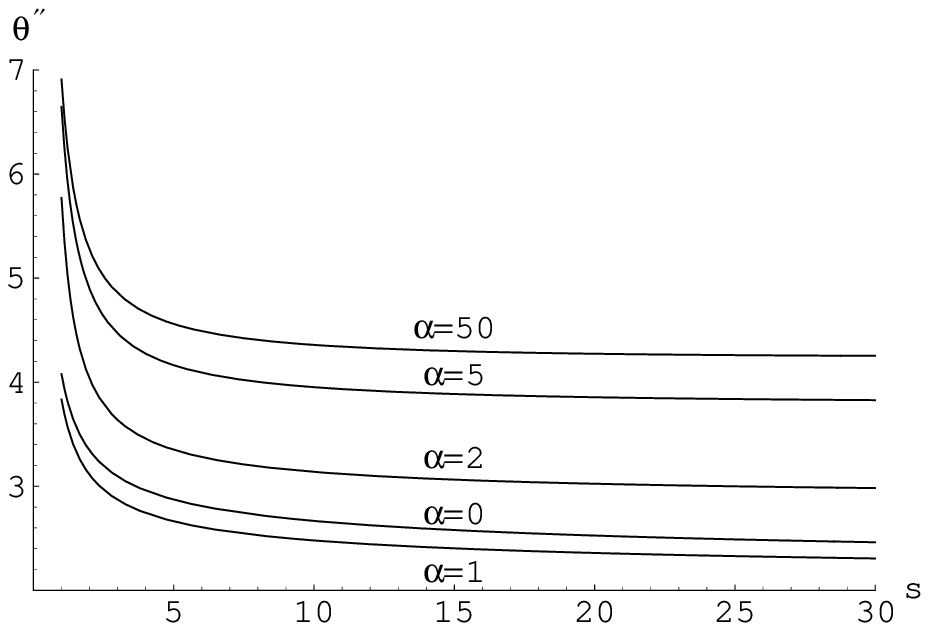}}
\centerline{(b)}
\end{minipage}
\vspace{0.3cm}
\caption{(a) $\theta'={\rm Re}\,\theta_1$, and (b) $\theta''={\rm Im}\,
\theta_1$ as functions of $s$ for various values of $\alpha$, using
the cutoff type B with the family of exponential shape functions (\ref{H6}) 
inserted.}  
\label{fi9}
\end{figure}
\renewcommand{\baselinestretch}{1.5}
\small\normalsize
\renewcommand{\baselinestretch}{1}
\small\normalsize
\begin{figure}[ht]
\begin{minipage}{7.9cm}
        \epsfxsize=7.9cm
        \epsfysize=5.2cm
        \centerline{\epsffile{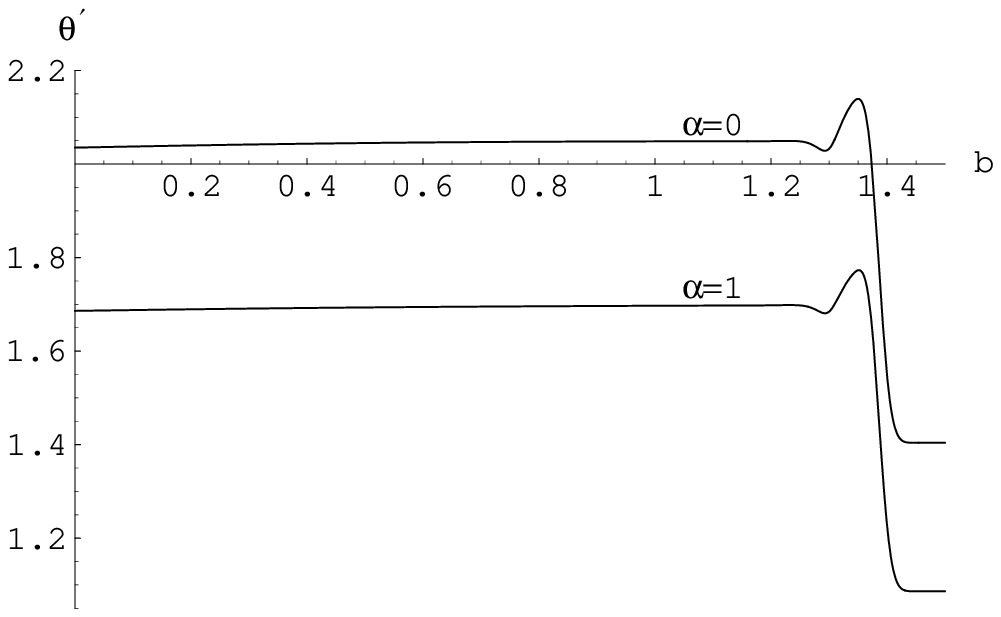}}
\centerline{(a)}
\end{minipage}
\hfill
\begin{minipage}{7.9cm}
        \epsfxsize=7.9cm
        \epsfysize=5.2cm
        \centerline{\epsffile{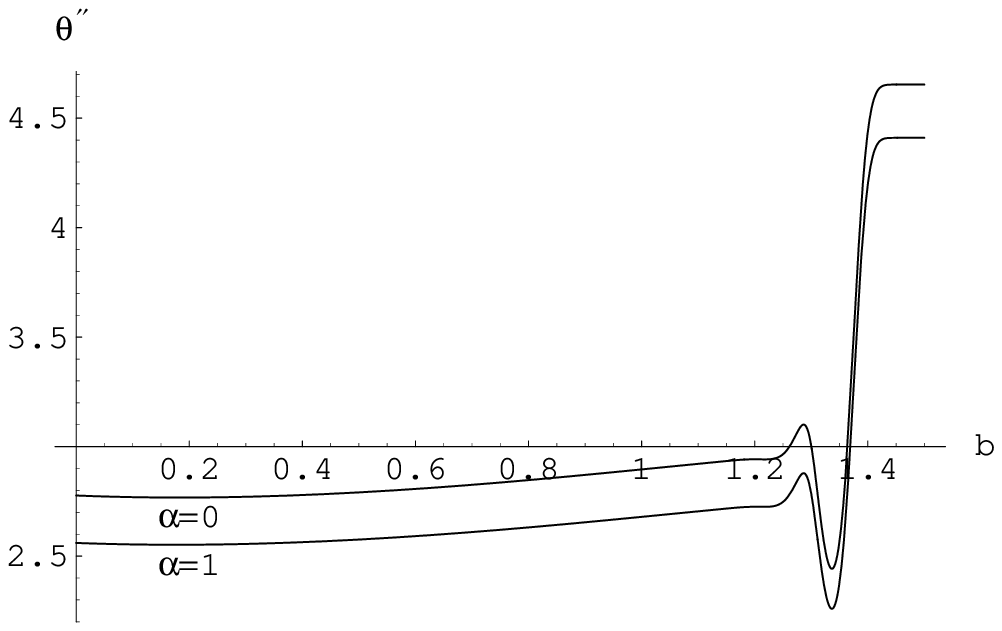}}
\centerline{(b)}
\end{minipage}
\vspace{0.3cm}
\caption{(a) $\theta'={\rm Re}\,\theta_1$, and (b) $\theta''={\rm Im}\,
\theta_1$ as functions of $b$ for $\alpha=0$ and $\alpha=1$, using
the cutoff type B with the family of shape functions with compact support 
(\ref{supp}) inserted.}  
\label{fi10}
\end{figure}
\renewcommand{\baselinestretch}{1.5}
\small\normalsize
\renewcommand{\baselinestretch}{1}
\small\normalsize
\begin{figure}[ht]
\begin{minipage}{7.9cm}
        \epsfxsize=7.9cm
        \epsfysize=5.2cm
        \centerline{\epsffile{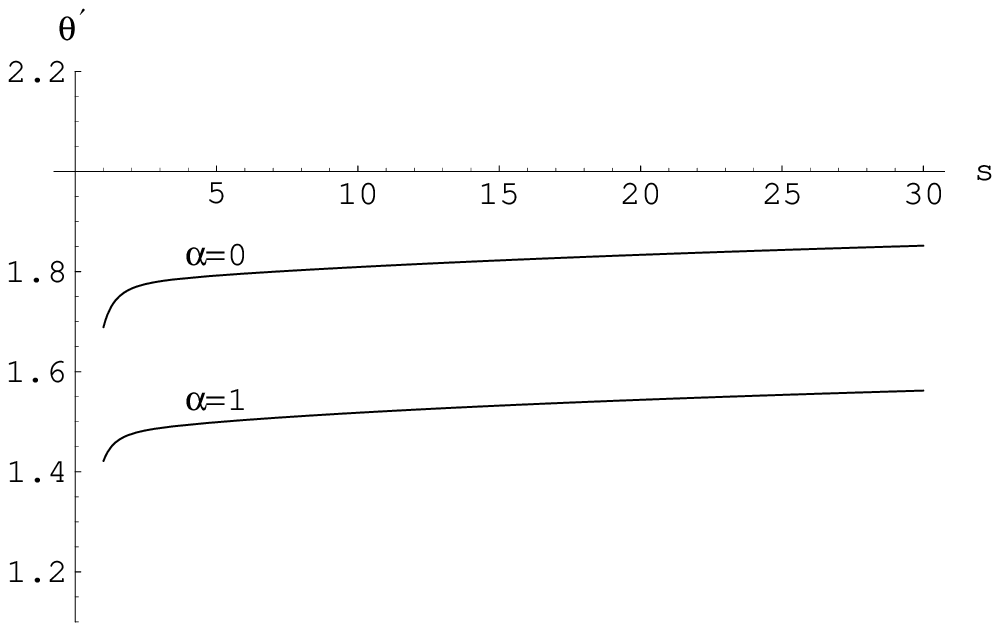}}
\centerline{(a)}
\end{minipage}
\hfill
\begin{minipage}{7.9cm}
        \epsfxsize=7.9cm
        \epsfysize=5.2cm
        \centerline{\epsffile{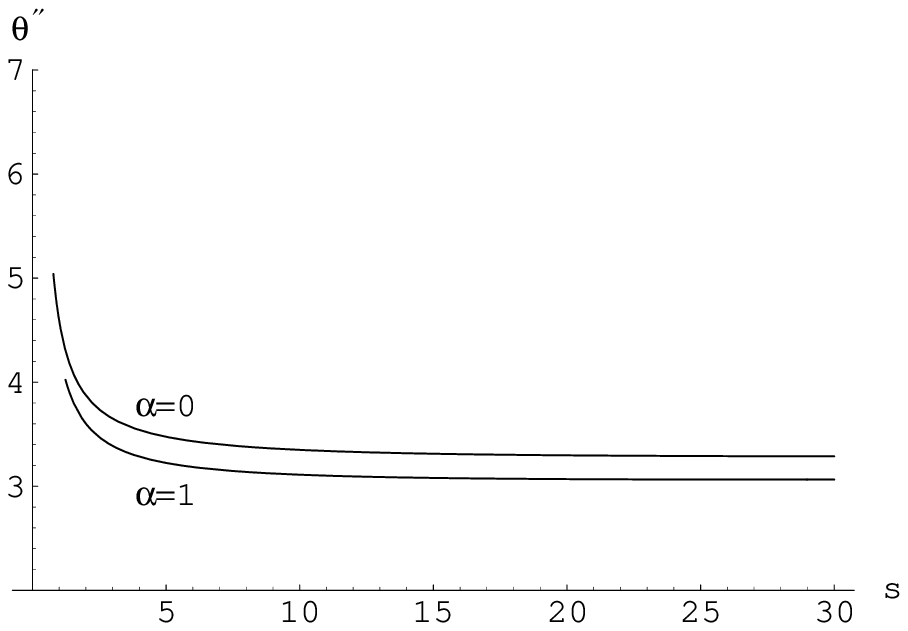}}
\centerline{(b)}
\end{minipage}
\vspace{0.3cm}
\caption{(a) $\theta'={\rm Re}\,\theta_1$, and (b) $\theta''={\rm Im}\,
\theta_1$ as functions of $s$ for $\alpha=0$ and $\alpha=1$, using
the cutoff type A with the family of exponential shape functions (\ref{H6}) 
inserted.}  
\label{fi11}
\end{figure}
\renewcommand{\baselinestretch}{1.5}
\small\normalsize

As discussed in subsection \ref{5A} the critical exponents are universal
in an exact treatment, in contrast to $g_*$ and $\lambda_*$.
However, in a truncated parameter space a scheme dependence is expected to
occur as an artifact of the truncation. Therefore we may use this scheme 
dependence of the critical exponents to judge the quality of our truncation. 
Also in this respect the Einstein-Hilbert
truncation yields satisfactory results, which we display in FIGS. 
\ref{fi9}-\ref{fi11}. First of all it should be noted that the critical
exponents obtained from our numerical analysis have a nonzero imaginary part
in any of the cases considered. FIGS. \ref{fi9}-\ref{fi11}(a) show their real 
part $\theta'$ while their imaginary part $\theta''$ is depicted in FIGS. 
\ref{fi9}-\ref{fi11}(b). FIGS. \ref{fi9} and \ref{fi10} are derived from the
$\mbox{\boldmath$\beta$}$-functions (\ref{001}) and (\ref{004}) based on the
cutoff B of the present paper, with the family of shape functions (\ref{H6}) 
(in FIG. \ref{fi9}) and (\ref{supp}) (in FIG. \ref{fi10}) inserted. 
For comparison, FIG. \ref{fi11} is obtained from the 
$\mbox{\boldmath$\beta$}$-functions of refs. \cite{sven,souma2} derived from
the cutoff type A, with the family of shape 
functions (\ref{H6}) inserted. The figures contain various plots for
distinct values of $\alpha$, which describe the $s$- or $b$-dependence
of $\theta'$ and $\theta''$. 

The $\theta'$-plots in FIGS. \ref{fi9}(a) and \ref{fi11}(a) for the cutoffs
B and A, respectively, exhibit a similar $s$-dependence, and the same holds 
true for the $\theta''$-plots in FIGS. \ref{fi9}(b) and \ref{fi11}(b).
Moreover, in the range $1\le s\le 5$ one recognizes a total
variation of both $\theta'$ and $\theta''$ which is approximately of the  
order of magnitude of $\theta'$ and $\theta''$, respectively. For $s>5$ there
remains only a rather weak dependence on $s$, such that the functions 
$\theta'(s)$ and $\theta''(s)$ develop a plateau-like shape. The 
$s$-dependence is slightly stronger for the cutoff type B than for type A, and
the positions of the ``plateaus'' are different for both cutoff types. For
$\alpha=1$, for instance, we have $\theta'(s=30)\approx 1.56$ and 
$\theta''(s=30)\approx 3.06$ using type A, while employing type B yields  
$\theta'(s=30)\approx 1.75$ and $\theta''(s=30)\approx 2.31$. These differences
have to be interpreted as truncation artifacts.

The above results may now be compared with those for the cutoff B with the 
shape functions with compact support, eq. (\ref{supp}),
inserted. Our results are shown in FIG. \ref{fi10}. Remarkably, $\theta'$ is
almost constant in the range $0\le b\le 1.2$. Furthermore, the 
change in $\theta''$ is only very weak in this region and occurs mainly for 
$b\gtrsim 0.4$. For $b\lesssim 0.4$ $\theta''$ is almost constant as well. 
In the region $1.2<b<1.5$ both $\theta'$ and $\theta''$ strongly vary with 
$b$ which is caused by the sharp cutoff limit as $b\rightarrow 1.5$, see 
above. The position of the plateaus is nearly identical with those obtained 
above by using the same cutoff B, but the shape function (\ref{H6}). For 
$\alpha=1$ we have $\theta'(b=0)\approx 1.69$ and 
$\theta''(b=0)\approx 2.56$.

As discussed above, $\alpha=0$ is assumed to be the physical value of the 
gauge parameter while in practical applications $\alpha=1$ is often preferred.
For a further justification of this approximation we compare the critical 
exponents obtained from 
$\alpha=1$ with those obtained from the (physical) gauge $\alpha=0$.
The results are qualitatively the same, but quantitatively we find a relative 
deviation of about 10 to 20 per-cent. This has to be kept in mind when 
calculations are performed with $\alpha=1$. 

{\bf To summarize:} For all admissible cutoffs, both of type A and type B, the
non-Gaussian fixed point is UV attractive in both directions of parameter
space. It is characterized by a pair of complex conjugate critical exponents
which leads to spiral-type trajectories in its vicinity. The exact critical
exponents are universal. Those obtained from the Einstein-Hilbert truncation
are approximately scheme independent, but their scheme dependence is stronger
than that of $g_*\lambda_*$. This might be related to the fact that the 
$\int\sqrt{g}R^2$-term neglected by the Einstein-Hilbert truncation is (very
weakly) relevant at the non-Gaussian fixed point. We shall come back to this
point in ref. \cite{LR3}.

\section{The graviton propagator at large momenta}
\renewcommand{\theequation}{6.\arabic{equation}}
\setcounter{equation}{0}
\label{S6}
The UV fixed point is characterized by an anomalous dimension $\eta\equiv
\eta_N(g_*,\lambda_*)=-2$. We can use this information in order to determine
the effective momentum dependence of the dressed graviton propagator for 
momenta in the fixed point regime, i.e. for $p^2\gg m_{\rm Pl}^2$. Expanding 
the truncated $\Gamma_k$ about flat space and omitting the standard tensor 
structures we find the inverse propagator $\widetilde{\cal G}_k(p)^{-1}\propto
Z_N(k)p^2$. The conventional dressed propagator $\widetilde{\cal G}(p)$ 
contained in $\Gamma\equiv\Gamma_{k=0}$ obtains from $\widetilde{\cal G}_k$ in
the limit $k\rightarrow 0$. Assuming that the actual physical cutoff scale is
the momentum $p^2$ itself (which is indeed true for $p^2>k^2\gg m_{\rm Pl}^2$),
the $k$-evolution of $\widetilde{\cal G}_k(p)$ stops at the threshold $k=
\sqrt{p^2}$. Therefore
\begin{eqnarray}
\label{gp1}
\widetilde{\cal G}(p)^{-1}\propto\;Z_N\left(k=\sqrt{p^2}\right)\,p^2\propto\; 
(p^2)^{1-\frac{\eta}{2}}
\end{eqnarray}
because $Z_N(k)\propto k^{-\eta}$ when $\eta\equiv-\partial_t\ln Z_N$ is 
(approximately) constant. In $d$ dimensions, and for $\eta\neq 2-d$, the
Fourier transform of $\widetilde{\cal G}(p)\propto 1/(p^2)^{1-\eta/2}$ yields 
the following Euclidean propagator in position space:
\begin{eqnarray}
\label{gp2}
{\cal G}(x;y)\propto\;\frac{1}{\left|x-y\right|^{d-2+\eta}}\;.
\end{eqnarray}
This is a standard result well known from the theory of critical phenomena, for
instance. In the latter case it applies to large distances, while in quantum
gravity we are interested in the extreme short distance regime governed by the
UV fixed point. However, as it stands eq. (\ref{gp2}) is not valid for the case
of interest, $d=4$ and $\eta=-2$. For $\eta=-2$ the dressed propagator is 
$\widetilde{\cal G}(p)=1/p^4$ which, in $d=4$ dimensions, has the following
representation in position space:
\begin{eqnarray}
\label{gp3}
{\cal G}(x;y)=-\frac{1}{8\pi^2}\,\ln\left(\mu\left|x-y\right|\right)\;.
\end{eqnarray}
Here $\mu$ is an arbitrary constant with the dimension of a mass. Usually in
ordinary matter field theories on flat space a $1/p^4$-propagator is considered
problematic because it is incompatible with a positive spectral density. It is
by no means clear, however, that the familiar notions of positivity, 
analyticity, and causality which are appropriate for Minkowski space are of 
relevance to the situation under consideration. For $k\rightarrow\infty$ the
``on shell'' spacetimes of $\Gamma_k$ have a large curvature $\propto k^2$,
after all. 

So let us assume that the improvement (\ref{gp1}) is indeed correct
and that in the fixed point regime, i.e. for distances much smaller than the 
Planck length, the effective graviton propagator has a logarithmic dependence 
on the distance. This result is quite remarkable because it implies a kind of 
dimensional
reduction from 4 to 2 dimensions. In fact, eq. (\ref{gp3}) is precisely what 
one obtains from a standard $1/p^2$-propagator in $d=2$ dimensions. This means
that, in a certain sense, spacetime appears to be two-dimensional when it is
probed by a very high energetic graviton.

Since, symbolically, $R=\partial\partial h$, the propagator (\ref{gp3}) yields
the curvature-curvature correlation function
\begin{eqnarray}
\label{gp4}
\left<{\bf R}(x)\,{\bf R}(y)\right>\propto\frac{1}{\left(x-y\right)^4}\;.
\end{eqnarray} 
Its short-distance singularity has to be contrasted with the $1/(x-y)^6$
behavior one finds at tree level. Here ${\bf R}$ stands for the curvature
scalar or for any component of the Riemann or Ricci tensor. 

Switching for a moment to spacetimes with a Lorentzian signature, it is 
interesting to look at the linearized gravitational field produced very close 
to a {\it static} source. Decomposing $x\equiv(x^0,{\bf x})$, the relevant
Green's function for static problems reads
\begin{eqnarray}
\label{gp5}
{\cal G}_{\rm stat}({\bf x};{\bf y})\equiv\int\limits_{-\infty}^\infty dy^0\,
{\cal G}\left(x^0,{\bf x};y^0,{\bf y}\right)\;.
\end{eqnarray} 
In our case this is the 3-dimensional Fourier transform of $1/
|{\bf p}|^4$, i.e.
\begin{eqnarray}
\label{gp6}
{\cal G}_{\rm stat}({\bf x};{\bf y})=\left<{\bf x}\left|\left(\nabla^2\nabla^2
\right)^{-1}\right|{\bf y}\right>=-\frac{1}{8\pi}\left|{\bf x}-{\bf y}\right|
\end{eqnarray}
provided $\left|{\bf x}-{\bf y}\right|\ll m_{\rm Pl}^{-1}$. In an, admittedly
somewhat naive, Newtonian language this result would mean that a point mass
located at ${\bf y}=0$ creates a gravitational potential which behaves as
$\Phi({\bf x})\propto |{\bf x}|$ as long as $|{\bf x}|$ is much smaller than 
the Planck length. Probably this linear potential is related to a 
similar phenomenon shown by the renormalization group improved Schwarzschild
black hole which has been constructed recently \cite{bh2}. The radial 
geodesics in this spacetime, in a fully relativistic treatment, experience a
linear\renewcommand{\baselinestretch}{1}\small\normalsize\footnote{In the 
notation of ref. \cite{bh2} the potential is linear if $\gamma=0$. This 
corresponds to the case $k\propto 1/r$ which ignores the 
impact the background curvature has on the cutoff identification, see eq. 
(4.35) of \cite{bh2}. This is consistent with the fact that eq. (\ref{gp3}) is
valid for a flat background.} 
\renewcommand{\baselinestretch}{1.5}\small\normalsize
repulsive ``potential'' close to the core of the
black hole.
\section{Summary and conclusion}
\renewcommand{\theequation}{7.\arabic{equation}}
\setcounter{equation}{0}
\label{S7}
On the formal side, the main result of this paper is the construction of a new
exact RG equation for the gravitational effective average action. It is 
formulated in terms of the component fields appearing in the TT-decomposition
of the metric. It is defined on a sufficiently large class of background 
spacetimes so as to facilitate the projection of the RG flow onto very general
truncated parameter spaces. It also helps in finding admissible IR cutoffs.
A formalism of this kind is mandatory for truncations including
higher-derivative invariants, matter fields, or a running gauge fixing, for
instance. In a forthcoming paper \cite{LR3} we shall use this technology in
order to explore a more general truncation with a $R^2$-term. In the present
paper we analyzed the Einstein-Hilbert truncation with the new equation and a
new cutoff.

After deriving the exact functional equation in section \ref{S2}, we discussed
in section \ref{S3} a general strategy for constructing an appropriate cutoff
operator in the context of a class of truncations which is still very general.
In section \ref{S4} we specialized for the Einstein-Hilbert truncation and 
derived the nonperturbative $\mbox{\boldmath$\beta$}$-functions which govern
the RG evolution of $g_k$ and $\lambda_k$.

On the applied side, most of our results concern the non-Gaussian fixed point 
of the $\lambda$-$g$ system in 4 dimensions which we analyzed in section 
\ref{S5}. If this fixed point is actually present in the exact theory, its
importance can hardly be overestimated. Its existence would imply that in spite
of its notorious perturbative nonrenormalizability quantum Einstein gravity is
most probably renormalizable at the nonperturbative level and thus qualifies
as a fundamental (microscopic) quantum theory of gravity. Clearly the crucial
question is whether the fixed point of the Einstein-Hilbert truncation is 
indeed genuine or merely a truncation artifact.

In order to get an impression of the reliability of the Einstein-Hilbert 
truncation we investigated how its predictions change when we vary the cutoff
which is built into the RG equations. We used both the original cutoff of 
type A, formulated in terms of $h_{\mu\nu}$, and the new cutoff of type B 
which is natural in the TT-approach; the cutoffs were equipped with two 
different one-parameter families of shape functions.

In 4 dimensions, we found that the Einstein-Hilbert truncation leads to a
non-Gaussian fixed point for all admissible cutoffs, both of type A and B. The
robustness of this prediction is a nontrivial result since in higher 
dimensions, for instance, where this truncation is less reliable, the fixed
point is present for some cutoffs but absent for others. 

Another consistency test successfully passed by the truncation is that all
cutoffs agree on {\it positive} values of $g_*$ and $\lambda_*$. A negative
$g_*$ would probably be unacceptable for stability reasons, but there is no
mechanism in the flow equation which would exclude it on general grounds. In
fact, $g_*<0$ is realized for $d<2$.

For all cutoffs of type A and B the non-Gaussian fixed point is found to be
UV attractive in both directions of the $\lambda$-$g$ plane. The linearized
flow in its vicinity is always characterized by a pair of complex conjugate
critical exponents leading to spiral-type trajectories which hit the fixed
point for $k\rightarrow\infty$. This is precisely the stability property 
needed for asymptotic safety.

By definition, universal quantities are scheme-, or cutoff-independent in an
exact calculation. Truncations lead to a scheme dependence, however. We can
use the degree of the scheme dependence as a measure for the reliability
of the truncation. The critical exponents and, as we argued, the product 
$g_*\lambda_*$ are universal quantities. The existence of fixed points is a
universal feature of the RG flow, but not their precise location in parameter
space.

The critical exponents were indeed found to be reasonably constant for a wide
range of shape parameters. The universality properties of $g_*\lambda_*$ are
much more impressive though. While we found a considerable scheme dependence
of $g_*$ and $\lambda_*$ separately, their product is scheme independent at a
rather amazing level of accuracy.

We believe that these results hardly can be a mathematical accident, and we
consider them a very nontrivial confirmation of the hypothesis that the region
of parameter space where the non-Gaussian fixed point is situated is well 
described by the Einstein-Hilbert truncation. As a consequence, the fixed 
point should exist in the exact theory, too. 

Apart from the renormalizability issue the nontrivial fixed point is also
intriguing from a ``phenomenological'' point of view. Its relevance for the
structure of black holes \cite{bh2} and the cosmology of the Planck era
\cite{cosmo} has been pointed out already. Moreover, we saw in section \ref{S6}
that the RG improvement of the graviton propagator suggests a kind of 
dimensional reduction from 4 to 2 dimensions when spacetime is probed at
sub-Planckian length scales.\\[24pt]
Acknowledgement: We would like to thank C. Wetterich for many helpful 
discussions.
\newpage
\begin{appendix}
\section{The TT-decomposition}
\renewcommand{\theequation}{A\arabic{equation}}
\setcounter{equation}{0}
\label{dec}
\subsection{Pseudo-projectors for the TT-decomposition}
In subsection \ref{2B} we introduced the TT-decomposition
\begin{eqnarray}
\label{Ap1}
h_{\mu\nu}=h_{\mu\nu}^T+h_{\mu\nu}^L+h_{\mu\nu}^{Tr}
\end{eqnarray}
valid for arbitrary symmetric rank 2 tensors defined on either closed or open,
asymptotically flat Riemannian $d$-spaces. Here $h^T_{\mu\nu}$, $h^L_{\mu\nu}$
and $h^{Tr}_{\mu\nu}$ represent the transverse traceless, longitudinal 
traceless and pure trace part, respectively, which are mutually orthogonal. 
According to eq. (\ref{f}) these parts may be expressed in terms of pure 
spin-2, spin-1 and spin-0 component fields $h^T_{\mu\nu}$, 
$\widehat{\xi}_\mu$, $\widehat{\sigma}$, $\phi$. In this subsection we show 
that the component fields can be obtained by applying certain operators ${\bf
\Pi}$ to the full field $h_{\mu\nu}$. In the following 
these operators will be termed pseudo-projectors.

As a first step we express the longitudinal-transverse and the pure trace part
as 
\begin{eqnarray}
\label{Ap2}
h_{\mu\nu}^L&=&(L\varepsilon)_{\mu\nu}\equiv\bar{D}_\mu\varepsilon_\nu
+\bar{D}_\nu\varepsilon_\mu-\frac{2}{d}\bar{g}_{\mu\nu}\bar{D}_\lambda
\varepsilon^\lambda\;,\nonumber\\
h_{\mu\nu}^{Tr}&=&\frac{1}{d}\bar{g}_{\mu\nu}\phi\;,\;\;\phi\equiv 
\bar{g}^{\mu\nu}h_{\mu\nu}^{Tr}\;.
\end{eqnarray}
Here the operator $L$ maps vectors onto longitudinal traceless tensors.
Given a tensor $h^L_{\mu\nu}$, the equation $(L\varepsilon)_{\mu\nu}
=h^L_{\mu\nu}$ is solved by
\begin{eqnarray}
\label{099}
\varepsilon_\mu\equiv\widehat{\xi}_\mu+\frac{1}{2}\bar{D}_\mu\widehat{\sigma}
\end{eqnarray}
where $\widehat{\xi}_\mu$ is a transverse vector and $\widehat{\sigma}$ a 
scalar. This solution is unique up to the addition of conformal Killing 
vectors (CKV's), as discussed in subsection \ref{2B}. We recover eq. (\ref{f})
by inserting eqs. (\ref{Ap2}), (\ref{099}) into eq. (\ref{Ap1}). Contrary to
$\varepsilon_\mu$, the scalar $\phi$ is uniquely determined by 
$h_{\mu\nu}$.

Now taking the covariant divergence of eq. (\ref{Ap1}) with eq. (\ref{Ap2}) 
inserted, and using the transversality requirement $\bar{D}^\mu h_{\mu\nu}^T
=0$ leads to
\begin{eqnarray}
\label{Ap3}
({\cal D}\varepsilon)_\mu=-\bar{D}^\nu\left(h_{\mu\nu}-\frac{1}{d}
\bar{g}_{\mu\nu}\phi\right)
\end{eqnarray}
with the operator ${\cal D}$ defined by
\begin{eqnarray}
\label{Ap4}
({\cal D}\varepsilon)_\mu\equiv -\bar{D}^\nu(L\varepsilon)_{\mu\nu}\;.
\end{eqnarray}
As shown in ref. \cite{York}, ${\cal D}$ is a positive definite, Hermitian 
operator mapping vectors onto vectors. Moreover, the equation $({\cal D}
\varepsilon)_\mu=u_\mu$ with an arbitrary given vector $u_\mu$ always 
possesses solutions $\varepsilon_\mu$ which are unique up to CKV's. However, 
even if these CKV's exist they cause no problems in solving eq. (\ref{Ap3}) 
for $\varepsilon_\mu$, see ref. \cite{York} for details. In order to determine
this solution we have to invert ${\cal D}$. For this purpose we assume that
${\cal D}$ (and any additional operator that needs to be inverted in the 
course of this discussion) has a complete set of orthogonal eigenfunctions 
and that the corresponding eigenvalues do not have zero as an accumulation 
point. Then ${\cal D}^{-1}$ exists and the solution to eq. (\ref{Ap3}) may be 
written as
\begin{eqnarray}
\label{Ap9}
\varepsilon_\mu=({\cal D}^{-1}\widetilde{\delta}h)_\mu\;.
\end{eqnarray}
Here the operator $\widetilde{\delta}$ maps tensors onto vectors according to
\begin{eqnarray}
\label{Ap8}
(\widetilde{\delta}h)_\mu=-\bar{D}^\nu{\Lambda_{\mu\nu}}^{\alpha\beta}
h_{\alpha\beta}
\end{eqnarray}
with
\begin{eqnarray}
\label{Ap5}
{\Lambda_{\mu\nu}}^{\alpha\beta}\equiv\frac{1}{2}\left(\delta^\alpha_\mu
\delta^\beta_\nu+\delta^\beta_\mu\delta^\alpha_\nu\right)-\frac{1}{d}
\bar{g}_{\mu\nu}\bar{g}^{\alpha\beta}
\end{eqnarray}
being the operator that projects symmetric tensors onto their traceless part:
\begin{eqnarray}
\label{Ap6}
{\Lambda_{\mu\nu}}^{\alpha\beta}h_{\alpha\beta}=h_{\mu\nu}-\frac{1}{d}
\bar{g}_{\mu\nu}\phi\;.
\end{eqnarray}
Taking the covariant divergence of the solution (\ref{Ap9}), inserting eq. 
(\ref{099}), and using $\bar{D}_\mu\widehat{\xi}^\mu=0$ yields 
\begin{eqnarray}
\label{Ap10}
\bar{D}^\mu\varepsilon_\mu=\frac{1}{2}\bar{D}^2\widehat{\sigma}=\bar{D}^\mu
({\cal D}^{-1}\widetilde{\delta}h)_\mu\;.
\end{eqnarray}
This leads to the final result for $\widehat{\xi}_\mu$ and $\widehat{\sigma}$
in terms of $h_{\mu\nu}$:
\begin{eqnarray}
\label{Ap11}
\widehat{\sigma}&=&2(\bar{D}^2)^{-1}\bar{D}^\mu({\cal D}^{-1}
\widetilde{\delta}h)_\mu\equiv \Xi h\;,\nonumber\\
\widehat{\xi}_\mu &=&({\cal D}^{-1}\widetilde{\delta}h)_\mu-\bar{D}_\mu
(\bar{D}^2)^{-1}\bar{D}^\nu({\cal D}^{-1}\widetilde{\delta}h)_\nu\equiv 
(\Omega h)_\mu\;.
\end{eqnarray}
Hence the pseudo-projectors $\bf \Pi$ which map $h_{\mu\nu}$ onto
the individual component fields are obtained as
\begin{eqnarray}
\label{Ap13}
h_{\mu\nu}^T&=&({\bf \Pi}_{TT}h)_{\mu\nu}\equiv(\Lambda h)_{\mu\nu}-
(L{\cal D}^{-1}\widetilde{\delta}h)_{\mu\nu}\;,\nonumber\\
\xi_\mu&=&({\bf \Pi}_{LT} h)_{\mu}\equiv\left(\sqrt{-\bar{D}^2
-\overline{\rm Ric}}\;\Omega h\right)_\mu\;,\nonumber\\
\sigma&=&{\bf \Pi}_{LL} h\equiv\sqrt{(\bar{D}^2)^2+\frac{d}{d-1}\bar{D}_\mu
\bar{R}^{\mu\nu}\bar{D}_\nu}\;\Xi h\;,\nonumber\\
\phi&=&{\bf \Pi}_{Tr} h\equiv \bar{g}^{\mu\nu}h_{\mu\nu}\;.
\end{eqnarray}
Here we defined ${\bf\Pi}_{LT}$ and ${\bf\Pi}_{LL}$ in terms of the redefined
fields $\xi_\mu$ and $\sigma$ which are related to $\widehat{\xi}_\mu$ and
$\widehat{\sigma}$ by eq. (\ref{y}).

Furthermore, the pseudo-projectors for the transverse decomposition of 
arbitrary vector fields can be inferred from eq. (\ref{Ap11}). For 
$C^\mu=C^{T\mu}+\bar{D}^\mu(-\bar{D}^2)^{-\frac{1}{2}}\eta$ with $\bar{D}_\mu
C^{T\mu}=0$ they are determined by
\begin{eqnarray}
\label{Ap7}
\eta&=&{\bf\Pi}_L C\equiv(-\bar{D}^2)^{-\frac{1}{2}}\bar{D}_\mu C^\mu
\;,\nonumber\\
C^{T\mu}&=&\left({\bf\Pi}_TC\right)^\mu\equiv C^\mu-\bar{D}^\mu
(-\bar{D}^2)^{-1}\bar{D}_\nu C^\nu\;.
\end{eqnarray}

Obviously ${\bf \Pi}_{LT}$ maps tensors onto vectors, ${\bf \Pi}_{LL}$ and 
${\bf\Pi}_{Tr}$ map tensors onto scalars, and ${\bf\Pi}_{L}$ maps vectors onto 
scalars. Hence these operators cannot be projection operators in the usual 
sense. However, projection operators ${\bf P}$ mapping arbitrary 
$h_{\mu\nu}$ onto $h^L_{\mu\nu}$ or $h^{Tr}_{\mu\nu}$, or arbitrary 
$\varepsilon_\mu$ onto their longitudinal component can be constructed from 
them \cite{York}. Since the ${\bf\Pi}$'s map vectors and symmetric tensors 
onto their component fields they 
generate a kind of projection in a wider sense of the word. Therefore we call
the ${\bf\Pi}$'s pseudo-projectors. Contrary to ${\bf \Pi}_{LT}$,
${\bf \Pi}_{LL}$, ${\bf\Pi}_{Tr}$ and ${\bf\Pi}_{L}$, the operators 
${\bf \Pi}_{TT}$ and ${\bf\Pi}_{T}$ are genuine projection operators mapping
symmetric tensors and vectors onto their $ST^2$ and $T$ component, 
respectively.
\subsection{Construction of the cutoff and the source terms}
In the present paper we need a formulation for $\Gamma_k$ which allows for a
description in terms of the fundamental as well as the component fields. The
translation between the two descriptions 
can be achieved by using the pseudo-projectors for the construction of the
cutoff and for an appropriate decomposition of the source terms.

Starting from the definition of the cutoff in terms of the fundamental fields,
eq. (\ref{v}), we choose the cutoff operators ${\cal R}_k^{\rm grav}$ and
${\cal R}_k^{\rm gh}$ as 
\begin{eqnarray}
\label{Ap15}
{\cal R}_k^{\rm grav}&=&\sum\limits_{\zeta_1,\zeta_2\in\{h^T,\xi,\sigma,\phi\}}
{\bf \Pi}_{\zeta_1}^\dagger\,({\cal R}_k)_{\zeta_1\zeta_2}\,
{\bf \Pi}_{\zeta_2}\;,\nonumber\\
{\cal R}_k^{\rm gh}&=&\sum\limits_{\bar{\vartheta}_1\in\{\bar{C}^T,
\bar{\eta}\}}\;\sum\limits_{\vartheta_2\in\{C^T,\eta\}}
{\bf \Pi}_{\bar{\vartheta}_1}^\dagger\,({\cal R}_k)_{\bar{\vartheta}_1
\vartheta_2}\,{\bf \Pi}_{\vartheta_2}\;.
\end{eqnarray}
Here $\left(({\cal R}_k)_{\bar{\vartheta}_1\vartheta_2}\right)_{
\bar{\vartheta}_1\in\{\bar{C}^T,
\bar{\eta}\},\vartheta_2\in\{C^T,\eta\}}$ represents a block of the more 
general matrix operator $\Big(({\cal R}_k)_{\psi_1\psi_2}\Big)_{\psi_1,\psi_2
\in\{\bar{C}^T,\bar{\eta},C^T,\eta\}}$ for which $({\cal R}_k)_{\psi_1
\psi_2}\equiv 0$ if both $\psi_1\in\{C^T,\eta\}$ and $\psi_2\in\{C^T,\eta\}$,
or if both $\psi_1\in\{\bar{C}^T,\bar{\eta}\}$ and $\psi_2\in\{\bar{C}^T,
\bar{\eta}\}$. The operators $({\cal R}_k)_{
\zeta_1\zeta_2}$ and $({\cal R}_k)_{\psi_1\psi_2}$ are required to satisfy the 
Hermiticity conditions $\left({\cal R}_k\right)_{\zeta_2\zeta_1}
=\left({\cal R}_k\right)_{\zeta_1\zeta_2}^\dagger$ and $\left({\cal R}_k
\right)_{\psi_2\psi_1}=-\left({\cal R}_k
\right)_{\psi_1\psi_2}^\dagger$. Furthermore,
we set ${\bf\Pi}_{h^T}\equiv{\bf\Pi}_{TT}$, ${\bf\Pi}_{\xi}\equiv
{\bf\Pi}_{LT}$, ${\bf\Pi}_{\sigma}\equiv{\bf\Pi}_{LL}$, ${\bf\Pi}_{\phi}
\equiv{\bf\Pi}_{Tr}$, ${\bf \Pi}_{\bar{C}^T}\equiv{\bf \Pi}_{C^T}\equiv
{\bf\Pi}_T$, ${\bf \Pi}_{\bar{\eta}}\equiv{\bf \Pi}_{\eta}\equiv{\bf\Pi}_L$, 
and ${\bf\Pi}_\zeta^\dagger$, ${\bf\Pi}_\psi^\dagger$
denote the Hermitian conjugates of ${\bf\Pi}_\zeta$, ${\bf\Pi}_\psi$. Hence 
the operators ${\bf\Pi}$ appearing in eq. (\ref{Ap15}) can be inferred from 
eqs. (\ref{Ap13}), (\ref{Ap7}). 
 
Inserting eq. (\ref{Ap15}) into eq. (\ref{v}) leads to
\begin{eqnarray}
\label{Ap17}
\Delta_k S[h,C,\bar{C};\bar{g}]&=&\frac{1}{2}\sum\limits_{\zeta_1,\zeta_2
\in\{h^T,\xi,\sigma,\phi\}}\left<{\bf\Pi}_{\zeta_1} h,({\cal R}_k)_{\zeta_1
\zeta_2}\,{\bf\Pi}_{\zeta_2}\,h\right>\nonumber\\
& &+\sum\limits_{\bar{\vartheta}_1\in\{\bar{C}^T,\bar{\eta}\}}\;
\sum\limits_{\vartheta_2\in\{C^T,\eta\}}\left<{\bf\Pi}_{\bar{\vartheta}_1}
\bar{C},({\cal R}_k)_{\bar{\vartheta}_1\vartheta_2}\,{\bf\Pi}_{\vartheta_2}\,C
\right>
\end{eqnarray}
which can be rewritten in terms of the component fields. In fact, applying eqs.
(\ref{Ap13}) and (\ref{Ap7}) and using the relation between the matrix 
operators $\left({\cal R}_k\right)_{\bar{\vartheta}_1\vartheta_2}$ and 
$\left({\cal R}_k\right)_{\psi_1\psi_2}$ stated above, we end up with eq. 
(\ref{013}).

Let us now consider the source terms in eq. (\ref{hh}). Decomposing the sources
for the ghost fields according to
\begin{eqnarray}
\label{Ap12}
K^\mu=K^\mu_{\bar{C}^T}+\bar{D}^\mu(-\bar{D}^2)^{-\frac{1}{2}}K_{\bar{\eta}}\;,
\;\;\bar{K}^\mu=\bar{K}^\mu_{C^T}+\bar{D}^\mu(-\bar{D}^2)^{-\frac{1}{2}}
\bar{K}_\eta
\end{eqnarray}
with $\bar{D}_\mu K_{\bar{C}^T}^\mu=\bar{D}_\mu\bar{K}_{C^T}^\mu=0$ yields
\begin{eqnarray}
\label{Ap18}
\left<\bar{K},C\right>+\left<K,\bar{C}\right>
=\sum\limits_{\psi\in\{C^T,\eta\}}\left<\bar{K}_{\psi},\psi\right>
+\sum\limits_{\psi\in\{\bar{C}^T,\bar{\eta}\}}\left<K_{\psi},\psi\right>\;.
\end{eqnarray}

The decomposition of $\left<J,h\right>$ is more involved. In analogy with 
$h_{\mu\nu}$ in eq. (\ref{Ap1}) we decompose the source $J_{\mu\nu}$ into its
orthogonal parts:
\begin{eqnarray}
\label{Ap24}
J_{\mu\nu}=J^T_{\mu\nu}+J^L_{\mu\nu}+J^{\rm Tr}_{\mu\nu}\;.
\end{eqnarray}
According to eqs. (\ref{Ap2}), (\ref{099}) we write $J^L_{\mu\nu}$ as
\begin{eqnarray}
\label{Ap20}
J^L_{\mu\nu}=J^{LT}_{\mu\nu}+J^{LL}_{\mu\nu}
\end{eqnarray}
with
\begin{eqnarray}
\label{Ap21}
J^{LT}_{\mu\nu}=(L\Theta)_{\mu\nu}=\bar{D}_\mu\Theta_\nu+\bar{D}_\nu\Theta_\mu
\;,\;\;
J^{LL}_{\mu\nu}=\frac{1}{2}(L\bar{D}\upsilon)_{\mu\nu}=\bar{D}_\mu\bar{D}_\nu
\upsilon-\frac{1}{d}\bar{g}_{\mu\nu}\bar{D}^2\upsilon\;.
\end{eqnarray}
Analogous to eq. (\ref{Ap2}), the vector $\Theta_\mu+\bar{D}_\mu\upsilon/2$ 
with $\bar{D}_\mu\Theta^\mu=0$ is unique up to CKV's.

According to eq. (\ref{08}) the inner product $\left<J,h\right>$ may now be 
decomposed as follows:
\begin{eqnarray}
\label{new}
\left<J,h\right>&=&\left<J^T,h^T\right>+\left<J^{L},h^{LT}\right>+\left<J^{L},
h^{LL}\right>+\left<J^{Tr},h^{Tr}\right>\;.
\end{eqnarray}
It is important to note that, for arbitrary nonvanishing $h^T_{\mu\nu}$,
$h^{LT}_{\mu\nu}$, $h^{LL}_{\mu\nu}$, $h^{Tr}_{\mu\nu}$, we have
\begin{eqnarray}
\label{Ap22}
\left({\bf\Pi}_{TT}\,h^T\right)_{\mu\nu}\neq 0\;,\;\;
\left({\bf \Pi}_{LT}\,h^{LT}\right)_\mu\neq 0\;,\;\;
{\bf \Pi}_{LL}\,h^{LL}\neq 0\;,\;\;{\bf \Pi}_{Tr}\,h^{Tr}\neq 0\;.
\end{eqnarray}
This implies that the operators ${\bf\Pi}_{TT}$, ${\bf\Pi}_{LT}$, 
${\bf\Pi}_{LL}$, ${\bf\Pi}_{Tr}$ are invertible if their action is restricted 
to fields of the type $h^T_{\mu\nu}$, $h^{LT}_{\mu\nu}$, $h^{LL}_{\mu\nu}$, 
$h^{Tr}_{\mu\nu}$, respectively. Therefore the inner product of eq. 
(\ref{new}) may be written as  
\begin{eqnarray}
\label{Ap19}
\left<J,h\right>&=&\left<J^T,{\bf \Pi}_{TT}^{-1}{\bf \Pi}_{TT}\,h^T
\right>+\left<J^{L},{\bf \Pi}_{LT}^{-1}{\bf \Pi}_{LT}\,h^{LT}\right>
+\left<J^{L},{\bf \Pi}_{LL}^{-1}{\bf \Pi}_{LL}\,h^{LL}\right>\nonumber\\
& &+\left<J^{Tr},{\bf \Pi}_{Tr}^{-1}{\bf \Pi}_{Tr}\,h^{Tr}\right>
\nonumber\\
&=&\sum\limits_{\zeta\in\{h^T,\xi,\sigma,\phi\}}\left<J_\zeta,\zeta\right>\;.
\end{eqnarray}
Here we used eq. (\ref{Ap13}) and introduced the source components 
\begin{eqnarray}
\label{Ap23}
J_{h^T}^{\mu\nu}&\equiv&\left(({\bf \Pi}_{TT}^{-1})^{\dagger}\,J^T
\right)^{\mu\nu}=J^{T\mu\nu}\;,\nonumber\\
J_{\xi}^\mu&\equiv&\left(({\bf \Pi}_{LT}^{-1})^{\dagger}\,J^{L}\right)^\mu
=2\sqrt{-D^2-{\rm Ric}}\,\theta^\mu+2\left(-D^2-{\rm Ric}\right)^{-\frac{1}{2}}
\left(D_\nu R^{\mu\nu}\right)\upsilon\;,
\nonumber\\
J_{\sigma}&\equiv&({\bf \Pi}_{LL}^{-1})^{\dagger}\,J^{L}=
2\left((D^2)^2+\frac{d}{d-1}D_\mu R^{\mu\nu}D_\nu\right)^{-\frac{1}{2}}
D_\alpha R^{\alpha\beta}\,\theta_\beta\nonumber\\
& &+\frac{d-1}{d}\sqrt{(D^2)^2
+\frac{d}{d-1}D_\mu R^{\mu\nu}D_\nu}\,\upsilon\;,\nonumber\\
J_{\phi}&\equiv&({\bf \Pi}_{Tr}^{-1})^{\dagger}\,J^{Tr}=.\frac{1}{d}\,
g^{\mu\nu}J^{Tr}_{\mu\nu}\;.
\end{eqnarray}
Then combining the results of eqs. (\ref{Ap18}) and (\ref{Ap19}) we eventually
arrive at eq. (\ref{071}).
\section{Evaluating the RHS of the truncated flow equation}
\renewcommand{\theequation}{B\arabic{equation}}
\setcounter{equation}{0}
In this section we present several rather lengthy calculations needed for the 
discussion
of the Einstein-Hilbert truncation in section \ref{S4}. In the following, all
calculations are performed with $g_{\mu\nu}=\bar{g}_{\mu\nu}$ where
$\bar{g}_{\mu\nu}$ is assumed to correspond to a spherical background and the
bars are omitted from the metric, the curvature and the operators.
\label{eval}
\subsection{Computation of the inverse operators $\left(\Gamma_k^{(2)}
+{\cal R}_k\right)^{-1}$ and $\left(S_{\rm gh}^{(2)}+{\cal R}_k\right)^{-1}$}
\label{inv}
In section \ref{S4} we derived explicit expressions for the kinetic operators
$\widetilde{\Gamma}_k^{(2)}\equiv\Gamma_k^{(2)}+{\cal R}_k$ and
$\widetilde{S}_{\rm gh}^{(2)}\equiv S_{\rm gh}^{(2)}+{\cal R}_k$. They may be
represented as matrix differential operators acting on the column vectors
$(\bar{h}^T,\bar{\xi},\bar{\phi}_0,\bar{\sigma},\bar{\phi}_1)^{\bf T}$ and
$(\bar{v}^T,v^T,\bar{\varrho},\varrho)^{\bf T}$, respectively. In this
representation they take the form
\begin{eqnarray}
\label{83}
\widetilde{\Gamma}_k^{(2)}[g,g]=
\left(\begin{array}{cccc}
\left(\widetilde{\Gamma}_k^{(2)}[g,g]\right)_{\bar{h}^T\bar{h}^T} & 0 & 0 &
0_{1\times 2}\\0 & \left(\widetilde{\Gamma}_k^{(2)}[g,g]\right)_{\bar{\xi}
\bar{\xi}} & 0 & 0_{1\times 2}\\
0 & 0 & \left(\widetilde{\Gamma}_k^{(2)}[g,g]\right)_{\bar{\phi}_0
\bar{\phi}_0} & 0_{1\times 2}\\
0_{2\times 1} & 0_{2\times 1} & 0_{2\times 1} & {\cal Q}_k
\end{array}\right)
\end{eqnarray}
and
\begin{eqnarray}
\label{84}
\widetilde{S}_{\rm gh}^{(2)}[g,g]=
\left(\begin{array}{cccc}
0 & \left(\widetilde{S}_{\rm gh}^{(2)}[g,g]\right)_{\bar{v}^Tv^T} & 0 & 0\\ 
\left(\widetilde{S}_{\rm gh}^{(2)}[g,g]\right)_{v^T\bar{v}^T} & 0 & 0 & 0\\
0 & 0 & 0 & \left(\widetilde{S}_{\rm gh}^{(2)}[g,g]\right)_{\bar{\varrho}
\varrho}\\
0 & 0 & \left(\widetilde{S}_{\rm gh}^{(2)}[g,g]\right)_{\varrho\bar{\varrho}}
& 0
\end{array}\right) 
\end{eqnarray}
where
\begin{eqnarray}
\label{85}
{\cal Q}_k\equiv\left(
\begin{array}{cc}
\left(\widetilde{\Gamma}_k^{(2)}[g,g]\right)_{\bar{\sigma}
\bar{\sigma}} & \left(\widetilde{\Gamma}_k^{(2)}[g,g]
\right)_{\bar{\phi}_1\bar{\sigma}}\\
\left(\widetilde{\Gamma}_k^{(2)}[g,g]\right)_{\bar{\phi}_1\bar{\sigma}} & 
\left(\widetilde{\Gamma}_k^{(2)}[g,g]\right)_{\bar{\phi}\bar{\phi}}
\end{array}\right)\;.
\end{eqnarray}
The entries of these matrices are given in eq. (\ref{50}). On the RHS of the 
flow equation
(\ref{032}) these operators appear in terms of their inverses, which are
determined in the following. At this point it is important to note that,
because of the maximally symmetric background, all covariant derivatives 
contained in the operators (\ref{83}) and (\ref{84}) appear as covariant 
Laplacians and that the various entries are $x$-independent otherwise. This 
implies that these entries are {\it commuting} differential operators which 
allows for particularly simple manipulations. Therefore it is not difficult to
verify that the inverse operators assume the form
\begin{eqnarray}
\label{86}
\left(\widetilde{\Gamma}_k^{(2)}[g,g]\right)^{-1}=
\left(\begin{array}{cccc}
\left[\left(\widetilde{\Gamma}_k^{(2)}[g,g]\right)_{\bar{h}^T\bar{h}^T}
\right]^{-1} & 0 &0 & 0_{1\times 2}\\
0 & \left[\left(\widetilde{\Gamma}_k^{(2)}[g,g]\right)_{\bar{\xi}\bar{\xi}}
\right]^{-1} & 0 & 0_{1\times 2}\\
0 & 0 &\left[\left(\widetilde{\Gamma}_k^{(2)}[g,g]\right)_{\bar{\phi}_0
\bar{\phi}_0}\right]^{-1} & 0_{1\times 2}\\
0_{2\times 1} & 0_{2\times 1} & 0_{2\times 1} & {\cal Q}_k^{-1}
\end{array}\right)
\end{eqnarray}
and
\begin{eqnarray}
\label{87}
\lefteqn{\left(\widetilde{S}_{\rm gh}^{(2)}[g,g]\right)^{-1}=}\nonumber\\
& &\left(\begin{array}{cccc}
0 & \left[\left(\widetilde{S}_{\rm gh}^{(2)}[g,g]\right)_{\bar{v}^Tv^T}
\right]^{-1} & 0 & 0\\ 
\left[\left(\widetilde{S}_{\rm gh}^{(2)}[g,g]\right)_{v^T\bar{v}^T}
\right]^{-1} & 0 & 0 & 0\\
0 & 0 & 0 & \left[\left(\widetilde{S}_{\rm gh}^{(2)}[g,g]\right)_{
\bar{\varrho}\varrho}\right]^{-1}\\
0 & 0 &\left[\left(\widetilde{S}_{\rm gh}^{(2)}[g,g]\right)_{\varrho
\bar{\varrho}}\right]^{-1} & 0
\end{array}\right) 
\end{eqnarray}
with
\begin{eqnarray}
\label{88}
{\cal Q}_k^{-1}&=&
\left[\left(\widetilde{\Gamma}_k^{(2)}[g,g]\right)_{\bar{\sigma}
\bar{\sigma}} 
\left(\widetilde{\Gamma}_k^{(2)}[g,g]\right)_{\bar{\phi}_1\bar{\phi}_1}
-\left(\widetilde{\Gamma}_k^{(2)}[g,g]\right)^2_{\bar{\phi}_1\bar{\sigma}}
\right]^{-1}\nonumber\\
& &\times\left(\begin{array}{cc}
\left(\widetilde{\Gamma}_k^{(2)}[g,g]\right)_{\bar{\phi}_1\bar{\phi}_1}
&-\left(\widetilde{\Gamma}_k^{(2)}[g,g]
\right)_{\bar{\phi}_1\bar{\sigma}}\\
-\left(\widetilde{\Gamma}_k^{(2)}[g,g]\right)_{\bar{\phi}_1\bar{\sigma}} & 
\left(\widetilde{\Gamma}_k^{(2)}[g,g]\right)_{\bar{\sigma}\bar{\sigma}} 
\end{array}\right)\;.
\end{eqnarray}
Inserting these expressions into the RHS of the flow equation (\ref{032})
leads to
\begin{eqnarray}
\label{061}
\lefteqn{{\cal S}_k(R)
=\frac{1}{2}{\rm Tr}'\left[\sum\limits_{\zeta\in\{\bar{h}^T,
\bar{\xi},\bar{\phi}_0\}}
\left[\left(\Gamma_k^{(2)}[g,g]
+{\cal R}_k[g]\right)_{\zeta\zeta}\right]^{-1}
\partial_t\left({\cal R}_k[g]\right)_{\zeta\zeta}\right]}
\nonumber\\
& &+\frac{1}{2}{\rm Tr}'\left[
\left\{\left(\widetilde{\Gamma}_k^{(2)}[g,g]\right)_{\bar{\sigma}\bar{\sigma}} 
\left(\widetilde{\Gamma}_k^{(2)}[g,g]\right)_{\bar{\phi}_1\bar{\phi}_1}
-\left(\widetilde{\Gamma}_k^{(2)}[g,g]\right)^2_{\bar{\phi}_1\bar{\sigma}}
\right\}^{-1}\right.\nonumber\\
& &\times\left\{\left(\Gamma_k^{(2)}[g,g]
+{\cal R}_k[g]\right)_{\bar{\sigma}\bar{\sigma}}
\partial_t\left({\cal R}_k[g]\right)_{\bar{\phi}_1\bar{\phi}_1}
+\left(\Gamma_k^{(2)}[g,g]+{\cal R}_k[g]\right)_{\bar{\phi}_1\bar{\phi}_1}
\partial_t\left({\cal R}_k[g]\right)_{\bar{\sigma}
\bar{\sigma}}\right.\nonumber\\
& &\left.\left.-2\left(\Gamma_k^{(2)}[g,g]+{\cal R}_k[g]
\right)_{\bar{\phi}_1\bar{\sigma}}
\partial_t\left({\cal R}_k[g]\right)_{\bar{\phi}_1\bar{\sigma}}
\right\}\right]\nonumber\\
& &-{\rm Tr}'\left[\sum\limits_{\psi\in\{v^T,\varrho\}}
\left[\left(S_{\rm gh}^{(2)}[g,g]
+{\cal R}_k\right)_{\bar{\psi}\psi}\right]^{-1}
\partial_t\left({\cal R}_k\right)_{\bar{\psi}\psi}\right]
\end{eqnarray}
where we used the relations
\begin{eqnarray}
\label{060}
{\left(\left(S_{\rm gh}^{(2)}\right)_{\bar{v}^Tv^T}
\right)^{\mu x}}_{\nu y}
&=&-{\left(\left(S_{\rm gh}^{(2)}\right)_{v^T\bar{v}^T}
\right)_{\nu y}}^{\mu x}
=\frac{1}{\sqrt{g(y)}}\frac{\delta}{\delta v^{T\nu}(y)}
\frac{1}{\sqrt{g(x)}}\frac{\delta S_{\rm gh}}{\delta\bar{v}^T_\mu(x)}
\nonumber\\
{\left(\left(S_{\rm gh}^{(2)}\right)_{\bar{\varrho}\varrho}
\right)^x}_y
&=&-{\left(\left(S_{\rm gh}^{(2)}\right)_{\varrho\bar{\varrho}}
\right)_y}^x
=\frac{1}{\sqrt{g(y)}}\frac{\delta}{\delta\varrho(y)}
\frac{1}{\sqrt{g(x)}}\frac{\delta S_{\rm gh}}{\delta\bar{\varrho}(x)}\;.
\end{eqnarray}
The trace of the $\phi_0$-term appearing in eq. (\ref{061}) may be easily 
evaluated since only the scalar eigenmodes  $T^{01}$ and $T^{1m}$ 
contribute. We obtain
\begin{eqnarray}
\label{100}
\lefteqn{\frac{1}{2}{\rm Tr}'\left[\left[\left(\Gamma_k^{(2)}[g,g]
+{\cal R}_k[g]\right)_{\phi_0\phi_0}\right]^{-1}
\partial_t\left({\cal R}_k[g]\right)_{\phi_0\phi_0}\right]}\nonumber\\
&=&\frac{1}{2Z_{Nk}}\sum\limits_{l=0}^{1}\sum\limits_{m=1}^{D_l(d,0)}
\int d^dx\,\sqrt{g(x)}\,T^{lm}(x)\left[P_k
+A_{S1}(d,\alpha) R+B_{S1}(d,\alpha)\bar{\lambda}_k\right]^{-1}\nonumber\\
& &\times\partial_t\left[Z_{Nk}k^2R^{(0)}(-D^2/k^2)\right]
T^{lm}(x)\nonumber\\
&=&\frac{1}{2Z_{Nk}}\sum\limits_{l=0}^{1}\left[D_l(d,0)\frac{\partial_t
\left[Z_{Nk}
k^2R^{(0)}(\Lambda_l(d,0)/k^2)\right]}{\Lambda_l(d,0)+k^2R^{(0)}
(\Lambda_l(d,0)/k^2)+A_{S1}(d,\alpha)R+B_{S1}(d,\alpha)\bar{\lambda}_k}\right]
\;.
\end{eqnarray}
Here $\Lambda_l(d,0)$ is the eigenvalue with respect to $-D^2$ corresponding
to $T^{lm}$. Inserting also the remaining operators given in eq. (\ref{50}) 
into eq. (\ref{061}) finally leads to eq. (\ref{52}).
\subsection{Evaluation of the traces}
\label{trace}
In this part of the appendix we evaluate the traces appearing in eq.
(\ref{55}) by applying the asymptotic heat kernel expansion. In its original 
form it has
often been used to compute traces of operators acting on unconstrained fields,
see e.g. refs. \cite{MPetal.}. For our purposes we need the corresponding
expansions for operators acting on constrained fields, i.e. fields satisfying
appropriate transversality conditions. In appendix \ref{heat} these expansions
are derived in detail for Laplacians $D^2$ acting on symmetric transverse
traceless tensors, on transverse vectors and on scalars, with the following 
results:
\begin{eqnarray}
\label{57}
{\rm Tr}_{(2ST^2)}\left[e^{-({\rm i} s-\varepsilon)D^2}\right]
&=&\left(\frac{{\rm i}}{4\pi (s+{\rm i}\varepsilon)}\right)^{d/2}\int d^dx\,
\sqrt{g}\left\{
\frac{1}{2}(d-2)(d+1)\right.\nonumber\\
& &\left.-\frac{(d+1)(d+2)(d-5+3\delta_{d,2})}{12(d-1)}
({\rm i} s-\varepsilon) R+{\cal O}(R^2)\right\}\;,
\end{eqnarray}
\begin{eqnarray}
\label{58}
{\rm Tr}_{(1T)}\left[e^{-({\rm i} s-\varepsilon)D^2}\right]
&=&\left(\frac{{\rm i}}{4\pi (s+{\rm i}\varepsilon)}\right)^{d/2}\int d^dx\,
\sqrt{g}\Bigg\{d-1\nonumber\\
& &-\frac{(d+2)(d-3)+6\delta_{d,2}}{6d}({\rm i}s-\varepsilon) R
+{\cal O}(R^2)\Bigg\}\;,
\end{eqnarray}
\begin{eqnarray}
\label{59}
{\rm Tr}_{(0)}\left[e^{-({\rm i} s-\varepsilon)D^2}\right]
=\left(\frac{{\rm i}}{4\pi (s+{\rm i}\varepsilon)}\right)^{d/2}\int d^dx\,
\sqrt{g}\left\{
1-\frac{1}{6}({\rm i} s-\varepsilon) R+{\cal O}(R^2)\right\}\;.
\end{eqnarray}

The next step is to consider an arbitrary function $W(z)$ with a Fourier
transform $\widetilde{W}(s)$. For such functions $W$, we may express the
trace of the operator $W(-D^2)$ that results from replacing the argument of $W$
with $-D^2$ in terms of $\widetilde{W}(s)$:
\begin{eqnarray}
\label{56}
{\rm Tr}\left[W(-D^2)\right]=\lim\limits_{\varepsilon\searrow 0}
\int\limits^{\infty}_{-\infty}ds\,\widetilde{W}(s)\,{\rm Tr}
\left[e^{-({\rm i} s-\varepsilon)D^2}\right]\;.
\end{eqnarray}
We obtain the asymptotic expansion of ${\rm Tr}[W(-D^2)]$ by inserting 
the heat kernel expansion for ${\rm Tr}[e^{-({\rm i}s-\varepsilon)D^2}]$ into
eq. (\ref{56}). For Laplacians acting on the constrained fields considered
here they read as follows:
\begin{eqnarray}
\label{60}
{\rm Tr}_{(2ST^2)}\left[W(-D^2)\right]&=&(4\pi)^{-d/2}\bigg\{
\frac{1}{2}(d-2)(d+1)\,Q_{d/2}[W]\,
\int d^dx\,\sqrt{g}\nonumber\\
& &+\frac{(d+1)(d+2)(d-5+3\delta_{d,2})}{12(d-1)}
\,Q_{d/2-1}[W]\,\int d^dx\,\sqrt{g}R\nonumber\\
& &+{\cal O}(r^{<d-2})\bigg\}\;,
\end{eqnarray}
\begin{eqnarray}
\label{61}
{\rm Tr}_{(1T)}\left[W(-D^2)\right]&=&(4\pi)^{-d/2}\bigg\{
(d-1)\,Q_{d/2}[W]\,
\int d^dx\,\sqrt{g}\nonumber\\
& &+\frac{(d+2)(d-3)+6\delta_{d,2}}{6d}
\,Q_{d/2-1}[W]\,\int d^dx\,\sqrt{g}R+{\cal O}(r^{<d-2})\bigg\}\;,
\end{eqnarray}
\begin{eqnarray}
\label{62}
{\rm Tr}_{(0)}\left[W(-D^2)\right]&=&(4\pi)^{-d/2}\bigg\{Q_{d/2}[W]\,
\int d^dx\,\sqrt{g}\nonumber\\
& &+\frac{1}{6}\,Q_{d/2-1}[W]\,\int d^dx\,\sqrt{g}R+{\cal O}(r^{<d-2})
\bigg\}\;.
\end{eqnarray}
Here the set of functionals $Q_n[W]$ is defined as
\begin{eqnarray}
\label{63}
Q_n[W]\equiv\lim\limits_{\varepsilon\searrow 0}\int
\limits^{\infty}_{-\infty} ds\,(-{\rm i} s+\varepsilon)^{-n}\,
\widetilde{W}(s)\;.
\end{eqnarray}
By virtue of the Mellin transformation we may now reexpress $Q_n$ in terms of
$W$ so that
\begin{eqnarray}
\label{64}
Q_n[W]&=&\frac{1}{\Gamma(n)}\int\limits_{0}^{\infty}dz\,z^{n-1}\,W(z)\;,\;\;
n>0\;;\nonumber\\
Q_n[W]&=&\frac{(-1)^m}{\Gamma(m+n)}\int\limits_{0}^{\infty}dz\,z^{m+n-1}\,
\frac{d^m W(z)}{dz^m}\;,\;\,n\le 0\;,\;\,m>-n\;,\;\,m\in 
|\!{\rm N}\;\;{\rm arbitrary}.
\end{eqnarray}
In particular we obtain $Q_0[W]=W(0)$.

Let us now consider the case where isolated eigenvalues have to be excluded
from ${\rm Tr}[W(-D^2)]$. According to appendix \ref{heat} such traces can be
expressed as the difference between the complete trace ${\rm Tr}[W(-D^2)]$
and a term of the form $\sum\limits_{l\in\{l_1,\ldots,l_n\}}D_l(d,s)$ $\times
W\left(
\Lambda_l(d,s)\right)$. Here $l_1,\ldots,l_n$ refer to the modes to be 
omitted and $\Lambda_l(d,s)$ and $D_l(d,s)$ denote the corresponding 
eigenvalues of $-D^2$ and their degrees of degeneracy, respectively. Since 
$\Lambda_l(d,s)\propto R$ we may view $W\left(\Lambda_l(d,s)\right)$ as a 
function of $R$. As outlined in section \ref{S4} such a function contributes
to the evolution of $Z_{Nk}$ and $\bar{\lambda}_k$ only for $d=2$, with the
contribution given by $W(0)$. Using the explicit expressions for $D_l(d,s)$ 
(see table 1 in appendix \ref{harm}) and applying eq. (\ref{035}) we therefore
obtain for the traces relevant to the flow equation
\begin{eqnarray}
\label{037}
{\rm Tr}_{(1T)}'[W(-D^2)]&=&{\rm Tr}_{(1T)}[W(-D^2)]
-\frac{3\delta_{d,2}}{8\pi}\,W(0)\,\int d^2x\,\sqrt{g}R+{\cal O}(r^{<d-2})\;,
\nonumber\\
{\rm Tr}_{(0)}''[W(-D^2)]&=&{\rm Tr}_{(0)}[W(-D^2)]
-\frac{\delta_{d,2}}{2\pi}\,W(0)\,\int d^2x\,\sqrt{g}R+{\cal O}(r^{<d-2})\;,
\nonumber\\
{\rm Tr}_{(0)}'[W(-D^2)]&=&{\rm Tr}_{(0)}[W(-D^2)]
-\frac{\delta_{d,2}}{8\pi}\,W(0)\,\int d^2x\,\sqrt{g}R+{\cal O}(r^{<d-2})\;,
\end{eqnarray}
where the primes have to be interpreted as in section \ref{S4}.

The next step is to insert the expansions of the traces into ${\cal S}_k(R)$,
eq. (\ref{55}), and to compare the coefficients of the operators
$\int d^dx\sqrt{g}$ and $\int d^dx\sqrt{g}R$ with those on the LHS, eq.
(\ref{040}). This leads to the following differential equations:
\begin{eqnarray}
\label{66}
\partial_t\left(Z_{Nk}\bar{\lambda}_k\right)
&=&(4\kappa^2)^{-1}(4\pi)^{-d/2}\bigg\{
\frac{1}{2}d(d-1)\,Q_{d/2}\left[\left({\cal A}_0-2\bar{\lambda}_k
\right)^{-1}{\cal N}\right]\nonumber\\
& &+d\,Q_{d/2}\left[\left({\cal A}_0-2\alpha\bar{\lambda}_k\right)^{-1}
{\cal N}\right]-2d\,Q_{d/2}\left[{\cal A}_0^{-1}{\cal N}_0\right]\bigg\}
\;,
\end{eqnarray}
\begin{eqnarray}
\label{67}
\partial_t Z_{Nk}&=&-(2\kappa^2)^{-1}(4\pi)^{-d/2}\bigg\{
c_1(d)\,Q_{d/2-1}\left[\left({\cal A}_0-2\bar{\lambda}_k
\right)^{-1}{\cal N}\right]\nonumber\\
& &+c_2(d)\,Q_{d/2-1}\left[\left({\cal A}_0
-2\alpha\bar{\lambda}_k\right)^{-1}{\cal N}\right]
+c_3(d)\,Q_{d/2}\left[\left({\cal A}_0-2\bar{\lambda}_k
\right)^{-2}{\cal N}\right]\nonumber\\
& &+c_4(d,\alpha)\,Q_{d/2}\left[\left({\cal A}_0-2\alpha
\bar{\lambda}_k\right)^{-2}{\cal N}\right]
-2c_2(d)\,Q_{d/2-1}
\left[{\cal A}_0^{-1}{\cal N}_0\right]\nonumber\\
& & +c_5(d)\,Q_{d/2}\left[{\cal A}_0^{-2}{\cal N}_0\right]
+3\delta_{d,2}\left[\frac{\partial_t(Z_{Nk}k^2)}{2Z_{Nk}(k^2
-2\bar{\lambda}_k)}-\frac{\partial_t(Z_{Nk}k^2)}{2Z_{Nk}(k^2
-2\alpha\bar{\lambda}_k)}\right]\bigg\}\;.
\end{eqnarray}
Here the coefficients $c_i$ are defined as in eq. (\ref{044}).

In eqs. (\ref{66}), (\ref{67}), the various $Q_n$ may now be expressed in
terms of the cutoff-dependent threshold functions $\Phi^p_n$ and 
$\widetilde{\Phi}^p_n$ introduced in eq. (\ref{68}). Using the relations
\begin{eqnarray}
\label{70}
Q_n\left[\left({\cal A}_0+c\right)^{-p}{\cal N}\right]
&=&k^{2(n-p+1)}\Phi^p_n(c/k^2)-\frac{1}{2}\eta_N(k)\,k^{2(n-p+1)}
\widetilde{\Phi}^p_n(c/k^2)\nonumber\\
Q_n\left[\left({\cal A}_0+c\right)^{-p}{\cal N}_0\right]
&=&k^{2(n-p+1)}\Phi^p_n(c/k^2)
\end{eqnarray}
we arrive at the differential equations (\ref{74}) and (\ref{75}).

At this stage the following point should be mentioned. In order to achieve
that the integrals in eq. (\ref{63}) actually converge we have to demand that
$R^{(0)}(y)$ rapidly decreases as $y\rightarrow \pm\infty$. However, since
from now on its form for $y<0$ does not play a role any more we identify
$R^{(0)}(y)$ with its part for nonnegative arguments and assume that
$R^{(0)}(y)$ is a smooth function defined only for $y\ge 0$ and endowed with
the properties stated in subsection \ref{3B}.
\section{Flow equations in four dimensions}
\renewcommand{\theequation}{C\arabic{equation}}
\setcounter{equation}{0}
\label{d=4}
In this section we compare our results to those derived in ref. \cite{DP97}
for the $d=4$ dimensional case. Inserting $d=4$ into eq. (\ref{52})
we obtain for the RHS ${\cal S}_k$ of the evolution equation
\begin{eqnarray}
\label{105}
{\cal S}_k(R)&=&\frac{1}{2}{\rm Tr}_{(2ST^2)}\left[\frac{\partial_t P_k}
{P_k+\frac{2}{3}R-2\bar{\lambda}_k}\right]
+\frac{1}{2}{\rm Tr}_{(1T)}'\left[\frac{\partial_t P_k}
{P_k+\frac{2\alpha-1}{4}R-2\alpha\bar{\lambda}_k}\right]
\nonumber\\
& &+\frac{1}{2}{\rm Tr}_{(0)}''\left[\frac{\partial_t P_k}{P_k+\frac{\alpha-1}
{2}R-2\alpha\bar{\lambda}_k}\right]
+\frac{1}{2}{\rm Tr}_{(0)}''\left[\frac{\partial_t P_k}{P_k
-2\bar{\lambda}_k}\right]\nonumber\\
& &+\frac{1}{2}\sum\limits_{l=0}^{1}\left[D_l(4,0)\frac{\partial_t P_k(
\Lambda_l(4,0))}{P_k(\Lambda_l(4,0))
-\frac{4\alpha}{3\alpha-1}\bar{\lambda}_k}\right]
-{\rm Tr}_{(1T)}\left[\frac{\partial_t P_k}
{P_k-\frac{R}{4}}\right]
-{\rm Tr}_{(0)}'\left[\frac{\partial_t P_k}
{P_k-\frac{R}{2}}\right]\nonumber\\
& &+\frac{\partial_t Z_{Nk}}{Z_{Nk}}\Bigg\{\frac{1}{2}{\rm Tr}_{(2ST^2)}\left[
\frac{P_k+D^2}{P_k+\frac{2}{3}R-2\bar{\lambda}_k}\right]
+\frac{1}{2}{\rm Tr}_{(1T)}'\left[\frac{P_k+D^2}
{P_k+\frac{2\alpha-1}{4}R-2\alpha\bar{\lambda}_k}\right]
\nonumber\\
& &+\frac{1}{2}\sum\limits_{l=0}^{1}\left[D_l(4,0)\frac{P_k(\Lambda_l(4,0))
-\Lambda_l(4,0)}{P_k(\Lambda_l(4,0))
-\frac{4\alpha}{3\alpha-1}\bar{\lambda}_k}\right]\nonumber\\
& &-\frac{1}{4\alpha}{\rm Tr}_{(0)}''\Bigg[
\bigg\{\left[P_k-2\bar{\lambda}_k\right]
\left[P_k+\frac{\alpha-1}{2}R-2\alpha\bar{\lambda}_k
\right]\bigg\}^{-1}\nonumber\\
& &\times\Bigg\{\left((1-3\alpha)\left[(3-\alpha)P_k
+\frac{\alpha-1}{2}R
\right]+4\alpha(\alpha+1)\bar{\lambda}_k\right)(P_k+D^2)\nonumber\\
& &-3(1-\alpha)^2\sqrt{P_k}\sqrt{P_k-\frac{R}{3}}\,\left[
\sqrt{P_k}\sqrt{P_k-\frac{R}{3}}-\sqrt{-D^2}\sqrt{-D^2-\frac{R}{3}}\right]
\Bigg\}\Bigg]\Bigg\}
\end{eqnarray}
with $P_k(\Lambda_l(4,0))\equiv\Lambda_l(4,0)
+k^2R^{(0)}(\Lambda_l(4,0)/k^2)$. Our result (\ref{105}) agrees with 
eq. (3.22) of ref. \cite{DP97}, up to a few (typographical) errors occuring in 
eq. (3.22). To be more precise, the prime at the ${\rm Tr}_{(1T)}'$-term in
lines 1 and 4 of eq. (\ref{105}), the factor $D_l(4,0)$ appearing in the first
term of lines 3 and 5, and the factor $1/2$ contained in the term $\propto R$ 
in line 7 are mistakenly left out in the corresponding equation of ref. 
\cite{DP97}.

Expanding the flow equations for $Z_{Nk}\bar{\lambda}_k$ and
$Z_{Nk}$, eqs. (\ref{74}) and (\ref{75}), with respect to $\bar{\lambda}_k$, 
using the relation $\partial_t\bar{\lambda}_k=Z_{Nk}^{-1}\partial_t(Z_{Nk}
\bar{\lambda}_k)+\bar{\lambda}_k\eta_N(k)$ and setting, as in \cite{DP97},
\begin{eqnarray}
\label{107}
\kappa_k\equiv2\kappa^2 Z_{Nk}\;,\;\;Z_k\equiv Z_{Nk}\;,\;\;\eta_k\equiv
\partial_t\ln Z_k\equiv -\eta_N(k)\;,\;\;q^p_n\equiv2\Phi^p_n(0)\;,
\widetilde{q}^p_n\equiv\widetilde{\Phi}^p_n(0)
\end{eqnarray}
leads to
\begin{eqnarray}
\label{109}
\partial_t\kappa_k&=&(4\pi)^{-2}k^2\left\{\frac{13}{24}q^1_1+\left(
\frac{55}{24}+\alpha\right)q^2_2
+\eta_k\left[\frac{1}{8}\widetilde{q}^1_1+\left(\frac{25}{24}
+\alpha\right)\widetilde{q}_2^2\right]\right\}+{\cal O}(\bar{\lambda}_k)\;,
\end{eqnarray}
\begin{eqnarray}
\label{110}
\partial_t\bar{\lambda}_k&=&\kappa_k^{-1}(4\pi)^{-2}\left\{
k^4\left[\frac{1}{2}q^1_2+\eta_k\frac{5}{2}\widetilde{q}^1_2\right]
+\bar{\lambda}_k k^2\left[-\frac{13}{24}q^1_1\right.\right.\nonumber\\
& &+\left.\left.\left(\frac{17}{24}+\alpha\right)q^2_2
+\eta_k\left(-\frac{1}{8}\widetilde{q}^1_1+\left(\frac{47}{24}
+\alpha\right)\widetilde{q}_2^2\right)\right]\right\}
+{\cal O}(\bar{\lambda}_k^2)\;.
\end{eqnarray}
In this form the flow equations for the couplings are suitable for a comparison
with the corresponding results in eqs. (4.6)-(4.9) of ref. \cite{DP97}. Apart 
from the contributions from matter fields, which are also considered there, 
the results differ by a factor $60/24$ in front of the $q_2^2$-terms. 
Presumably, this deviation can be explained by a wrong sign introduced in 
\cite{DP97} for a certain term. This term is a contribution from the ghosts  
produced by heat kernel expanding the last two terms in line 3 of eq. 
(\ref{105}), and it carries the prefactor $30/24$. 
We may conclude that, apart from these corrections, our results agree with 
those in \cite{DP97}, which is as it should be since the same cutoff is used.
\section{Tensor Spherical harmonics}
\renewcommand{\theequation}{D\arabic{equation}}
\setcounter{equation}{0}
\label{harm}
In this section we introduce the spherical harmonics $T^{lm}_{\mu\nu}$, 
$T^{lm}_\mu$ and $T^{lm}$ for symmetric transverse traceless $(ST^2)$ 
tensors $h^T_{\mu\nu}$, transverse $(T)$ vectors $\xi_\mu$, and scalars $\phi$
on a $d$-dimensional spherical background $S^d$. These harmonics form 
complete sets of orthogonal eigenfunctions with respect to the covariant 
Laplacians acting on $ST^2$ tensors, $T$ vectors and scalars, i.e. they satisfy
\begin{eqnarray}
\label{094}
-\bar{D}^2\,T^{lm}_{\mu\nu}(x)&=&\Lambda_l(d,2)\,T^{lm}_{\mu\nu}(x)\;,
\nonumber\\
-\bar{D}^2\,T^{lm}_\mu(x)&=&\Lambda_l(d,1)\,T^{lm}_\mu(x)\;,\nonumber\\
-\bar{D}^2\,T^{lm}(x)&=&\Lambda_l(d,0)\,T^{lm}(x)
\end{eqnarray}
and, after proper normalization,
\begin{eqnarray}
\label{093}
\delta^{lk}\,\delta^{mn}&=&\int d^dx\,\sqrt{\bar{g}}\,
\left({\bf 1}_{(2ST^2)}\right)^{\mu\nu\rho\sigma}\,T^{lm}_{\mu\nu}\,
T^{kn}_{\rho\sigma}
=\int d^dx\,\sqrt{\bar{g}}\,\left({\bf 1}_{(1T)}\right)^{\mu\nu}
\,T^{lm}_\mu\,T^{kn}_\nu\nonumber\\
&=&\int d^dx\,\sqrt{\bar{g}}\,T^{lm}\,T^{kn}\;.
\end{eqnarray}
Here $\left({\bf 1}_{(2ST^2)}\right)^{\mu\nu\rho\sigma}=(d-2)/2d
\left(\bar{g}^{\mu\rho}\bar{g}^{\nu\sigma}+\bar{g}^{\mu\sigma}\bar{g}^{\nu
\rho}\right)$ and $\left({\bf 1}_{(1T)}\right)^{\mu\nu}=(d-1)/d\,
\bar{g}^{\mu\nu}$ are the unit matrices in the spaces of $ST^2$ tensors and
transverse vectors, respectively. In eq. (\ref{094}) 
the $\Lambda_l(d,s)$'s denote the eigenvalues with respect to
$-\bar{D}^2$ where $s$ refers to the spin of the field under consideration
and $l$ takes the values $s,s+1,s+2,\cdots$. Furthermore, the second upper 
index  at $T^{lm}_{\mu\nu}$, $T^{lm}_\mu$ and $T^{lm}$, $m$, takes the
degeneracy of the eigenvalues into account. It assumes values from one to
$D_l(d,s)$ with $D_l(d,s)$ the degree of degeneracy. In ref. \cite{OR} 
explicit expressions for $\Lambda_l(d,s)$ and $D_l(d,s)$ are derived
which can be found in table 1. The eigenvalues are expressed in terms of the
curvature scalar ${\bar R}=d(d-1)/r^2$ of the sphere with radius $r$. In ref. 
\cite{OR} it is also shown that the 
spherical harmonics $T^{lm}_{\mu\nu}$, $T^{lm}_\mu$ and $T^{lm}$ span the 
spaces of $ST^2$ tensors, $T$ vectors and scalars so that we may expand 
arbitrary functions $h^T_{\mu\nu}$, $\xi_\mu$ and $\phi$ according to
\begin{eqnarray}
\label{067}
h_{\mu\nu}^T(x)&=&
\sum\limits_{l=2}^{\infty}\sum\limits_{m=1}^{D_l(d,2)}h^T_{lm}\,
T^{lm}_{\mu\nu}(x)\;,\nonumber\\
\xi_\mu(x)&=&
\sum\limits_{l=1}^{\infty}\sum\limits_{m=1}^{D_l(d,1)}\xi_{lm}\,
T^{lm}_\mu(x)\;,\nonumber\\
\phi(x)&=&\sum\limits_{l=0}^{\infty}\sum\limits_{m=1}^{D_l(d,0)}\phi_{lm}\,
T^{lm}(x)\;.
\end{eqnarray}
Here the coefficients $\{h^T_{lm}\}$, $\{\xi_{lm}\}$ and $\{\phi_{lm}\}$ are 
countably infinite
sets of constants that are uniquely determined by $h^T_{\mu\nu}$, 
$\xi_\mu$ and $\phi$. Eq. (\ref{067}) may now be used to expand also any 
symmetric non-$T^2$ tensor and nontransverse vector in terms of spherical 
harmonics
since they may be expressed in terms of $ST^2$ tensors, $T$ vectors and 
scalars by using the decompositions (\ref{f}), (\ref{p}), see e.g.
\cite{OR,Al86,FT84,TV90}. 

In this context it is important to note that
the $D_1(d,1)=d(d+1)/2$ modes $T^{1,m}_\mu$ and the $D_1(d,0)=d+1$ modes 
$T^{1,m}$ satisfy the Killing equation (\ref{09}) and the scalar equation 
(\ref{010}), respectively, and that $T^{0,1}=\rm const$. As discussed in 
subsection \ref{2B}, arbitrary symmetric rank 2 tensors receive no 
contribution from these modes. In the case of arbitrary vectors it is the 
constant scalar mode that does not contribute. Such modes have no physical 
meaning and have to be omitted therefore.   

\vspace{1cm} 
\begin{tabular}
{|c|c|c|c|c|}
\hline\multicolumn{5}{|c|}{Table 1: Eigenvalues of $-\bar{D}^2$ and their 
degeneracy on the $d$-sphere}\\
\hline Eigenfunction & Spin $s$ & Eigenvalue $\Lambda_l(d,s)$ & Degeneracy 
$D_l(d,s)$ & \\
\hline $T_{\mu\nu}^{lm}(x)$ & 2 & $\frac{l(l+d-1)-2}{d(d-1)}\bar{R}$
& $\frac{(d+1)(d-2)(l+d)(l-1)(2l+d-1)(l+d-3)!}{2(d-1)!(l+1)!}$ &
$l=2,3,\ldots $ \\
\hline $T_\mu^{lm}(x)$ & 1 & $\frac{l(l+d-1)-1}{d(d-1)}\bar{R}$ &
$\frac{l(l+d-1)(2l+d-1)(l+d-3)!}{(d-2)!(l+1)!}$ & $l=1,2,\ldots$
\\ \hline $T^{lm}(x)$ & 0 & $\frac{l(l+d-1)}{d(d-1)}\bar{R}$ & 
$\frac{(2l+d-1)(l+d-2)!}{l!(d-1)!}$ & $l=0,1,\ldots$ \\
\hline\end{tabular}
\vspace{0.5cm}
\section{Heat kernel coefficients for differentially constrained fields}
\renewcommand{\theequation}{E\arabic{equation}}
\setcounter{equation}{0}
\label{heat}
In this part of the appendix we supply the tools necessary for the evaluation
of functional traces and we derive the heat kernel expansions for 
Laplacians acting on differentially constrained fields.

As a first step we consider a functional trace of the form
\begin{eqnarray}
\label{hk14}
{\rm Tr}_{(s\,[C])}[f(-D^2)]&=&\int d^dx\,\sqrt{g}\,{\left<x\left|f(-D^2)
\right|x\right>_{\mu_1\ldots\mu_s}}^{\mu_1\ldots\mu_s}\nonumber\\
&=&\int d^dx\,\left[\left(f(-D^2)\right)^{\mu_1\ldots\mu_s\nu_1\ldots\nu_s}
\,\left({\bf 1}_{(s[C])}\right)_{\nu_1\ldots\nu_s\mu_1\ldots\mu_s}\delta^d(x-y)
\right]_{x=y}
\end{eqnarray}
Here $f$ is an arbitrary smooth function whose argument is replaced with the
covariant Laplacian defined on the space of spin-$s$ fields with a possible
symmetry and/or transversality constraint $C$, as indicated by the subscript 
$(s[C])$ at the trace. Note that $f$ inherits the matrix structure from the 
corresponding Laplacian. Furthermore, ${\bf 1}_{(s[C])}$
denotes the unit matrix in the space of independent field components. Given
a closed\renewcommand{\baselinestretch}{1}\small\normalsize\footnote{The 
restriction to closed Riemannian manifolds is done for the sake of notational
simplicity only. In principle, the results extend to noncompact asymptotically
flat Riemannian manifolds.}
\renewcommand{\baselinestretch}{1.5}\small\normalsize Riemannian manifold 
$({\cal M},g)$ we now assume that $\{U^k_{\mu_1\ldots\mu_s}(x)\}$ is a 
complete set of orthonormal functions on $({\cal M},g)$ spanning the space of 
fields under consideration. Then by making use of the completeness relation
\begin{eqnarray}
\label{hk19}
\left({\bf 1}_{(s[C])}\right)_{\nu_1\ldots\nu_s\mu_1\ldots\mu_s}\,
\frac{\delta^d(x-y)}{\sqrt{g(x)}}=\frac{1}{2}
\sum\limits_k\left(U^k_{\nu_1\ldots\nu_s}(x)\,U^k_{\mu_1\ldots\mu_s}(y)
+U^k_{\mu_1\ldots\mu_s}(x)\,U^k_{\nu_1\ldots\nu_s}(y)\right)
\end{eqnarray}
eq. (\ref{hk14}) can be written as
\begin{eqnarray}
\label{hk20}
{\rm Tr}_{(s[C])}[f(-D^2)]=\int d^dx\,\sqrt{g}\,\sum\limits_k
U^k_{\mu_1\ldots\mu_s}(x)\left(f(-D^2)\right)^{\mu_1\ldots\mu_s\nu_1\ldots
\nu_s}U^k_{\nu_1\ldots\nu_s}(x)\;.
\end{eqnarray}
Clearly if $\{U^k_{\mu_1\ldots\mu_s}(x)\}$ is taken to be a complete set of 
orthonormal eigenfunctions with respect to the covariant Laplacian such that 
$-D^2\,U^k_{\mu_1\ldots\mu_s}(x)=\widetilde{\Lambda}_k\,U^k_{\mu_1\ldots\mu_s}
(x)$ eq. (\ref{hk20}) boils down to
\begin{eqnarray}
\label{096}
{\rm Tr}_{(s[C])}[f(-D^2)]=\sum\limits_{k}f\left(\widetilde{\Lambda}_k
\right)\;.
\end{eqnarray}

In general the evaluation of such traces is a formidable task and one has to
resort to approximations. The most familiar such approximation is the ``early
time'' expansion of the diagonal heat kernel $\left<x\left|e^{-{\rm i}{\sf t}
D^2}\right|x\right>$ for $|{\sf t}|\rightarrow 0$. (We write
${\sf t}\equiv s+{\rm i}\varepsilon$ with ${\rm Im}({\sf t})=\varepsilon>0$.) 
It has been discussed in many references \cite{MPetal.}:
\begin{eqnarray}
\label{hk22}
\left<x\left|e^{-{\rm i}{\sf t}\left(D^2+Q\right)}\right|x\right>
=\left(\frac{{\rm i}}{4\pi {\sf t}}\right)^{\frac{d}{2}}
\left\{a_0(x;Q)-{\rm i}{\sf t}\,a_2(x;Q)-{\sf t}^2\,a_4(x;Q)+{\cal O}
\left({\sf t}^3
\right)\right\}\;.
\end{eqnarray}
Here $Q$ is an arbitrary smooth matrix potential and the $a_n$'s are tensor
polynomials proportional to the $n/2$-th power of the curvature and endowed 
with the same matrix structure as $D^2+Q$.
They depend on the space of fields under consideration. For operators $D^2+Q$
acting on unconstrained fields, i.e. fields with
independent components, the first three coefficients take the 
form\renewcommand{\baselinestretch}{1}\small\normalsize\footnote{In the 
context of the Einstein-Hilbert truncation performed in the present paper we
only need the coefficients $a_0\propto r^0$ and $a_2\propto r^{-2}$. However,
for truncations containing invariants quadratic in the curvature we shall also 
need $a_4\propto r^{-4}$, see \cite{LR3}.}
\renewcommand{\baselinestretch}{1.5}\small\normalsize
\begin{eqnarray}
\label{hk4}
a_0(x;Q)&\equiv& a_0={\bf 1}\;,\nonumber\\
a_2(x;Q)&=&P\;,\nonumber\\
a_4(x;Q)&=&\frac{1}{180}\left(R_{\mu\nu\alpha\beta}R^{\mu\nu\alpha\beta}
-R_{\mu\nu}R^{\mu\nu}+D^2 R\right){\bf 1}+\frac{1}{2}P^2
+\frac{1}{12}{\cal R}_{\mu\nu}{\cal R}^{\mu\nu}+\frac{1}{6}D^2P\;.
\end{eqnarray}
Here ${\bf 1}$ is the unit matrix in the space of field components, and
\begin{eqnarray}
\label{hk5}
P\equiv Q+\frac{1}{6}R\,{\bf 1}\;.
\end{eqnarray}
Furthermore, ${\cal R}_{\mu\nu}$ is the curvature operator defined as the
commutator
\begin{eqnarray}
\label{hk6}
{\cal R}_{\mu\nu}=\left[D_\mu,D_\nu\right]\equiv D_\mu D_\nu-D_\nu D_\mu\;.
\end{eqnarray}

Inserting the asymptotic expansion (\ref{hk22}) into eq. (\ref{hk14}) we 
obtain for the functional trace of the heat kernel
\begin{eqnarray}
\label{hk3}
{\rm Tr}\left[e^{-{\rm i}{\sf t}\left(D^2+Q\right)}\right]=\left(\frac{{\rm i}}
{4\pi {\sf t}}
\right)^{\frac{d}{2}}\!\int d^dx\,\sqrt{g}\left\{{\rm tr}\,a_0-{\rm i}{\sf t}\,
{\rm tr}\,a_2(x;Q)-{\sf t}^2\,{\rm tr}\,a_4(x;Q)+{\cal O}\left({\sf t}^3\right)
\right\}
\end{eqnarray}
where ${\rm tr}$ is the matrix trace with respect to the tensor or spinor
indices.

Let us have a closer look  at the coefficients $a_n$. For arbitrary scalars 
$\phi$, ${\cal R}_{\mu\nu}$ vanishes since $D_\mu D_\nu\phi=D_\nu D_\mu\phi$.
In the case of arbitrary spin-$s$ fields $F_{\mu_1\ldots\mu_s}(x)$ with 
integer spin $s\ge 1$ we obtain
\begin{eqnarray}
\label{hk7}
{\cal R}_{\alpha\beta}\,F_{\mu_1\ldots\mu_s}(x)=
\sum\limits_{i=1}^s {R_{\alpha\beta\mu_i}}^\mu\,F_{\mu_1\ldots\mu_{i-1}\mu
\mu_{i+1}\ldots\mu_s}(x)\;.
\end{eqnarray}
Hence, for scalars, $a_4$ receives no contribution from 
${\cal R}_{\alpha\beta}{\cal R}^{\alpha\beta}$,  while for fields with nonzero
spin it amounts to a nonvanishing contribution with a $(d^s\times d^s)$-matrix
structure. For arbitrary vectors and rank-2 tensors we find, respectively,
\begin{eqnarray}
\label{hk8}
\left({\cal R}_{\alpha\beta}{\cal R}^{\alpha\beta}\right)_{\mu\nu}
&=&-R_{\alpha\beta\gamma\mu}{R^{\alpha\beta\gamma}}_\nu\;,\nonumber\\
\left({\cal R}_{\alpha\beta}{\cal R}^{\alpha\beta}\right)_{\mu\nu\rho\sigma}&=&
-R_{\alpha\beta\gamma\mu}{R^{\alpha\beta\gamma}}_\rho\,g_{\nu\sigma}
-R_{\alpha\beta\gamma\nu}{R^{\alpha\beta\gamma}}_\sigma\,g_{\mu\rho}
+2R_{\alpha\beta\mu\rho}{R^{\alpha\beta}}_{\nu\sigma}\;.
\end{eqnarray}

From now on we restrict our considerations to matrix potentials $Q$ of the 
form $Q=qR{\bf 1}_{(s)}$ with $q$ a real constant, and we assume that the 
metric corresponds to a maximally symmetric background $S^d$. Setting 
$a_n(x;qR{\bf 1}_{(s)})\equiv
a_n(q)$ we obtain $a_0(q)\equiv a_0={\bf 1}_{(s)}$ and $a_2(q)=(1+6q)/6 R
{\bf 1}_{(s)}$, independently of the spin $s$ of the field. Here the
dependence on $s$ is totally encrypted in the unit matrix ${\bf 1}_{(s)}$.
This is not the case for the coefficient $a_4$. It is given by  
\begin{eqnarray}
\label{hk9}
a_4(q)=\frac{1}{360}\left[\frac{5d^2-7d+6}{d(d-1)}+60q+180q^2\right]\,
{\bf 1}_{(0)}\,R^2\;,
\end{eqnarray}
\begin{eqnarray}
\label{hk10}
\left[a_4(q)\right]_{\mu\nu}=\frac{1}{360}\left[\frac{5d^3-7d^2+6d-60}
{d^2(d-1)}+60q+180q^2\right]\left({\bf 1}_{(1)}\right)_{\mu\nu}\,R^2\;,
\end{eqnarray}
\begin{eqnarray}
\label{hk11}
\left[a_4(q)\right]_{\mu\nu\rho\sigma}
&=&\frac{1}{360}\left[\frac{5d^3-7d^2+6d-120}{d^2(d-1)}+60q+180q^2\right]
\left({\bf 1}_{(2)}\right)_{\mu\nu\rho\sigma}\,R^2\nonumber\\
& &+\frac{1}{3d^2(d-1)^2}\left(g_{\mu\nu}g_{\rho\sigma}-g_{\mu\sigma}
g_{\nu\rho}\right)R^2
\end{eqnarray}
for scalars, vectors and rank-2 tensors, respectively, with ${\bf 1}_{(0)}=1$,
$\left({\bf 1}_{(1)}\right)_{\mu\nu}=g_{\mu\nu}$ and
$\left({\bf 1}_{(2)}\right)_{\mu\nu\rho\sigma}=g_{\mu\rho}g_{\nu\sigma}$.

Up to this point we considered only unconstrained fields. For fields subject
to constraints like transverse vectors $\xi_\mu$ and $ST^2$ tensors
$h^T_{\mu\nu}$ we cannot directly apply eqs. (\ref{hk4}) or (\ref{hk10}), 
(\ref{hk11}). However, the heat kernel coefficients for $\xi_\mu$ and
$h^T_{\mu\nu}$ can be computed from those of the unconstrained fields using
the decompositions (\ref{p}), (\ref{f}) for arbitrary vectors
$\varepsilon_\mu$ and arbitrary symmetric tensors $h_{\mu\nu}$. 

From appendix \ref{harm} we can infer that the sets of orthonormal 
$-D^2$-eigenfunctions
\begin{eqnarray}
\label{hk23}
& &\left\{\left. T^{lm}_\mu\right|m\in\{1,\ldots,D_l(d,1)\},\;l=1,2,\ldots
\right\}\nonumber\\
\bigcup & &\left\{\left. (\Lambda_l(d,0))^{-\frac{1}{2}}\,D_\mu T^{lm}
\right|m\in\{1,\ldots,D_l(d,0)\},\;l=1,2,\ldots
\right\}
\end{eqnarray}
and
\begin{eqnarray}
\label{hk21}
& &\left\{\left. T^{lm}_{\mu\nu}\right|m\in\{1,\ldots,D_l(d,2)\},\;l=2,3,\ldots
\right\}\nonumber\\
\bigcup& & \left\{\left.\left(2\left(\Lambda_l(d,1)-\frac{R}{d}\right)
\right)^{-\frac{1}{2}}\left(D_\mu T^{lm}_\nu+D_\nu T^{lm}_{\mu}\right)
\right|m
\in\{1,\ldots,D_l(d,1)\},\;l=2,3,\ldots\right\}\nonumber\\
\bigcup& &
\Bigg\{\left(\Lambda_l(d,0)\left(\frac{d-1}{d}\,
\Lambda_l(d,0)-
\frac{R}{d}\right)\right)^{-\frac{1}{2}}
\left(D_\mu D_\nu-\frac{1}{d}g_{\mu\nu}\,D^2\right)T^{lm}
\Bigg|m\in\{1,\ldots,D_l(d,0)\},\nonumber\\
& &l=2,3,\ldots\Bigg\}\bigcup 
\left\{\frac{1}{\sqrt{d}}g_{\mu\nu}T^{lm}|m\in\{1,\ldots,D_l(d,0)\},\;l=0,1,
\ldots\right\}
\end{eqnarray}
span the spaces of {\it all} vectors $\varepsilon_\mu$ and of {\it all} 
symmetric tensors $h_{\mu\nu}$, respectively. Here the $T^{lm}$'s are the 
normalized spherical harmonics of eq. (\ref{093}). 

Now we insert these 
eigenfunctions into the trace formula (\ref{hk20}) with $f$ taken to be an 
exponential. Then we use the commutation relations of appendix 
\ref{commut} in order to pull the $D_\mu$'s from the TT-decomposition through
the exponentials and to combine them to Laplacians. This leads to the 
following decomposition for the traced heat kernels of the unconstrained 
vectors and symmetric tensors in terms of the heat kernels for the 
differentially constrained fields: 
\begin{eqnarray}
\label{hk2}
{\rm Tr}_{(1)}\left[e^{-{\rm i}{\sf t}\left(D^2+q\,R\right)}\right]=
{\rm Tr}_{(1T)}\left[e^{-{\rm i}{\sf t}\left(D^2+q\,R\right)}\right]
+{\rm Tr}_{(0)}\left[e^{-{\rm i}{\sf t}\left(D^2+\frac{dq+1}{d}R\right)}\right]
-e^{-{\rm i}{\sf t}\,\frac{dq+1}{d}R}\;,
\end{eqnarray}
\begin{eqnarray}
\label{hk1}
{\rm Tr}_{(2S)}\left[e^{-{\rm i}{\sf t}(D^2+qR)}\right]&=&
{\rm Tr}_{(2ST^2)}\left[e^{-{\rm i}{\sf t}(D^2+qR)}\right]
+{\rm Tr}_{(1T)}\left[e^{-{\rm i}{\sf t}\left(D^2+\left(\frac{d+1}{d(d-1)}
+q\right)R\right)}\right]\nonumber\\
& &+{\rm Tr}_{(0)}\left[e^{-{\rm i}{\sf t}\left(D^2+\left(\frac{2}{d-1}
+q\right)R\right)}\right]+{\rm Tr}_{(0)}\left[e^{-{\rm i}{\sf t}(D^2+qR)}
\right]
-e^{-{\rm i}{\sf t}\left(\frac{2}{d-1}+q\right)R}\nonumber\\
& &-(d+1)\,e^{-{\rm i}{\sf t}\left(\frac{1}{d-1}+q\right)R}
-\frac{d(d+1)}{2}\,e^{-{\rm i}{\sf t}\left(\frac{2}{d(d-1)}+q\right)R}\;.
\end{eqnarray}

The last term of eq. (\ref{hk2}) and the last three terms of eq. (\ref{hk1})
arise from those spherical harmonics $T^{lm}$ and $T^{lm}_\mu$ which are not 
contained in the sets of eigenfunctions (\ref{hk23}) and (\ref{hk21}). To be
more precise, the last term in eq. (\ref{hk2}) comes from the constant 
eigenmode $T^{0,1}$ of the operator $D^2+(dq+1)/d\,R$. Furthermore, the last
but second and the last but first term in eq. (\ref{hk1}) take account of the 
eigenmodes $T^{0,1}={\rm const}$ and $T^{1,m}$ of the operator $D^2+(2/(d-1)
+q)R$, respectively. As discussed in subsection \ref{2B} the $T^{1,m}$'s 
satisfy the scalar equation (\ref{010}) and are therefore in a one-to-one
correspondence with the PCKV's of $S^d$. The last term in eq. 
(\ref{hk1}) comes from the eigenmodes $T^{1,m}_\mu$ of the operator $D^2
+((d+1)/(d(d-1))+q)R$, which are the KV's of $S^d$. 

These subtraction terms 
compensate for the corresponding unphysical contributions contained in the 
{\it complete} traces for the constrained fields on the RHS of eqs.
(\ref{hk2}), (\ref{hk1}). This can be seen as follows. Consider the functional
trace of eq. (\ref{hk20}). Omitting the contributions from the modes 
$U_{k_1},\ldots,U_{k_n}$, we denote the functional trace involving only the
remaining modes with ${\rm Tr}^{\prime\ldots\prime}_{(s[C])}[f(-D^2)]$. Then 
eq. (\ref{096}) implies the following relation between ${\rm Tr}^{\prime\ldots
\prime}_{(s[C])}[f(-D^2)]$ and the {\it complete} trace ${\rm Tr}_{(s[C])}
[f(-D^2)]$:
\begin{eqnarray}
\label{097}
{\rm Tr}^{\prime\ldots\prime}_{(s[C])}[f(-D^2)]={\rm Tr}_{(s[C])}[f(-D^2)]
-\sum\limits_{k\in\{k_1,\ldots,k_n\}}f\left(\widetilde{\Lambda}_k\right)\;.
\end{eqnarray}
This rule indeed yields the last term in eq. (\ref{hk2}) and the last three 
terms in eq. (\ref{hk1}).

As the next step we insert the asymptotic expansion (\ref{hk3}) into both sides
of eqs. (\ref{hk2}) and (\ref{hk1}) and compare the coefficients of $R$. This 
leads to the following Seeley coefficients for the constrained fields:
\begin{eqnarray}
\label{hk17}
\left.{\rm tr}\,a_0\right|_{(1T)}&=&\left.{\rm tr}\,a_0\right|_{(1T)}
-\left.{\rm tr}\,a_0\right|_{(0)}\;,\nonumber\\
\left.{\rm tr}\,a_2(q)\right|_{(1T)}&=&\left.{\rm tr}\,a_2(q)\right|_{(1)}
-\left.{\rm tr}\,a_2\left(\frac{dq+1}{d}\right)\right|_{(0)}
+\frac{1}{2}\delta_{d,2}\,R\;,\nonumber\\
\left.{\rm tr}\,a_4(q)\right|_{(1T)}&=&\left.{\rm tr}\,a_4(q)\right|_{(1)}
-\left.{\rm tr}\,a_4\left(\frac{dq+1}{d}\right)\right|_{(0)}
+\frac{1}{4}\delta_{d,2}
\left(1+2q\right)\,R^2+\frac{1}{24}\delta_{d,4}\,R^2\;,
\end{eqnarray}
\begin{eqnarray}
\label{hk18}
\left.{\rm tr}\,a_0\right|_{(2ST^2)}&=&\left.{\rm tr}\,a_0\right|_{(2S)}
-\left.{\rm tr}\,a_0\right|_{(1T)}
-2\left.{\rm tr}\,a_0\right|_{(0)}\;,\nonumber\\
\left.{\rm tr}\,a_2(q)\right|_{(2ST^2)}
&=&\left.{\rm tr}\,a_2(q)\right|_{(2S)}-\left.{\rm tr}\,
a_2\left(\frac{d+1}{d(d-1)}+q\right)\right|_{(1T)}-\left.{\rm tr}\,
a_2\left(\frac{2}{d-1}+q\right)\right|_{(0)}\nonumber\\
& &-\left.{\rm tr}\,a_2(q)\right|_{(0)}+\frac{7}{2}\delta_{d,2}\,R\;,
\nonumber\\
\left.{\rm tr}\,a_4(q)\right|_{(2ST^2)}
&=&\left.{\rm tr}\,a_4(q)\right|_{(2S)}
-\left.{\rm tr}\,a_4\left(\frac{d+1}{d(d-1)}+q\right)\right|_{(1T)}
-\left.{\rm tr}\,a_4\left(\frac{2}{d-1}+q\right)\right|_{(0)}\nonumber\\
& &-\left.{\rm tr}\,a_4(q)\right|_{(0)}+\delta_{d,2}\left(
4+\frac{7}{2}q\right)\,R^2+\frac{2}{3}\delta_{d,4}\,R^2\;.
\end{eqnarray}
The terms proportional to the $\delta$'s originate from the subtraction terms 
on the RHS of eqs. (\ref{hk2}), (\ref{hk1}) which are due to the unphysical 
eigenmodes. 
These terms have an expansion of the form $\sum\limits_{m=0}^\infty 
b_{2m}r^{-2m}$, while the terms of the heat kernel expansion are of the form
 $\int d^dx\sqrt{g}\,{\rm tr}\,a_n\propto r^{d-n}$. Comparing powers of 
$R\propto 1/r^2$, only under the condition $-2m=d-n$ a given term 
$b_{2m}r^{-2m}$ contributes to $\int d^dx\sqrt{g}\,{\rm tr}\,a_n$. 
Hence for $n$, $m$ fixed, the Seeley coefficients ${\rm tr}\,a_n$ for the 
differentially constrained fields receive a contribution from a term of the 
form $b_{2m}r^{-2m}/(\int d^dx\sqrt{g})$ at most for one specific value of the 
dimensionality $d$. In particular, 
the subtraction terms in eqs. (\ref{hk2}), (\ref{hk1}) do not contribute to 
${\rm tr}\,a_0$, while ${\rm tr}\,a_2$ and ${\rm tr}\,a_4$ on the LHS of eqs. 
(\ref{hk17}), (\ref{hk18}) receive contributions from terms of the form 
$\delta_{d,2}\,b_0/r^2$ and 
$\delta_{d,2}\,b_2/r^4$, $\delta_{d,4}\,b_0/r^4$, respectively.

The matrix traces on the RHS of eqs. (\ref{hk17}), (\ref{hk18}) can now be
evaluated by using the heat kernel coefficients for the (differentially)
unconstrained fields. For scalars we have ${\rm tr}\,a_n(q)=a_n(q)$ 
(${\bf 1}_{(0)}=1$). For vectors the traces are evaluated according to
${\rm tr}\,a_n(q)=g^{\mu\nu}[a_n(q)]_{\mu\nu}$ so that we obtain from eqs. 
(\ref{hk4}), (\ref{hk10}) 
\begin{eqnarray}
\label{hk12}
\left.{\rm tr}\,a_0\right|_{(1)}&=&d\nonumber\\
\left.{\rm tr}\,a_2(q)\right|_{(1)}&=&\frac{1+6q}{6}d\,R\nonumber\\
\left.{\rm tr}\;a_4(q)\right|_{(1)}&=&\frac{1}{360}
\left[\frac{5d^3-7d^2+6d-60}{d(d-1)}+60dq+180dq^2\right]\,R^2\;.
\end{eqnarray}

In order to determine ${\rm tr}\,a_n$ for symmetric tensor fields we have to
symmetrize the heat kernel for unconstrained rank-2 tensors according to
\begin{eqnarray}
\label{hk15}
& &\left.\left<x\left|e^{-{\rm i}s\left(D^2+q R\right)}\right|x
\right>^{\mu\nu\rho\sigma}\right|_{(2S)}
=\frac{1}{4}\Bigg\{
\left.\left<x\left|e^{-{\rm i}s\left(D^2+q R\right)}\right|x\right>^{\mu\nu
\rho\sigma}\right|_{(2)}
+\left.\left<x\left|e^{-{\rm i}s\left(D^2+q R\right)}\right|x\right>^{\nu\mu
\rho\sigma}\right|_{(2)}\nonumber\\
& &+\left.\left<x\left|e^{-{\rm i}s\left(D^2+q R\right)}\right|x\right>^{\mu\nu
\sigma\rho}\right|_{(2)}
+\left.\left<x\left|e^{-{\rm i}s\left(D^2+q R\right)}\right|x\right>^{\nu\mu
\sigma\rho}\right|_{(2)}\Bigg\}
\end{eqnarray}
before we can apply eq. (\ref{hk3}) with eqs. (\ref{hk4}) and (\ref{hk11}). 
This leads to
\begin{eqnarray}
\label{hk13}
\left.\left[a_n(q)\right]_{\mu\nu\rho\sigma}\right|_{(2S)}&=&
\frac{1}{4}\Big(\left.\left[a_n(q)\right]_{\mu\nu\rho\sigma}\right|_{(2)}
+\left.\left[a_n(q)\right]_{\nu\mu\rho\sigma}\right|_{(2)}\nonumber\\
& &+\left.\left[a_n(q)\right]_{\mu\nu\sigma\rho}\right|_{(2)}
+\left.\left[a_n(q)\right]_{\nu\mu\sigma\rho}\right|_{(2)}\Big)
\end{eqnarray}
and in particular $({\bf 1}_{(2S)})_{\mu\nu\rho\sigma}=
\left(g_{\mu\rho}g_{\nu\sigma}+g_{\mu\sigma}g_{\nu\rho}\right)/2$. For tensors
the matrix traces are computed according to  ${\rm tr}\,a_n=g^{\mu\rho}
g^{\nu\sigma}[a_n(q)]_{\mu\nu\rho\sigma}$ which yields
\begin{eqnarray}
\label{hk16}
\left.{\rm tr}\,a_0\right|_{(2S)}&=&\frac{1}{2}d(d+1)\nonumber\\
\left.{\rm tr}\,a_2(q)\right|_{(2S)}&=&\frac{1+6q}{12}d(d+1)\,R\nonumber\\
\left.{\rm tr}\,a_4(q)\right|_{(2S)}&=&\frac{1}{720}
\left[\frac{5d^4-2d^3-d^2-114d-240}{d(d-1)}
+60q(1+3q)d(d+1)\right]\,R^2\;.
\end{eqnarray}

Finally we insert the matrix traces eqs. (\ref{hk12}), (\ref{hk16}) and 
$\left.{\rm tr}\,a_n\right|_{(0)}=\left.a_n\right|_{(0)}$ into eqs. 
(\ref{hk17}), (\ref{hk18}) and determine the heat kernel coefficients for 
transverse vectors and $ST^2$ tensors. The results are summarized in table 2.

Let us add a final remark concerning the applicability of the asymptotic
expansion (\ref{hk22}). Since it is valid only in the limit
$|{\sf t}|\rightarrow 0$ it is clear that it cannot be integrated 
over ${\rm Re}({\sf t})=s$ or ${\rm Im}({\sf t})=\varepsilon$ term by term,
in general. However, this is possible if the heat kernel is integrated against
a ``test'' function which suppresses large values of $s$ or $\varepsilon$. 
This is indeed the case for our application of the asymptotic expansion 
presented in subsection \ref{trace}. 
\vspace{1cm}

\begin{tabular}{|c|c|c|c|}
\hline\multicolumn{4}{|c|}{Table 2: Heat Kernel Coefficients}\\
\hline\rule[-4mm]{0mm}{10mm} field &${\rm tr}\;a_0$&${\rm tr}\;a_2(q)$&
${\rm tr}\;a_4(q)$\\\hline
\raisebox{-0.8ex}[0.8ex]{$ST^2$}& & 
$\frac{(d+1)(d+2)(d-5+3\,\delta_{d,2})}{12(d-1)}R$ & 
\rule[5mm]{0mm}{2mm}$\frac{(d+1)(5d^4-22d^3-83d^2-392d-228+1440\,\delta_{d,2}
+3240\,\delta_{d,4})}
{720d(d-1)^2}R^2$\\
\raisebox{0.8ex}[-0.8ex]{tensor}&\raisebox{2.2ex}[-2.2ex]
{$\frac{(d-2)(d+1)}{2}$}& +$q\,\frac{(d-2)(d+1)}{2}R$&
\rule[-4mm]{0mm}{10mm}$+q\,\frac{(d+1)(d+2)(d-5+3\,\delta_{d,2})}{12(d-1)}R^2
+q^2\,\frac{(d-2)(d+1)}{4}R^2$\\
\hline \raisebox{-0.8ex}[0.8ex]{$T$} & &
\raisebox{-0.4ex}[0.4ex]{$\frac{(d+2)(d-3)+6\,\delta_{d,2}}{6d}R$} & 
\rule[5mm]{0mm}{2mm}$\frac{5d^4-12d^3-47d^2-186d+180+360\,\delta_{d,2}
+720\,\delta_{d,4}}{360d^2(d-1)}R^2$  \\
\raisebox{0.8ex}[-0.8ex]{vector} & \raisebox{2.2ex}[-2.2ex]
{$d-1$} &\raisebox{0.4ex}[-0.4ex]{$+q(d-1)R$} &
$\rule[-4mm]{0mm}{10mm}+q\,\frac{(d+2)(d-3)+6\,\delta_{d,2}}{6d}R^2+q^2\,
\frac{d-1}{2}R^2$  \\
\hline \rule[-9mm]{0mm}{19mm}scalar& $1$ &$\frac{1+6q}{6}R$&$\frac{5d^2-7d+6}
{360d(d-1)}R^2+\frac{1}{6}q\,R^2+\frac{1}{2}q^2\,R^2$\\
\hline\end{tabular}

\vspace{1cm}
\section{Various coefficients}
\renewcommand{\theequation}{F\arabic{equation}}
\setcounter{equation}{0}
\label{coeff}
In this appendix we define the coefficient functions which appear in eqs. 
(\ref{44})-(\ref{55}) of subsection \ref{4C} and in eq. (\ref{100})
of appendix \ref{inv}.
\begin{eqnarray}
\label{40}
& &A_T(d)\equiv\frac{d(d-3)+4}{d(d-1)}\;,\;\;
A_V(d,\alpha)\equiv\frac{\alpha(d-2)-1}{d}\;,\;\;
A_{S1}(d,\alpha)\equiv\frac{\alpha(d-4)}{2\alpha(d-1)-(d-2)}\;,\nonumber\\
& &A_{S2}(d,\alpha)\equiv-\frac{\alpha(d-2)-2}{\alpha(d-2)-2(d-1)}\;,\;\;
A_{S3}(d)\equiv\frac{d-4}{d}\;,\;\;
A_{S4}(d,\alpha)\equiv\frac{\alpha(d-2)-2}{d}\;,\nonumber\\
& &A_{S5}\equiv\frac{(d-2)(d+2)\alpha^2+(d^2-10d+8)\alpha+2(d-2)}{d^2\alpha}
\;,\nonumber\\
& &B_{S1}(d,\alpha)\equiv-\frac{2\alpha d}{2\alpha(d-1)-(d-2)}\;,\;\;
B_{S2}(d,\alpha)\equiv\frac{2\alpha d}{\alpha(d-2)-2(d-1)}\;,\nonumber\\
& &C_{S1}(d,\alpha)\equiv-\frac{2\alpha(d-1)-(d-2)}{4(d-1)-2\alpha (d-2)}
\frac{d-2}{d-1}\;,\;\;
C_{S2}(d,\alpha)\equiv\frac{d-1}{d^2}\frac{2(d-1)-\alpha(d-2)}{\alpha}\;,
\nonumber\\
& &C_{S3}(d,\alpha)\equiv\frac{(d-2)(\alpha-1)}{\alpha (d-2)-2(d-1)}\;,\;\;
E_S(d,\alpha)\equiv-\frac{d-2}{4\alpha d^2}\left(2\alpha(d-1)-(d-2)\right)
\;,\nonumber\\
& &F_{S1}(d,\alpha)\equiv-\frac{2\left[\alpha(d-2)-2(d-1)\right]
\left[2\alpha(d-1)-(d-2)\right]}{d^2\alpha}\;,\nonumber\\
& &F_{S2}(d,\alpha)\equiv\frac{4(d-2)(d-1)(\alpha-1)^2}{d^2\alpha}
\end{eqnarray}

\section{Commutation relations for a maximally symmetric background} 
\renewcommand{\theequation}{G\arabic{equation}}
\setcounter{equation}{0}
\label{commut}
In the following we summarize the commutation relations which were used in 
order to derive eq. (\ref{44}) of subsection \ref{4C}, and eqs. 
(\ref{hk2}) and (\ref{hk1}) of appendix \ref{heat}. They are valid for
the class of maximally symmetric backgrounds.
\begin{eqnarray}
\label{ApG1}
& &\bar{D}_\mu\bar{D}_\nu\widehat{\xi}^\mu=\frac{\bar{R}}{d}\widehat{\xi}_\nu\\
& &\bar{D}^2\left(\bar{D}_\mu\widehat{\xi}_\nu+\bar{D}_\nu\widehat{\xi}_\mu
\right)=\bar{D}_\mu\left(\bar{D}^2+\frac{(d+1)\bar{R}}{d(d-1)}\right)
\widehat{\xi}_\nu+\bar{D}_\nu\left(\bar{D}^2+\frac{(d+1)\bar{R}}{d(d-1)}\right)
\widehat{\xi}_\mu\\
& &\left(\bar{D}_\mu\widehat{\xi}_\nu+\bar{D}_\nu\widehat{\xi}_\mu\right)
\left(\bar{D}^\mu\widehat{\xi}^\nu+\bar{D}^\nu\widehat{\xi}^\mu\right)
=-2\widehat{\xi}_\mu\left(\bar{D}^2+\frac{\bar{R}}{d}\right)\widehat{\xi}^\mu
+{\rm cov.\;divergence}\\
& &\bar{D}^2\bar{D}_\mu\widehat{\sigma}=\bar{D}^\nu\bar{D}_\mu\bar{D}_\nu
\widehat{\sigma}
=\bar{D}_\mu\left(\bar{D}^2+\frac{\bar{R}}{d}\right)\widehat{\sigma}\\
& &\bar{D}^2\left(\bar{D}_\mu\bar{D}_\nu-\frac{1}{d}\bar{g}_{\mu\nu}
\bar{D}^2\right)\widehat{\sigma}=\left(\bar{D}_\mu\bar{D}_\nu-\frac{1}{d}
\bar{g}_{\mu\nu}\bar{D}^2\right)\left(\bar{D}^2+\frac{2\bar{R}}{d-1}
\right)\widehat{\sigma}\\
& &\left(\bar{D}_\mu\bar{D}_\nu\widehat{\sigma}-\frac{1}{d}\bar{g}_{\mu\nu}
\bar{D}^2\widehat{\sigma}\right)
\left(\bar{D}^\mu\bar{D}^\nu\widehat{\sigma}-\frac{1}{d}\bar{g}^{\mu\nu}
\bar{D}^2\widehat{\sigma}\right)\nonumber\\
& &\;\;\;\;\;=\frac{d-1}{d}\widehat{\sigma}\bar{D}^2\left(\bar{D}^2+
\frac{\bar{R}}{d-1}\right)\widehat{\sigma}+{\rm cov.\;divergence}\\
& &\left(\bar{D}_\mu\widehat{\xi}_\nu+\bar{D}_\nu\widehat{\xi}_\mu\right)
\exp\left(\bar{D}^2\right)\left(\bar{D}^\mu\widehat{\xi}^\nu+\bar{D}^\nu
\widehat{\xi}^\mu\right)\nonumber\\
& &\;\;\;\;\;=-2\widehat{\xi}_\mu\left(\bar{D}^2+\frac{\bar{R}}{d}\right)
\exp\left(\bar{D}^2+\frac{(d+1)\bar{R}}{d(d-1)}\right)\,\widehat{\xi}^\mu
+{\rm cov.\;divergence}\\
& &\left(\bar{D}_\mu\widehat{\sigma}\right)\,\exp\left(\bar{D}^2\right)\,
\bar{D}^\mu\widehat{\sigma}=-\widehat{\sigma}\,\bar{D}^2\,\exp\left(\bar{D}^2
+\frac{\bar{R}}{d}\right)\,\widehat{\sigma}+{\rm cov.\;divergence}\\
& &\left[\left(\bar{D}_\mu\bar{D}_\nu-\frac{1}{d}\bar{g}_{\mu\nu}
\bar{D}^2\right)\widehat{\sigma}\right]\exp\left(\bar{D}^2\right)
\left(\bar{D}^\mu\bar{D}^\nu
-\frac{1}{d}\bar{g}^{\mu\nu}\bar{D}^2\right)\widehat{\sigma}\nonumber\\
& &\;\;\;\;\;=\frac{d-1}{d}\widehat{\sigma}\bar{D}^2\left(\bar{D}^2+
\frac{\bar{R}}{d-1}\right)\exp\left(\bar{D}^2+\frac{2\bar{R}}{d-1}\right)\,
\widehat{\sigma}+{\rm cov.\;divergence}
\end{eqnarray}
\section{Approximate solutions for the fixed point} 
\renewcommand{\theequation}{H\arabic{equation}}
\setcounter{equation}{0}
\label{approx}
In the following we determine the approximate formula for the position of the 
non-Gaussian fixed point discussed in subsection \ref{5C}. 
In a first approximation
we set $\lambda_k=\lambda_*=0$ and determine $g_*$ from the condition 
$\eta_{N*}=2-d$ alone. Solving this equation for $g_*$ leads to
\begin{eqnarray}
\label{H0}
g_*=\frac{2-d}{B_1(\lambda_*;\alpha,d)-(d-2)B_2(\lambda_*;\alpha,d)}
\end{eqnarray}
which, for $\lambda_*=0$, boils down to
\begin{eqnarray}
\label{H1}
g_*&=&\frac{2-d}{4}(4\pi)^{\frac{d}{2}-1}\Bigg\{k_1(d)\,\Phi^1_{d/2-1}(0)
+k_2(d)\,\widetilde{\Phi}^1_{d/2-1}(0)+k_3(d,\alpha)\,\Phi^2_{d/2}(0)
\nonumber\\
& &+k_4(d,\alpha)\,\widetilde{\Phi}^2_{d/2}(0)+3\delta_{d,2}\left[
\frac{1}{1-2\lambda_k}-\frac{1}{1-2\alpha\lambda_k}\right]\Bigg\}^{-1}\;.
\end{eqnarray}
Here $k_1,\ldots,k_4$ are $d$- and $\alpha$-dependent coefficients defined as
\begin{eqnarray}
\label{H2}
& &k_1(d)=\frac{d^4-4d^3-9d^2-12}{12d(d-1)}\;,\;\;
k_2(d)=\frac{(d-2)(d^4-13d^2-24d+12)}{24d(d-1)}\;,\nonumber\\
& &k_3(d,\alpha)=-\frac{d^4-4d^3+9d^2-8d-2}{2d(d-1)}-(d-2)\alpha\;,\nonumber\\
& &k_4(d,\alpha)=-\frac{(d-2)(d^4-4d^3+5d^2-8d+2)}{4d(d-1)}-\frac{(d-2)^2}{2}
\alpha\;.
\end{eqnarray}
Employing the exponential shape function (\ref{H6}) with $s=1$, and setting 
$d=4$ and $\alpha=1$, for instance, eq. (\ref{H1}) yields $g_*\approx 0.590$.
Here we used that for this shape function $\Phi^1_1(0)=\pi^2/6$,
$\Phi^2_2(0)=1$, $\widetilde{\Phi}^1_1(0)=1$,
$\widetilde{\Phi}^2_2(0)=1/2$.

As a different approximation scheme, we determine $(\lambda_*,g_*)$ from a 
set of Taylor-expanded $\mbox{\boldmath $\beta$}$-functions. Using 
\begin{eqnarray}
\label{gfp1}
\frac{d}{d\lambda_k}\Phi^p_n(-2\alpha\lambda_k)&=&2\alpha p\,\Phi^{p+1}_n(-2
\alpha\lambda_k)\;,\nonumber\\
\frac{d}{d\lambda_k}\widetilde{\Phi}^p_n(-2\alpha\lambda_k)&=&2\alpha p\,
\widetilde{\Phi}^{p+1}_n(-2\alpha\lambda_k)\;,
\end{eqnarray}
we expand the $\mbox{\boldmath $\beta$}$-functions (\ref{001}) and (\ref{004})
about $g_k=\lambda_k=0$ and obtain
\begin{eqnarray}
\label{gfp2}
\mbox{\boldmath $\beta$}_\lambda(\lambda_k,g_k;\alpha,d)&=&-2\lambda_k+\nu_d
\,d\;g_k+\left[2d(d-1+2\alpha)(4\pi2)^{1-\frac{d}{2}}\,\Phi^2_{d/2}(0)-(d-2)\,
\omega_d\right]\lambda_k\,g_k\nonumber\\
& &+\frac{1}{2}d(d+1)(d-2)(4\pi)^{1-\frac{d}{2}}\omega_d\,
\Phi^{1}_{d/2}(0)\,g_k^2+{\cal O}\left({\rm g}^3\right)\;,\nonumber\\
\mbox{\boldmath $\beta$}_g(\lambda_k,g_k;\alpha,d)&=&(d-2)\,g_k-(d-2)
\omega_d\,g_k^2+{\cal O}\left({\rm g}^3\right)\;.
\end{eqnarray}
Here $\nu_d$ and $\omega_d$ are defined as in eqs. (\ref{gfp3}) and 
(\ref{gfp13}), and ${\cal O}\left({\rm g}^3\right)$ stands for terms of third
and higher orders in the couplings ${\rm g}_1(k)=\lambda_k$ and ${\rm g}_2(k)
=g_k$. Now $g_*$ is obtained as the nontrivial solution to $\mbox{\boldmath
$\beta$}_g=0$, which reads
\begin{eqnarray}
\label{H4}
g_*=\omega_d^{-1}=\frac{2-d}{4}(4\pi)^{\frac{d}{2}-1}\left\{k_1(d)\,
\Phi^1_{d/2-1}(0)+k_3(d,\alpha)\,\Phi^2_{d/2}(0)\right\}^{-1}\;.
\end{eqnarray}
Inserting eq. (\ref{H4}) into $\mbox{\boldmath $\beta$}_\lambda$ and 
neglecting also the terms quadratic in the couplings the condition
$\mbox{\boldmath $\beta$}_\lambda=0$ leads to
\begin{eqnarray}
\label{H5}
\lambda_*=\frac{\nu_d\,d}{2\omega_d}=-\frac{d(d-2)(d-3)}{8}\,
\Phi^1_{d/2}(0)\left\{k_1(d)\,
\Phi^1_{d/2-1}(0)+k_3(d,\alpha)\,\Phi^2_{d/2}(0)\right\}^{-1}\;.
\end{eqnarray}
Using the shape function (\ref{H6}) with $s=1$ we obtain from eqs. (\ref{H4})
and (\ref{H5}) in $d=4$ dimensions
\begin{eqnarray}
\label{H7}
g_*&=&\left(\frac{13\pi}{144}+\frac{55}{24\pi}+\frac{\alpha}{\pi}\right)^{-1}
\;,\nonumber\\
\lambda_*&=&\zeta(3)\left(\frac{13\pi^2}{144}+\frac{55}{24}+\alpha\right)^{-1}
\;,
\end{eqnarray}
which yields $(\lambda_*,g_*)=(0.287,0.751)$ for $\alpha=1$, for instance.
\end{appendix}

\end{document}